\documentclass{aa}
\usepackage{psfig}
\usepackage{graphicx}
\usepackage{amsmath}
\usepackage{amssym}
\usepackage{psfrag}
\usepackage{natbib}
\bibpunct{(}{)}{;}{a}{}{,}

\def\M21{\hbox{Mrk~421} }

\def\etal{et al.\/ }

\def\ltsima{$\; \buildrel < \over \sim \;$}
\def\simlt{\lower.5ex\hbox{\ltsima}}            % < over ~
\def\gtsima{$\; \buildrel > \over \sim \;$}
\def\simgt{\lower.5ex\hbox{\gtsima}}            % > over ~
\def\eg{e.g., }
\def\ie{i.e., }
\def\sun{\odot}
\def\ms{M_{\sun}}
\def\ergsec{erg\,s$^{-1}$}
\def\grcm3{g\,cm$^{-3}$}
\def\gra{^{\circ}}
\newcommand{\e}[1]{\cdot 10^{#1}}

\long\def\symbolfootnote[#1]#2{\begingroup%
\def\thefootnote{\fnsymbol{footnote}}\footnote[#1]{#2}\endgroup}

\begin{document}
\title{Relativistic outflows from remnants of compact object mergers
and their viability for short gamma-ray bursts}
\author{M.A. Aloy\inst{1}, H.-T. Janka\inst{1} and E. M\"uller\inst{1}}
\offprints{MAA, e-mail: maa@mpa-garching.mpg.de}
\institute{Max--Planck--Institut f\"ur Astrophysik,
  Postfach 1317, D-85741 Garching, Germany}
\date{Received ?; accepted ?}
 \abstract{
We present the first general relativistic hydrodynamic models of the
launch and evolution of relativistic jets and winds, driven by thermal
energy deposition, possibly due to neutrino--antineutrino
annihilation, in the close vicinity of black hole--accretion torus
systems. The latter are considered to be the remnants of compact
object mergers.  Our two-dimensional simulations establish the link
between models of such mergers and future observations of short
gamma-ray bursts by the SWIFT satellite. They show that
ultrarelativistic outflow with maximum terminal Lorentz factors around
1000 develops for polar energy deposition rates above some
$10^{48}\,$erg$\,$s$^{-1}$ per steradian, provided the merger
environment has a sufficiently low baryon density. By the interaction
with the dense accretion torus the ultrarelativistic outflow with
Lorentz factors $\Gamma$ above 100 is collimated into a sharp-edged
cone that is embedded laterally by a wind with steeply declining
Lorentz factor. The typical semi-opening angles of the $\Gamma > 100$
cone are $5\gra-10\gra$, corresponding to about $0.4 - 1.5\%$ of the
hemisphere and apparent isotropized energies (kinetic plus internal)
up to $\approx 10^{51}\,$erg although at most $10 - 30\%$ of the
deposited energy is transferred to the outflow with $\Gamma >
100$. The viability of post-merger black hole--torus systems as
engines of short, hard gamma-ray bursts is therefore confirmed. The
annihilation of neutrino--antineutrino pairs radiated from the hot
accretion torus appears as a suitable energy source for powerful axial
outflow even if only $\approx 10^{49}\,$erg are deposited within a
cone of $45\gra$ half-opening angle around the system axis. Although
the torus lifetimes are expected to be only between some 0.01 seconds
and several 0.1 seconds, our models can explain the durations of all
observed short gamma-ray bursts, because different propagation
velocities of the front and rear ends will lead to a radial stretching
of the ultrarelativistic fireball before transparency is reached.  The
ultrarelativistic flow reveals a highly non-uniform structure caused
by the action of Kelvin-Helmholtz instabilities that originate at the
fireball-torus interface. Large radial variations of the baryon
density (up to several orders of magnitude) are uncorrelated with
moderate variations of the Lorentz factor (factors of a few) and
fluctuations of the gently declining radiation-dominated pressure. In
the angular direction the Lorentz factor reveals a nearly flat
plateau-like maximum with values of several hundreds, that can be
located up to $7\gra$ off the symmetry axis, and a steep decrease to
less than 10 for polar angles larger than $15\gra$--$20\gra$. Lateral
expansion of the ultrarelativistic core of the flow is prevented by a
subsonic velocity component of about $0.05c$ towards the symmetry
axis, whereas the moderately relativistic wings show a subsonic
sideways inflation with less than $0.07c$ (measured in the frame
comoving with the radial flow).
}

\titlerunning{Production of short GRBs}
\authorrunning{Aloy, Janka and M\"uller}
\maketitle

\section{Introduction}
\label{sec:introduction}

Gamma-ray bursts (GRBs) are transient astrophysical events which
are characterized by the release of enormous amounts of energy in
non-thermal, highly variable (in time and event by
event) emission on timescales between some milliseconds and about a
thousand seconds. The distribution of burst durations reveals two 
classes, ``short bursts'' which last less than about
2 seconds and have typical duration of 0.3--0.5 seconds,
and ``long bursts'' with typically softer spectra and a mean
durations of about 30--60 seconds \citep{Ketal93}. 

So far only long bursts could be sufficiently well localized on the
sky to enable multi-wavelength observations.  In many such cases
afterglow emission was detected, the host galaxies were discovered,
and the associated measurement of spectral line redshifts confirmed
the cosmological distances to the burst sources
\citep[e.g.,][]{CX97,PX98,PG97,BK98,FK97}. Observational evidence also
indicates that the bursts originate from star formation regions,
suggesting a possible link to massive star explosions. This idea was
nourished by several cases where a bump-like re-brightening phase of
the decaying afterglow emission was interpreted as a superimposed
supernova lightcurve. The link was directly confirmed by two cases
where the spatial and temporal coincidence of a GRB with a peculiar
kind of supernova explosion was found, GRB980425 with SN1998bw
\citep{GV98}, which was still ambiguous, and GRB030329 with SN2003dh
\citep{Stanek03,Hjorth03}. For short bursts such constraining
observations do not exist.

According to the most widely accepted theoretical concept GRBs are
understood as the consequence of an explosive release of energy
associated with the accretion of matter at huge rates (up to a several
solar masses per second) onto stellar-mass black holes, e.g., as a
consequence of the birth of such a black hole (BH) by the catastrophic
collapse of a massive star in a so-called ``collapsar'' event
\citep{Wo93,MW99,Alo00,MWH01} or by a merger of compact objects in a
close binary \citep{Pa86,Go86,Ei89,MH93}.  The nascent black hole is
initially girded by a thick gas torus from which it swallows matter at
a hypercritical rate, i.e., at a rate far in excess of the Eddington
limit. In this case radiation is advected inward with the accretion
flow and the cooling is dominated by the emission of neutrinos
\citep{PWF99}, a situation for which the term ``neutrino-dominated
accretion disk'' (NDAF) was coined \citep{NPK01}. The annihilation of
neutrino-antineutrino pairs in the immediate vicinity of the black
hole-torus system is considered to create an e$^+$e$^-$-pair
plasma-photon fireball which could power an ultrarelativistic outflow
of baryons (with typical Lorentz factors $> 10^2$) provided the baryon
loading remains sufficiently low \citep{CR78}. Alternatively,
magnetohydrodynamic processes might accelerate matter to
ultrarelativistic velocities by tapping the gravitational binding
energy of the accretion flow \cite[e.g.,][]{Drenkhahn02,DS02} or the
rotational energy of the black hole \citep{BZ77}. When these highly
relativistic ejecta finally reach optically thin conditions at much
larger radii ($\sim 10^{14}\,$cm) the kinetic energy of the outflow is
dissipated and the observable GRB is produced mainly by synchrotron
radiation and inverse Compton scattering (e.g., see the recent review
by \citealp{Me02}).

A large number of theoretical studies focused on the physics of the
relativistic fireball and the formation of the observable properties
of the GRB and of the afterglow emission \citep[for reviews, see
e.g.,][]{Me02,Piran99}, assuming that the energy release by some
compact central engine has been able to produce the ultrarelativistic
outflow. However, only few of the suggested astrophysical scenarios or
concepts have so far been investigated by detailed numerical
simulations which attempt to develop a quantitative picture of the
physical processes that play a role at the source of the energy and
that are responsible for driving the highly relativistic ejection of
matter.

Hypercritical steady-state accretion from accretion disks and tori was
first discussed with semi-analytic models by \cite{PWF99}, and later
reconsidered by \cite{NPK01}, \cite{DMPN02}, and \cite{KM02}.  These
works have brought a theoretical understanding of the conditions that
are present in the vicinity of the accreting BH and which determine
the efficiency of energy loss by neutrino emission and the efficiency
of energy conversion by neutrino-antineutrino annihilation. To this
end, three-dimensional hydrodynamic simulations were performed by
\cite{RJ99} and more recently by \cite{SRJ04} for the {\em
time-dependent} accretion in BH-tori systems as the remnants of
neutron star--neutron star (NS+NS) and NS+BH mergers. In this case the
torus is not fed by an external mass reservoir as in case of a
collapsar, and the timescale of BH accretion is determined by viscous
transport in the torus, not by the collapse timescale of a massive
star.  The torus is very compact. Typically it contains a few
hundredth up to some 0.1 solar masses and extends to 15--20
Schwarzschild radii in the equatorial plane and about half that
distance in the vertical direction. Its density and accretion rate are
therefore high so that the torus can become partly opaque to
neutrinos. With neutrino luminosities in excess of
$10^{53}\,$erg$\,$s$^{-1}$ $\nu\bar\nu$-annihilation can deposit
energy in the close vicinity of the black hole at rates between
several $10^{50}\,$erg$\,$s$^{-1}$ and more than
$10^{51}\,$erg$\,$s$^{-1}$ \citep{RJ99,Jetal99}.

The early phase of the propagation of jets in collapsars has been
studied by nonrelativistic \citep{MW99,MWH01} as well as special
\citep{ZWM03} or general relativistic \citep{Alo00,AM02} hydrodynamic
simulations in two dimensions (i.e., assuming axial symmetry). Also
first axisymmetric magnetohydrodynamic simulations have been done for
the formation of polar outflows from the central non-rotating BH in a
collapsar in both Newtonian (\citealp{Petal03}; using a
pseudo-Newtonian potential, \citealp{PW80}) and general relativity
(\citealp{Mizuetal04}; assuming a fixed Schwarzshild
metric). Numerical modeling of compact object mergers with some
realism of the microphysics input (equation of state, neutrino
processes) has made progress in case of NS+NS binaries
\citep{RJS96,Retal97,RJ01,RL03,RRD03} and NS+BH systems
\citep{Jetal99,JR02,RSW04}, but detailed hydrodynamic simulations of
the formation of the relativistic outflow from the accreting
post-merging BH have not been undertaken prior to the present work.

In this paper we present results of the first general relativistic (a
fix Schwarzschild metric is assumed), axisymmetric models for the
acceleration and expansion of ultrarelativistic jets and winds, which
originate from the pair-plasma-photon fireball that is produced by the
deposition of thermal energy above the poles of a stellar-mass BH that
accretes matter from a thick accretion torus as obtained in NS+NS or
NS+BH merger simulations \citep{RJ99,Jetal99}. We use the
high-resolution shock-capturing code GENESIS \citep{Alo99a} to
integrate the general relativistic hydrodynamic equations in ``2.5
dimensions'' (i.e., in spherical ($r, \theta$) coordinates assuming
that there is no azimuthal ($\phi$) dependence of physical quantities
although the $\phi$-components of vectors can have non-zero values).

Our simulations attempt to answer questions about the collimation
mechanism of the polar outflow, the opening angle of the
ultrarelativistic ejecta, the maximum and mean Lorentz factors that
can be attained, the internal structure of the outflow, the duration
of a possible GRB event, and the isotropic equivalent energy which an
observer would infer by assuming the source to expand isotropically.

These questions may at most be guessed but they cannot be reliably
answered on grounds of merger models and a consideration of their
energy release by neutrino emission and the subsequent conversion of
some of this energy by $\nu\bar\nu$-annihilation to e$^+$e$^-$-pairs
\citep{RJ99,Jetal99,RR02,RRD03}. We stress here that the fraction of
the latter energy which ends up in ultrarelativistic outflow with low
baryon loading {\em cannot} be deduced or ``extrapolated'' from known
spatial distributions of mass and $\nu\bar\nu$-energy deposition at
any instant of a merger or accretion simulation without
self-consistently time-dependent hydrodynamic modeling. The
relativistic outflow develops in a complex hydrodynamic interaction
with the accretion torus, cleaning its own axial funnel such that
later energy deposition encounters a much reduced baryon pollution.

Answers to the questions mentioned above are of crucial importance for
discussing compact binary mergers as potential sources of observed
short GRBs and the observational properties of those.  Current work on
these issues \citep[e.g.,][]{RR03} is much handicapped by the lack of
quantitative information from detailed numerical
simulations. Conclusions in the latter paper, for example, depend
sensitively on the collimation of the ultrarelativistic outflow. In
this context work by \cite{LE00} is referred to, who suggest a
hydrodynamic collimation of the relativistic fireball by a surrounding
baryonic wind emanating from the accretion torus and calculate the
corresponding opening angle by analytic solutions. The discussion by
these authors, however, is based on a number of simplifications and
assumptions and therefore awaits verification by more complete
models. Our simulations, for example, reveal that the vertical
extension of the thick accretion torus cannot be ignored but plays a
decisive role with respect to outflow collimation, baryon entrainment
and structure development by Kelvin-Helmholtz instabilities at the
jet-torus interface.

Our paper is organized as follows. In Sect.~2 we shall describe in
some more detail our model setup, the initial conditions, the
numerical methods, and the parameters and assumptions which are varied
systematically in our studies. Section~3 will provide a discussion of
our results for two types of model sequences with different
assumptions about the conditions for the deposition of thermal energy
in the vicinity of the BH-torus system. We distinguish ``type-A''
models (in Sect.~3.1) with fairly high density in the surroundings of
the torus, and ``type-B'' models with low environmental density
(Sect.~3.2). In Sect.~\ref{sec:switchoff} we shall also consider the
evolution of type-B models for a period of 0.5 seconds after the onset
of the energy release. In some of these models the energy deposition
was taken constant up to 0.1 seconds and then switched off. In other
models the energy deposition was modeled more ``burst-like'', reaching
maximum values for a period of only 20 milliseconds and then decaying
with time like a $t^{-3/2}$ power law \citep[cf.,][]{SRJ04} or
modulated during the whole evolution (including the decay phase) with
a certain time variability of the energy released per unit of time. A
summary of our main results and conclusions will follow in
Sect.~\ref{sec:summary}.

\section{The Model}
\smallskip

Short GRBs (or, at least, a subclass of these events) might be
produced by the systems resulting from mergers of pairs of compact
objects (either two neutron stars or a neutron star and a black
hole). The situation that can arise from the mentioned merging
processes consists of a central black hole of a few solar masses
girded by a thick accretion torus with mass between $0.05$ and
$\approx 0.3 \ms$ \citep[see, e.g.,][]{RJ99,Jetal99}. Once the thick
disk has formed, neutrino--antineutrino annihilation may release up to
$\sim 10^{51}\,$ergs above the poles of the black hole in a region
that contains less than $10^{-5} \ms$ of baryonic matter. This may
lead to the acceleration of this matter to ultrarelativistic speeds,
possibly accounting for a GRB with observed properties. If the
duration of the event is related to the lifetime of the system
\citep{SP97} this kind of events can only belong to the class of short
GRBs because the expected timescale on which the black hole engulfs
the disk is at most of a few hundred milliseconds
\citep{RJ99,Jetal99,LR02,SRJ04}.
\begin{figure*}[hbt]
\begin{center}
\includegraphics[width=0.95\textwidth]{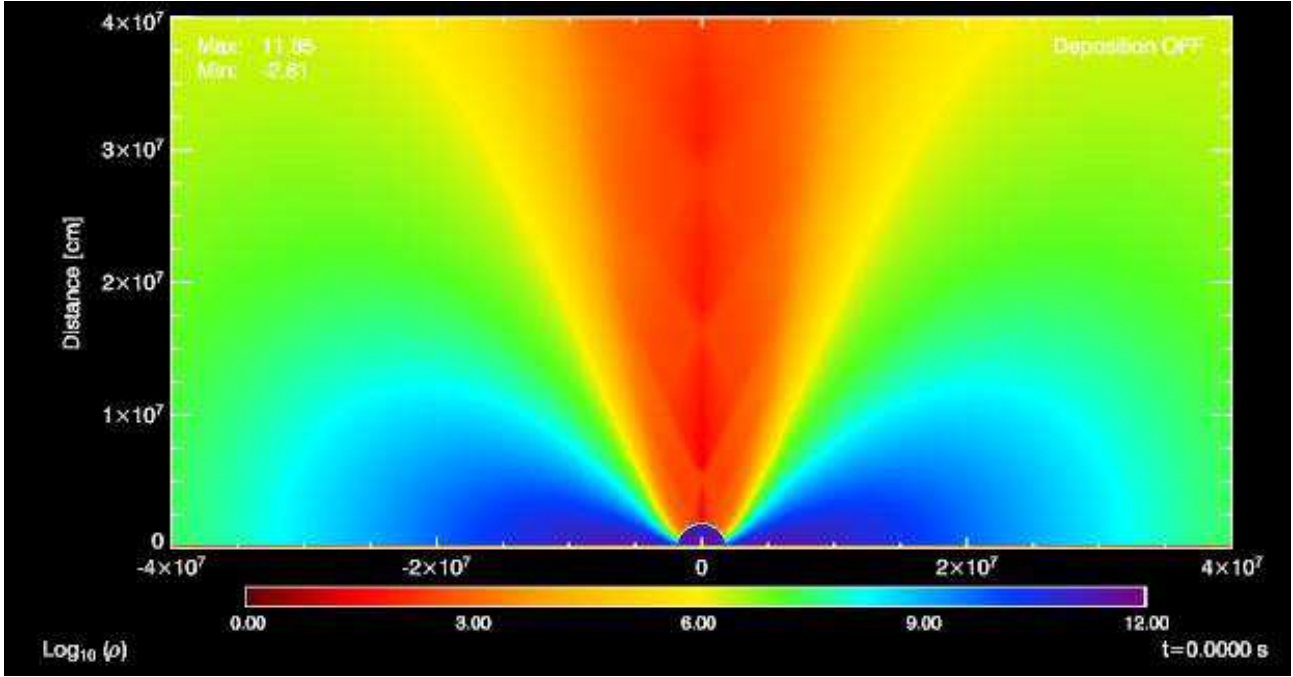}
\end{center}
\caption{Logarithm of the rest-mass density of the initial model of
type-A simulations showing the structure of the relaxed accretion
torus. The scale of the plot is limited in order to be able to
distinguish properly the toroidal structure. In the upper left corner
the maximum and minimum values of the displayed variable are
reported.}
\label{fig:rho_initial_type-A}
\end{figure*}

In order to test the viability of post-neutron star merger systems as
likely progenitors of short GRBs we have used general relativistic,
hydrodynamic simulations for studying the formation, acceleration and
early propagation of the relativistic outflow that is driven by the
deposition of thermal energy near the accretion torus around a
stellar--mass black hole. In such a black hole--torus configuration a
jet is expected to be launched by any process which gives rise to a
local deposition of energy and/or momentum, \eg $\nu
\bar\nu$--annihilation, or magneto-hydrodynamic processes. We mimic
this process by releasing thermal energy in a prescribed cone around
the rotational axis of the system.

\begin{figure}[hbt]
\begin{center}
\includegraphics[width=0.999\columnwidth]{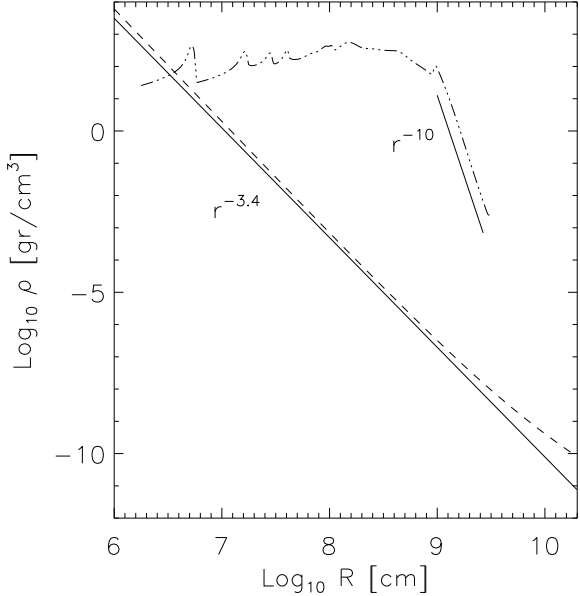}
\end{center}
\caption{Density vs radius along the symmetry axis of the two types of
initial models. The dashed-dotted (dashed) line corresponds to initial
models of type-A (type-B). The solid straight lines display power-law
fits to both density distributions.}
\label{fig:rhoaxis}
\end{figure}

The initial model is set up to reproduce the expected state of a
post-neutron star merger accretion torus around a central black hole
with a mass in the range $2.44\,\ms - 3\,\ms$. Thus, the initial model
is not the result of a merger simulation of two compact objects (at
least one of them being a neutron star) but instead, we use two
different {\em ad hoc} procedures to construct it. In the first one, a
toroidal distribution of matter and angular momentum is placed in our
computational domain, embedded by a larger, essentially spherically
symmetric dilute gas cloud. Such a configuration is suggested by the
previous results of \cite{RJ99} and \cite{Jetal99} where the gas cloud
is produced by mass shed off the hot massive merger remnant. Of
course, a more refined setup would require to take into account
asphericities due to the angular momentum in the system. For the time
being, however, we constrain ourselves to spherically symmetric
external gas clouds and avoid further refinements.  This initial
configuration is not in equilibrium and we need to evolve it for some
tenths of a second in order to obtain a relaxed initial state. The
relaxation of the model leads to the shedding of mass from the torus
that expands into the environment. Hence, the relaxed model has a
relatively high environmental density of $\simeq 10^2\,$g/cm$^{-3}$
(Fig.~\ref{fig:rhoaxis}) between $r \simeq 10^7\,$cm and $r \simeq
10^9\,$cm, declining steeply down to $\simeq 2.5 \cdot
10^{-3}\,$g/cm$^{-3}$ at $r \simeq 3\cdot10^9\,$cm. We will refer to
this kind of initial models as type-A models
(Fig.\ref{fig:rho_initial_type-A}). The torus and black hole masses of
type-A models are $0.17\,\ms$ and $3\,\ms$ after the relaxation time,
respectively. The inner grid boundary is located in type-A models at 4
gravitational radii ($R_g = GM/c^2$; $G$, $M$ and $c$ being the
gravitational constant, the mass of the black hole and the speed of
light in vacuum, respectively), i.e., at $\approx 1.78\e{6}\,$cm from
the center.
\begin{figure*}[hbt]
\begin{center}
\includegraphics[width=0.95\textwidth]{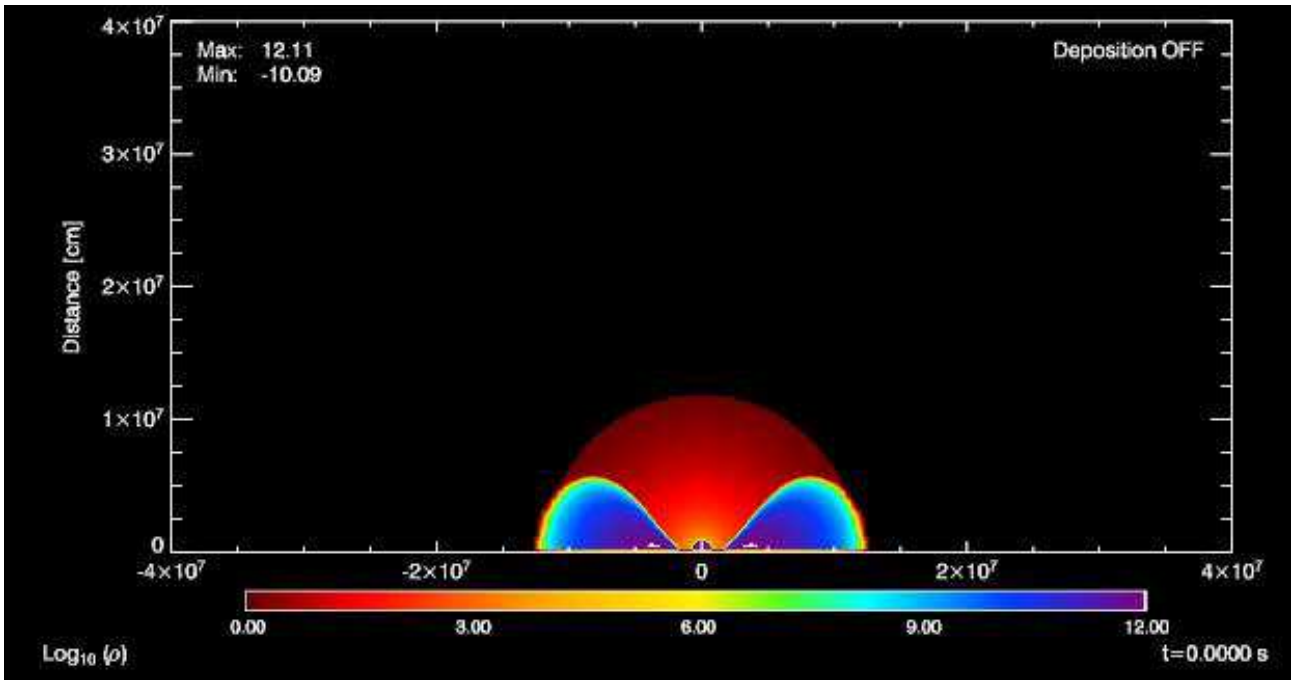}
\end{center}
\caption{Logarithm of the rest-mass density of the initial model of
type-B simulations showing the structure of the accretion torus. The
scale of the plot is limited in order to be able to distinguish
properly the toroidal structure. In the upper left corner the maximum
and minimum values of the displayed variable are reported. Note that
in contrast to Fig.~\ref{fig:rho_initial_type-A} the torus is much
denser and more compact.}
\label{fig:rho_initial_type-B}
\end{figure*}

 The production of type-B initial models uses a prescription very
close to that of \cite{FD02} in order to build an equilibrium torus
around a Schwarzschild black hole. Differences with respect to
\cite{FD02} arise from the fact that we do not use a barotropic
equation of state (EoS) and, therefore, it is not possible to obtain
from the specific enthalpy distribution (which is the primary
thermodynamical variable that is obtained from the Euler equation)
unambiguously all other thermodynamic variables in the equilibrium
torus. Thus, the adiabatic index ($\gamma$) at every point has to be
guessed, i.e., the initial model is not in perfect equilibrium
(although the initial state is much closer to equilibrium than in the
previous case), and it also requires some transient period until the
initial model evolves to a relaxed configuration. However, as the
deviation from equilibrium is small we can start releasing energy
right from the beginning of the simulation because whatever the
changes due to the relaxation are, they are relatively small and they
propagate radially at a speed smaller than any ultrarelativistic
outflow. On the other hand, a neutron star merger remnant originates
from a violent event and, therefore, both the emergent accretion torus
and its environment must naturally be expected to exhibit some
variability in time. Thus, the initial models of type-B, although not
fully relaxed, are not in obvious conflict with the situation that is
expected from merger simulations
(Fig.\ref{fig:rho_initial_type-B}). The type-B models have a much
smaller environmental density than those of type-A for radii larger
than $\approx 10^7\,$cm and their density gradient (in most of the
computational domain) is also steeper than in the other case ($\rho
\sim r^{-3.4}$; Fig.~\ref{fig:rhoaxis}). Models of type-B have a torus
and a black hole mass of $0.13\,\ms$ and $2.44\,\ms$,
respectively. The inner grid radial boundary is located in type-B
models at $2R_g$, i.e., at $\approx 7.57\e{5}\,$cm from the center.

\subsection{Numerical details}

In the radial direction the region of energy deposition extends from
the inner to outer grid boundaries. In the angular direction, the
half-opening angle of the deposition cone ($\theta_0$) around the
system axis spans a range which was varied between $30^{\circ}$ and
$75^{\circ}$. From the results of \cite{RJ99} and \cite{Jetal99} we
have inferred a power--law decline for the energy deposition rate per
unit of volume by $\nu \bar\nu$--annihilation in the direction
perpendicular to the equatorial plane of the accretion torus. We have
chosen an explicit form of the energy deposition law in terms of
observer frame coordinates as
\begin{equation}
\dot{q} = \dot{q_0} (z_0/z)^{n},
\label{q}
\end{equation}
where $z$ is the distance along the rotation axis, $n$ is the
power--law index, $z_0 =2R_g$ is the minimum vertical distance above
the equator where energy is deposited, and $\dot{q_0}$ is the
normalization factor that we use to set the total energy deposition
rate $\dot E$ as measured by a static observer at infinity. In this
paper we have considered only the case $n=5$, because this value
complies best with the data in \cite{Jetal99}. We assume energy
deposition rates which vary in the range from $10^{49}\,$erg\,s$^{-1}$
to $5 \cdot 10^{51}\,$erg\,s$^{-1}$ (for the summed values of both
hemispheres) and which are constant in time. We have also included two
cases where the energy release is chosen to be time dependent (the
explicit time dependence of models B07 and B08 is provided in
Eqs.~\ref{eq:B07} and \ref{eq:B08}, respectively).  The considered
values roughly bracket the expected energy deposition rates from
$\nu\bar\nu$--annihilation in the context of post--neutron star merger
accretion tori.

The simulations were performed with the multidimensional relativistic
hydrodynamic code GENESIS \citep{Alo99a} using 2D spherical
coordinates ($r, \theta$). We assume equatorial symmetry and we cover
the $\theta$-direction with 200 uniform zones between $0\gra$ and
$90\gra$. In the $r$--direction the computational grid consists of 400
or 500 zones. These are spaced logarithmically between the inner
boundary located at $4R_g$ (type-A models) or $2R_g$ (type-B models)
and an outermost radius of $R_{\rm max} = 3 \cdot 10^9\,$cm (type-A
models) or $R_{\rm max} = 2 \cdot 10^{10}\,$cm (type-B models). Even
for this moderate resolution, the typical number of time steps per
model in order to reach a time of 0.5\,s is $\approx 10^{6-7}$.

 The first set of models (type-A), \ie those having the smaller
$R_{\rm max}$, is employed to study only the generation phase of a
relativistic fireball by following the evolution of the resulting
outflow in a medium of relatively high density up to $100\,$ms. The
second set of models, \ie those having the larger $R_{\rm max}$, start
from non-relaxed type-B initial models and, as the computational
domain extends up to a larger radius, allows us to study the evolution
of the fireballs in relatively low--density environments until $\sim
0.5\,$s. We have also extended the domain of one of the most energetic
type-A models in order to follow the evolution of fireballs in
relatively high--density environments until $\sim 0.4\,$s.

The space-time around the non-rotating black hole is described by the
steady spherical Schwarzschild metric.  Effects on the dynamics due to
the self-gravity of the accretion torus or the external gas are
neglected, \ie we consider only the gravitational potential of the
black hole. The values of the energy deposition rate that we will use
throughout the paper are measured by an observer at
infinity. Therefore, the energy deposition rate per unit of coordinate
volume $\dot q$ measured by such an observer has to be blue-shifted in
order to determine the local value of the energy deposition rate in
the close vicinity of the BH. This value has to be further transformed
to the local comoving fluid frame because it is numerically included
in the algorithm as a source of the comoving energy density.  The
amount of energy released per unit of coordinate time and proper
volume in the comoving frame is
\begin{eqnarray}
\dot{q'} = \frac{\dot q}{ (1-2R_g/r) ^{1/2} (\gamma(\Gamma^2-1)+1) }\: .
\end{eqnarray}
where $\Gamma$ is the bulk Lorentz factor of the moving fluid.

The EoS assumes nuclei to be disintegrated to free, non-relativistic
nucleons, treated as a mixture of Boltzmann gases, and includes the
contribution from radiation, and an approximate correction due to
$e^+e^-$--pairs as described in \cite{WJT94}. Complete ionization is
assumed, and the effects due to possible degeneracy are neglected.
\begin{figure}[hbt]
\begin{center}
\includegraphics[width=0.95\columnwidth]{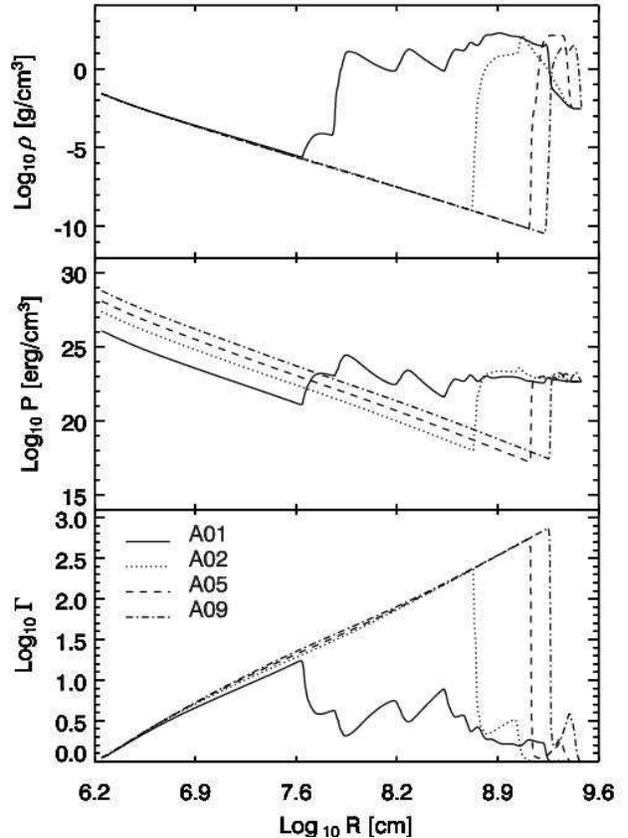}
\end{center}
\caption{Logarithm of the rest-mass density (top panel), of the
pressure (middle panel) and of the fluid Lorentz factor (bottom panel)
vs radius along the symmetry axis for the models A01 (solid lines),
A02 (dashed lines), A05(dashed-dotted lines) and A09 (long-dashed
lines) after 100\,ms. The displayed time is larger than the
transition time in models A01 and A02 and, therefore, the
recollimation of the initial ultrarelativistic wind solution yields
relativistic jets in these two cases. }
\label{fig:r-p-W-axis}
\end{figure}

\section{Results}

Given the fact that our models do not include a fully consistent
treatment of all relevant physics and our initial conditions are set
up {\it ad hoc} instead of resulting from merger simulations of
compact binary stars, we have performed a series of simulations trying
to cover a physically meaningful parameter space. Nonetheless, we
point out here that, even if we had done consistent simulations, the
variety of resulting initial configurations could be very large
because the properties of the forming black hole--accretion torus
system and its environment are sensitive to a number of parameters
which are incompletely known or variable (the masses and spins of the
binary objects, the neutron star equation of state, general
relativistic effects in the binary merging, etc.).

According to \cite{RJ99}, the accretion torus might be surrounded by a
halo of relatively high density (a by--product of the merger) that
could extend up to a few hundred kilometers from the center of the
merger. On the other hand, merger events should mostly occur in the
interstellar medium and, therefore, a merger and its halo should be
surrounded by a very low--density environment. The size and the exact
density profile of the halo can strongly depend on, e.g., the physical
parameters of the merging progenitors, the details of the merger
dynamics, the time delay of the black hole formation, and the
neutrino--driven baryonic outflow from a transient, massive, hot
post--merging neutron star. A high density halo would have to be
expected if, for example, the collapse to a black hole were delayed
due to the effects of very rapid (differential) rotation or viscous
heating \citep[e.g.,][]{Duezetal04,Morrisonetal04}, in which case the
hot neutron star would radiate neutrinos and a neutrino--driven wind
\cite[e.g.,][]{DWS86} would lead to a dense, expanding baryonic cloud
around the merger site.  It may also be possible to find situations
where the accretion torus is surrounded by a thin, dilute halo, in
particular, if the BH forms during the merger or within a few
dynamical timescales afterwards
\citep[e.g.,][]{SU00,STU03,Oechslinetal04}.  In regard of these
considerations we have performed two series of simulations. The first
one (models A01 to A09 of Table~\ref{tab:models}) uses an initial
model of type-A and corresponds to a case where the density of the
halo is high and the density decline with distance is moderate
(Fig.~\ref{fig:rhoaxis}). The second set of models (B1 to B8) start
from an initial model of type-B and, therefore, they correspond to the
low--density halo class.

\subsection{The case of a high density halo}
\label{sec:type-A}

For models of type-A, the parameter study comprises three different
aspects. The first one is the study of the effects associated with a
variation of the total energy deposition rate from $\dot E =
10^{49}\,$erg\,s$^{-1}$ to $\dot E = 5 \cdot 10^{51}\,$erg\,s$^{-1}$,
fixing the value of the opening half--angle of the deposition region
at $\theta_0 = 30^{\circ}$ and using a power law index of the energy
deposition profile of $n=5$. The models computed for this purpose are
A01, A02, A05 and A09 (Table~\ref{tab:models}). The second aspect is
the dependence of the results on $\theta_0$ while keeping the rest of
the parameters fixed ($\dot E = 2 \cdot 10^{50}\,$erg\,s$^{-1}$,
$n=5$). The models used in this study are A02, A03 and A04
(Table~\ref{tab:models}). Finally, we have made two experiments where
the energy released per unit of volume and time (the {\em deposition
rate density}, hereafter) was kept constant but both $\dot E$ and
$\theta_0$ were changed (models A03 and A08).
\begin{table*}
 \caption{The first three columns show the model name and the two main
 parameters which characterize the energetics of the models, namely
 the energy deposition rate $\dot E$ (which represents the summed
 value for both hemispheres) and the half-opening angle of the cone
 over which the energy is released, respectively.  For the models B07
 and B08 entry $\dot E$ gives the peak values of the energy deposition
 rate ($\dot{E}_0$, Eq.~\ref{eq:B07}). Columns four, five and six give
 some kinematic and dynamic properties of the outflow: the velocity
 ($v_p$) of the propagation of the fireball in the radial direction
 (\ie the velocity of the leading, radial edge of the outflow), the
 maximum Lorentz factor ($\Gamma_{\rm max}$) attained in the outflow
 and, the wind (or jet) half-opening angle ($\theta_w$) of the outflow
 after 10\,ms or 100\,ms, respectively. The last column gives the mass
 of the outflow ($M_f$) after 100\,ms. For models A04 and A06, there
 are data only up to 10\,ms.  The half-opening angle of the outflow is
 computed as the maximum $\theta$--coordinate of all computational
 cells in where the fluid Lorentz factor exceeds a value of 10 and the
 radial velocity is positive. The mass of the fireball is computed by
 adding up the mass in all computational cells that match the same
 criterion as the one used to compute $\theta_w$.  The choice of a
 relatively small value of the Lorentz factor ($\Gamma = 10$) in our
 criterion to separate the fireball matter from the external medium is
 motivated by the still ongoing acceleration of the fluid in the
 fireball that takes place at the expense of the large internal energy
 of the outflow.}
 \label{tab:models}
 \centering
 \begin{tabular}{@{}l|c|c|cc|cc|cc|c}
 \hline 
Model& $\dot{E}$ [\ergsec]  & $\theta_0$&
\multicolumn{2}{c|}{$\displaystyle{\phantom{\int}}$ $v_p$ [$c$] $\displaystyle{\phantom{\int}}$}  & 
\multicolumn{2}{c|}{$\Gamma_{\rm max}$} &
\multicolumn{2}{c|}{$\theta_w$} & $M_f$ [g]\\                                  
     &   &   & (10\,ms) & (100\,ms) &(10\,ms)&(100\,ms)& (10\,ms) &
 (100\,ms) & (100\,ms) \\  \hline 
 A01 &$\phantom{\int^{10}}$ $10^{49}$ $\phantom{\int^{10}}$  &  $30^{\circ}$ & 0.67 & 0.62 & 18 &
 18  & $<1\gra$ & $ <1\gra$    &$4.0\e{15}$\\  % Ae49
 A02 &$2 \cdot 10^{50}$&  $30^{\circ}$ & 0.63 & 0.63 & 81 &
 232 & $11.3\gra$ & $6\gra$   &$8.8\e{23}$\\  % 2e50
 A03 &$2 \cdot 10^{50}$&  $45^{\circ}$ & 0.80 & 0.67 & 11 &
 27 & $9.5\gra$ & $3.9\gra$  &$4.5\e{24}$\\  % 2F50
 A04 &$2 \cdot 10^{50}$&  $75^{\circ}$ & 0.67 &  -   & 7 &   -
 & $8.5\gra$ &  -   & -\\ % 2E50
 A05 &    $10^{51}$    &  $30^{\circ}$ &  0.99 & 0.82 & 84 &
 562 & $15.0\gra$ & $15\gra$  &$3.5\e{25}$ \\ % 1e51
 A06 &    $10^{51}$    &  $45^{\circ}$ &  0.97 &  -   & 80 & -
  & $15.8\gra$ & -   & -\\ % AB51
 A07 &    $10^{51}$    &  $75^{\circ}$ &  0.90 & 0.60 & 13 &
 37 & $12.5\gra$ &  $8.13\gra$& $2.4\e{25}$\\  % Ae51
 A08 &    $10^{50}$    &$31.4^{\circ}$ &  0.83 & 0.70 & 20 &
 20  & $3.8\gra$ &  $2.9\gra$  & $1.4\e{22}$ \\  % 1A50
 A09 &$5 \cdot 10^{51}$&  $30^{\circ}$ &  0.70 & 0.97 & 91 &
 748 & $23\gra$ & $26\gra$    & $3.3\e{26}$ \\  % 5e51
%
% A10 &  $2 \cdot 10^{50}$  &  $90^{\circ}$ & $0^{\circ}$  \\  %
%
 B01  &  $2 \cdot 10^{50}$  &  $45^{\circ}$ & 0.995 & 0.99994 &
 33 & 247 & $36\gra$ & $30\gra$    &$5.4\e{24}$ \\ % TES2 %done
 B02  &  $2 \cdot 10^{50}$  &  $60^{\circ}$ & 0.999 & 0.99995&
 40 & 274 & $35\gra$ & $21\gra$ & $5.0\e{24}$ \\ % B02n   %done
 B03  &  $2 \cdot 10^{50}$  &  $75^{\circ}$ & 0.97 &  0.998 & 
 17 & 17 & $9.4\gra$ & $2.3\gra$ & $6.2\e{22}$ \\  % B03n %done
 B04  &      $10^{49}$      &  $45^{\circ}$ & 0.96 & 0.99991 &
 30 & 244 & $30\gra$ & $18\gra$ & $3.2\e{23}$ \\  % B04n % don
 B05  &      $10^{51}$      &  $45^{\circ}$ & 0.999 & 0.99997 &
 33 & 232 & $34\gra$ & $28\gra$ & $2.8\e{25}$ \\  % B05n % done
 B06  &  $10^{50}$  &  $41.4^{\circ}$ &  0.9991 & 0.99992 &
 40 & 238 & $30\gra$ & $23\gra$ & $2.8\e{24}$ \\  % B06n %done
 B07  &  $2.35 \cdot 10^{50}$  &  $45^{\circ}$ & 0.995 & 0.99996 &
 34 & 238 & $35\gra$ & $28\gra$ & $4.0\e{24}$ \\  % B07n  %done
 B08  &  $2.35 \cdot 10^{50}$  &  $45^{\circ}$ & 0.999 & 0.99996 &
 34 & 253 & $33\gra$ & $24\gra$ & $3.8\e{24}$ \\  % B08n  %done
 \hline
 \end{tabular}
\end{table*}

As a general conclusion, we find that, for energy deposition rates
larger than a certain threshold, all the models lead to a relativistic
or ultrarelativistic outflow (\ie a fireball). This threshold is
caused by the need to overcome the ram pressure $p_{\rm ram}$ that the
external medium exerts on the nascent fireball close to its initiation
site.  If the amount of energy per unit of time and per unit of volume
that is pumped into the region extending up to the scale hight of the
torus ($H\sim 10^7\,$cm; note that about $97\%$ of the energy is
released in this region due to the large value of $n$ in Eq.~\ref{q})
in a time interval of about the free--fall time of fluid elements at a
distance of the order of $H$, is less than $p_{\rm ram}$, then the
fireball is unable to expand, and is swallowed by the BH. The precise
value of this threshold is model dependent. The threshold is higher
for halos of higher density than for dilute ones. The shape of the
fireball and its dynamical features depend very strongly on the type
of initial model that is considered as will be discussed below.

\subsubsection{Dependence on the total energy deposition rate}
\label{sec:depwithedot}

As we have already pointed out, there is a threshold energy deposition
rate below which a relativistic fireball is not initiated. For models
of type-A, the threshold is $\dot{E}_{\rm thr} \simeq
10^{49}\,$\ergsec (for $\theta_0=30\gra$), while for models of type-B
$\dot{E}_{\rm thr} < 10^{48}\,$\ergsec (for $\theta_0=45\gra$). The
difference between both thresholds comes from the fact that the higher
halo density of models of type-A leads to a higher ram pressure than
in models of type-B.

In models of type-A, when $\dot E$ is above $\dot{E}_{\rm thr}$, the
resulting outflow consists of either (i) a stratified (both in radial
and polar directions) wind of approximately conical shape or (ii) a
relativistic jet. Let us note that, by definition, the difference
between a wind and a jet is that in the latter the lateral boundaries
are causally connected while they are not so in case of a wind. This
distinction is relevant when the stability properties of both kind of
solutions are considered\footnote{For example, a jet can be destroyed
by Kelvin Helmholtz instabilities that originate at its surface, while
a wind only develops a turbulent layer which does not affect the whole
solution.}.
\begin{figure*}[hbt]
\begin{center}
\includegraphics[width=0.95\textwidth]{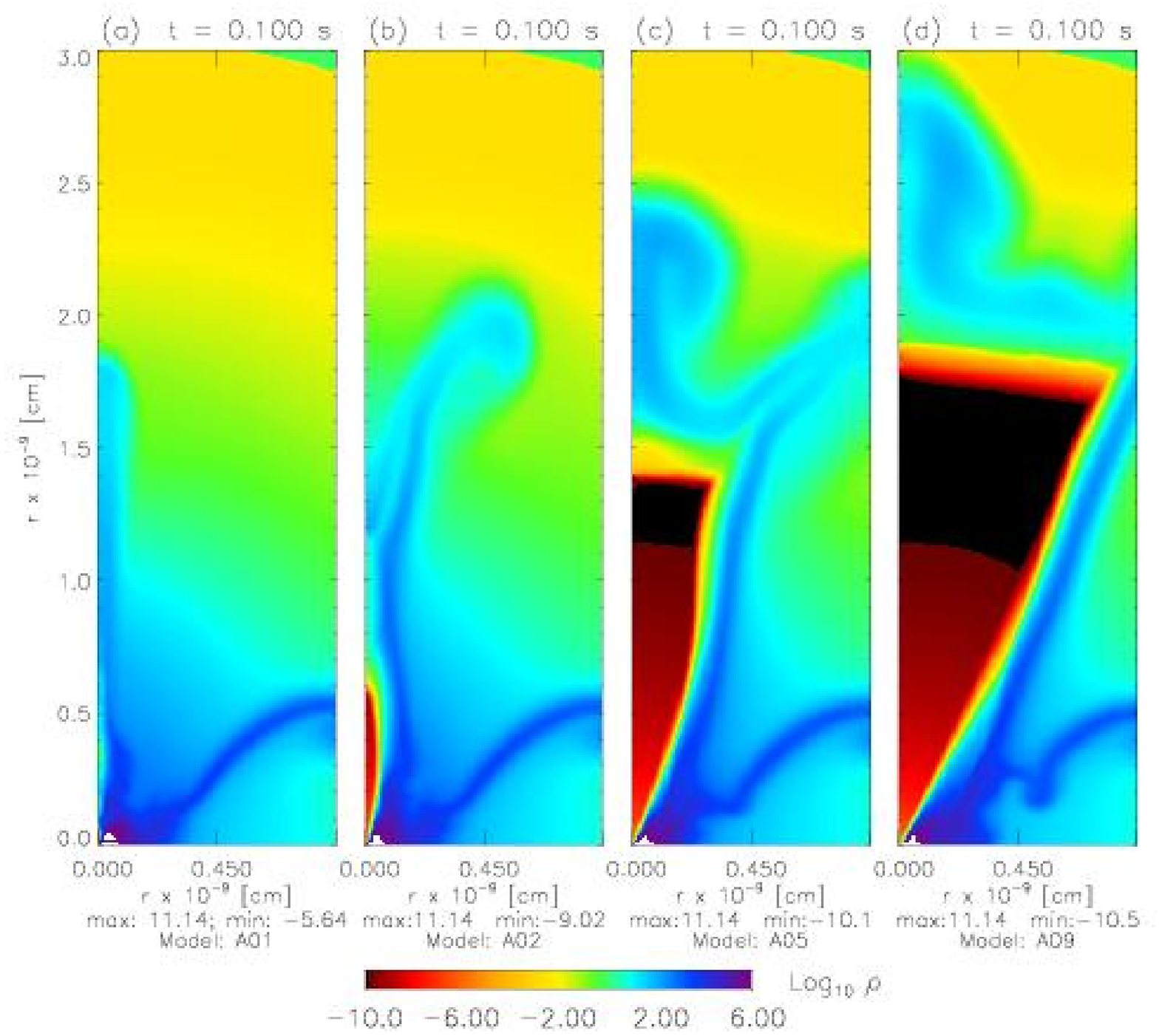}
\end{center}
\caption{Logarithm of the rest-mass density for models A01 (panel a),
A02 (panel b), A05 (panel c) and A09 (panel d) after 100\,ms of energy
deposition.  The color scale is limited in order to enhance the
details of the outflow region. The maximum and minimum density values
(in logarithmic scale) of each model are given below the panels.}
\label{fig:rho-Amodels}
\end{figure*}

\begin{figure*}[hbt]
\begin{center}
\includegraphics[width=0.95\textwidth]{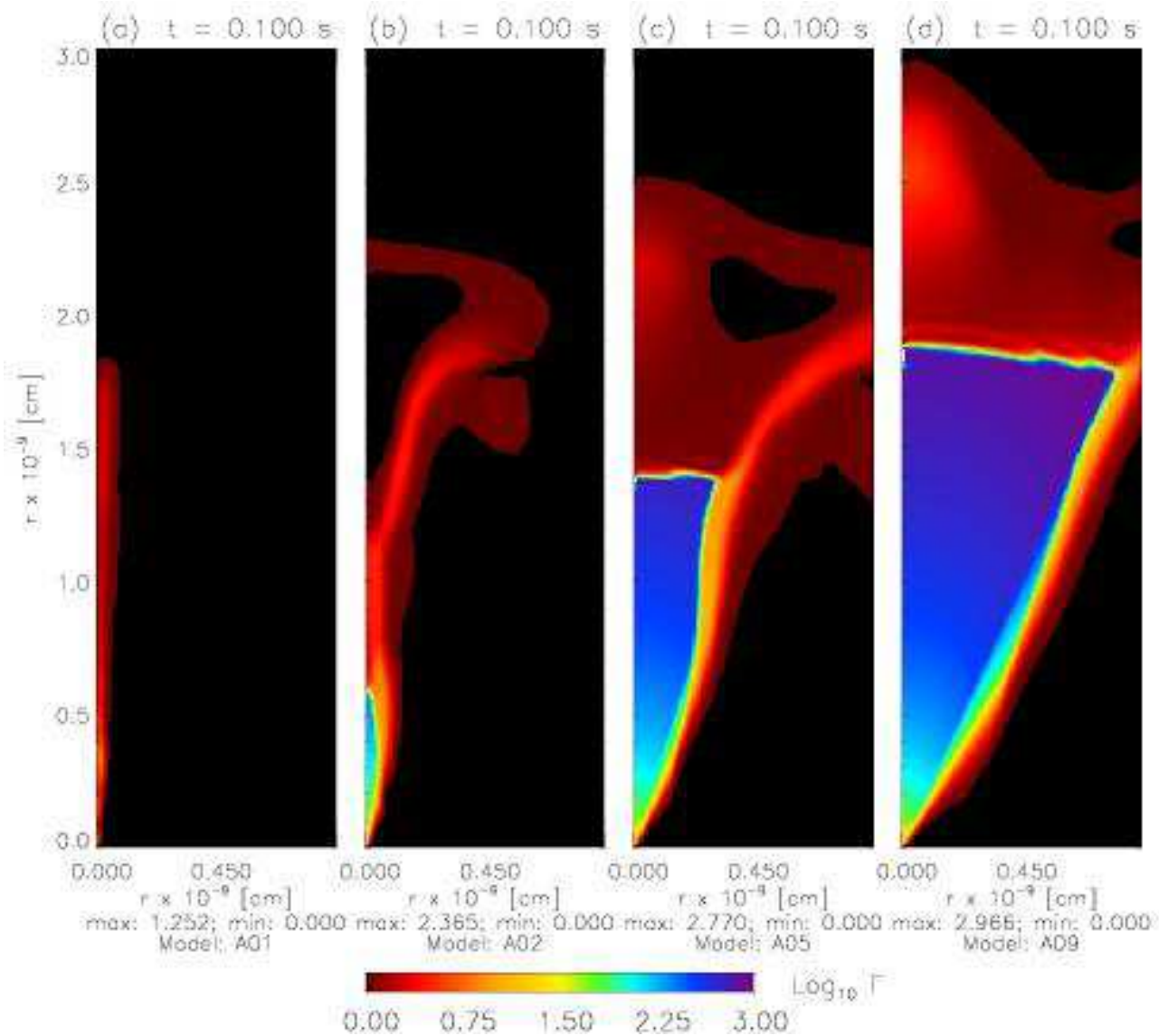}
\end{center}
\caption{Logarithm of the Lorentz factor for models A01 (panel a), A02
(panel b), A05 (panel c) and A09 (panel d) after 100\,ms of energy
deposition. The color scale is limited in order to enhance the details
of the outflow region. The maximum and minimum density values (in
logarithmic scale) for each model are given below the panels.}
\label{fig:lor-Amodels}
\end{figure*}

 A wind results for high energy deposition rates per solid angle
(models A05 and A09,
Figs.~\ref{fig:rho-Amodels},~\ref{fig:lor-Amodels}). In the polar
direction, the wind is confined in a cone the angle of which is model
dependent (see below). Its structure consists of a sheared wind with a
central, low density, unshocked, ultrarelativistic part (near the
symmetry axis) surrounded by a slow (mildly relativistic), denser,
shocked layer. Surfaces of constant physical variables are almost
spherical in the ultrarelativistic part of the wind. In the radial
direction, the wind extends radially from a fiducial point (located at
$\sim 3 \cdot 10^6\,$cm) to a terminal shock where it interacts with
the external medium. Due to the larger density of the external medium,
two shocks form: a reverse and a bow shock. Through the reverse shock
the wind decelerates from ultrarelativistic velocities to mildly
relativistic ones (between $0.6c$ and $0.97c$, see $v_p$ in
Table~\ref{tab:models}). The propagation speed of the bow shock is
smaller than that of the fluid in the wind
(Fig.~\ref{fig:r-p-W-axis}). This explains why the radial expansion of
the wind is only moderately relativistic.

In a conically expanding relativistic wind the physical variables
change with the distance $z$ along the axis of rotational symmetry
according to simple power laws \citep[\eg][]{LE00}. For example, the
pressure, the density and the Lorentz factor of a fluid whose
adiabatic index is $\gamma = 4/3$ follow the relations:
\begin{eqnarray}
p/p_0           = (z/z_0)^{-4},\,\, 
\rho/\rho_0     = (z/z_0)^{-3},\,\, 
\Gamma/\Gamma_0 =  z/z_0,
\label{eq:power-laws-anal}
\end{eqnarray}
where $p_0$, $\rho_0$ and $\Gamma_0$ are the values of the pressure,
of the density and of the Lorentz factor at the fiducial point of the
wind located at a distance $z_0$. In our models
(Fig.~\ref{fig:r-p-W-axis}), we obtain for the unshocked region of the
outflow (the shocked one does not follow a power law)
\begin{eqnarray}
p/p_0           = (z/z_0)^{-3.77},\\ \nonumber
\rho/\rho_0     = (z/z_0)^{-2.98},\\ \nonumber 
\Gamma/\Gamma_0 = (z/z_0)^{0.925}.
\label{eq:power-laws}
\end{eqnarray}
The discrepancy in the power-law indices arises from the fact that the
value of $\gamma$ is slightly smaller than $4/3$ in the wind of our
models due to the presence of $e^+e^-$--pairs in the fluid. Since for
type-A models the pressure and the density in the wind decrease faster
than in the external medium up to $z \sim 10^9\,$cm
(Fig.~\ref{fig:rhoaxis}) the outflow turns into a jet, because the
wind pressure eventually becomes equal to that of the external
medium. At that point, a recollimation shock forms starting from the
wind surface and propagating towards the axis.  Since the fiducial
values of the pressure increase with an increasing $\dot E$, a higher
$\dot E$ reduces the possibility of matching the pressure of the
external medium. Therefore, the structure of the wind tends to remain
conical when the energy deposition rate is increased. If the
recollimation shock does not form before the fireball reaches the part
of the environment where a steep density decline sets in
(\simgt$10^9\,$cm; Fig.~\ref{fig:rhoaxis}), it might not form at all
(because for radial distances larger than $\sim 10^9\,$cm the rate of
decline of the pressure in the external gas exceeds that in the
ultrarelativistic wind).

When the energy deposition rate is only slightly (model A01) or
moderate (model A02) above the threshold, the conical wind converts to
a relativistic jet due to the recollimation shock
(Fig.~\ref{fig:rho-Amodels}). The outflow then possesses all the
morphological elements known for light relativistic jets, \ie a
knotty, unsteady beam that ends in a Mach disc surrounded by an
over-pressured cocoon. The outflow begins as a transient, expanding,
over pressured, conical wind. But after a transition time ($\Delta
t_{\rm trans} \approx 10\,$ms for model A01 and $\Delta t_{\rm trans}
\approx 30\,$ms for model A02) a recollimation shock forms due to the
pressure confinement of the external medium reducing the opening angle
to values less than $1^{\circ}$. The formation of the recollimation
shock has two main consequences. First, it separates the outflow in
two parts: an unshocked inner flow (close to the deposition region;
the smooth region in Fig.~\ref{fig:r-p-W-axis}), and an outher shocked
flow where the pressure and the density have substantially larger
values (Fig.~\ref{fig:r-p-W-axis}). Second, the recollimation shock
effectively stops the linear increase (Eq.~\ref{eq:power-laws}) of the
beam Lorentz factor observed in the more energetic models (A05 and
A09; see Fig.~\ref{fig:r-p-W-axis}) even at large distances. Hence,
the propagation speed of the head of the outflow is about $35\%$
smaller in models A01 and A02 than that in models with higher energy
deposition rates (Table~\ref{tab:models}) after the initial transient
phase. As the fiducial pressure of the fireball grows with $\dot E$,
while the pressure decrease with distance is approximately the same
for the wind outflow in all models (Eq.~\ref{eq:power-laws}), the
distance where the pressure of the wind and the external medium match
increases with growing $\dot E$ (the same trend holds for $\Delta
t_{\rm trans}$). Whether pressure matching occurs or not depends on
the value of the fiducial pressure and on the pressure gradient in the
halo. Within $100\,$ms this happens in none of the more energetic
models (A05 and A09).

Increasing the energy deposition rate in models of type-A leads to a
progressive widening of the resulting fireball
(Fig.~\ref{fig:lor-Amodels}).  For models that produce an
ultrarelativistic wind (but not a relativistic jet),
Table~\ref{tab:models} shows that the half-opening angle of the wind is
smaller than the half-opening angle $\theta_0$ of the energy
deposition cone (in case of a jet the half-opening angle is set by the
interaction with the external medium and the cocoon dynamics). Thus,
the opening angle is not determined by our {\em ad-hoc} choice of
$\theta_0$. Instead, the opening angle is defined by the angular
coordinate where the pressure of the fireball equals that of the
torus. This angle depends on two factors: (i) the torus pressure at a
given radial distance, increases as one moves from the axis to the
equator and, (ii) the pressure of the fireball, at a given radial
distance, is an increasing function of the energy deposition
rate. Hence, as we increase $\dot E$ the location where the pressure
of the fireball matches that of the torus shifts to larger angles, \ie
the opening angle of the outflow grows.

Within an outflow crossing time of the torus in the axial direction
(\simgt$1\,$ms), the fluid accelerates up to Lorentz factors of
$\approx 10$ (Fig.~\ref{fig:r-p-W-axis}) and, if it does not form a
jet, the fluid becomes ultrarelativistic, \ie any motion except in the
radial direction is prevented. Hence, a sideways expansion is
drastically reduced, and the initially imprinted opening angle remains
unchanged. After $10\,$ms the maximum Lorentz factor rises to $\sim
100$ in the case of winds while it is only $\sim 18$ for jets.

The amount of mass in the fireball (Table~\ref{tab:models}), defined as
the sum of the masses of each computational cell in the outflow that
moves out radially with a $\Gamma>10$, increases almost
linearly with $\dot E$ during the period in which energy is released
in the system.  The mass is mainly concentrated in the cocoon that
surrounds either the outflowing jet or wind. In the case of winds,
only a small fraction of the mass moves at Lorentz factors larger than
100.

\subsubsection{Dependence on the half-opening angle of the energy deposition cone}
\label{sec:depwithangle-A}

\begin{figure*}
\begin{center}
\includegraphics[width=0.75\textwidth]{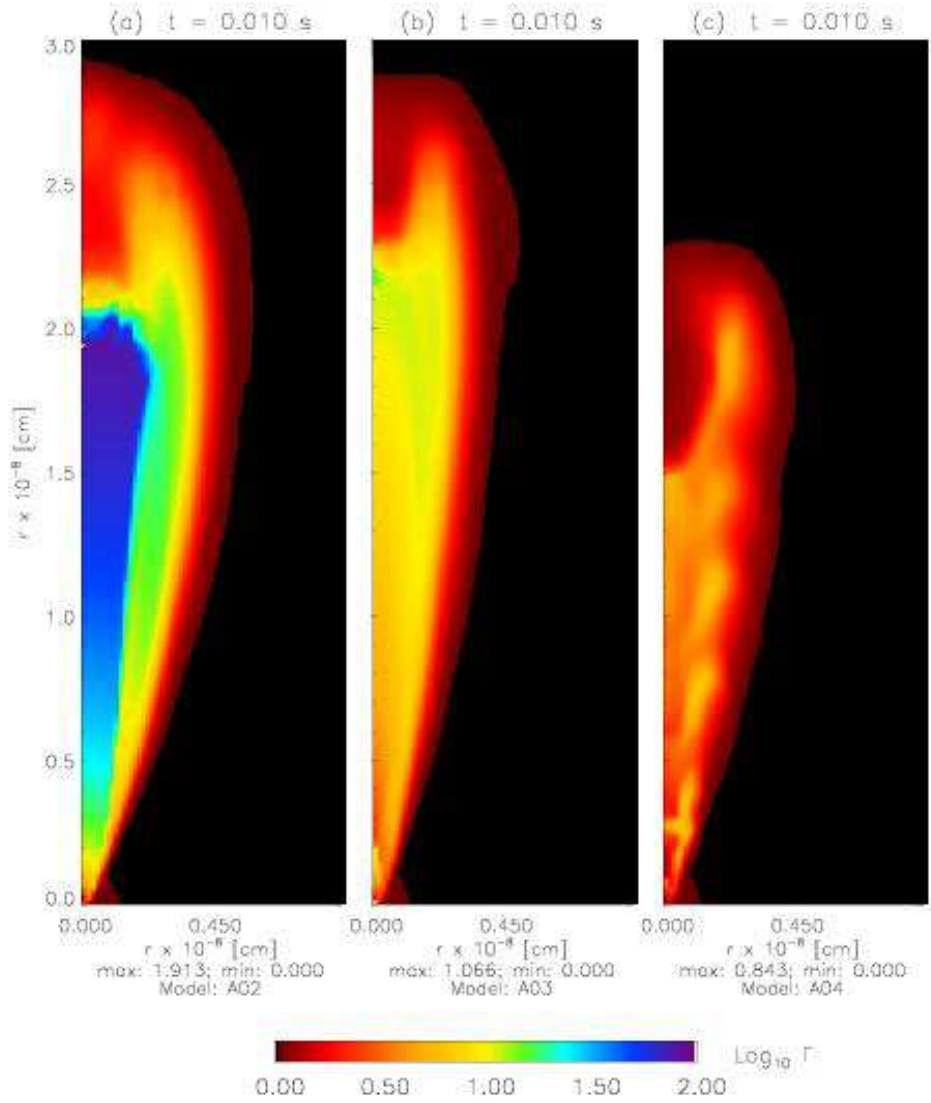}
\end{center}
\caption{Logarithm of the Lorentz factor for models A02 (panel a), A03
(panel b) and A04 (panel c) after 10\,ms of energy deposition. The
color scale is limited in order to enhance the details of the outflow
region. The maximum and minimum Lorentz factor values (in logarithmic
scale) of each model are given below the panels.}
\label{fig:lor-Amodels-isoangle1}
\end{figure*}

\begin{figure*}
\begin{center}
\includegraphics[width=0.75\textwidth]{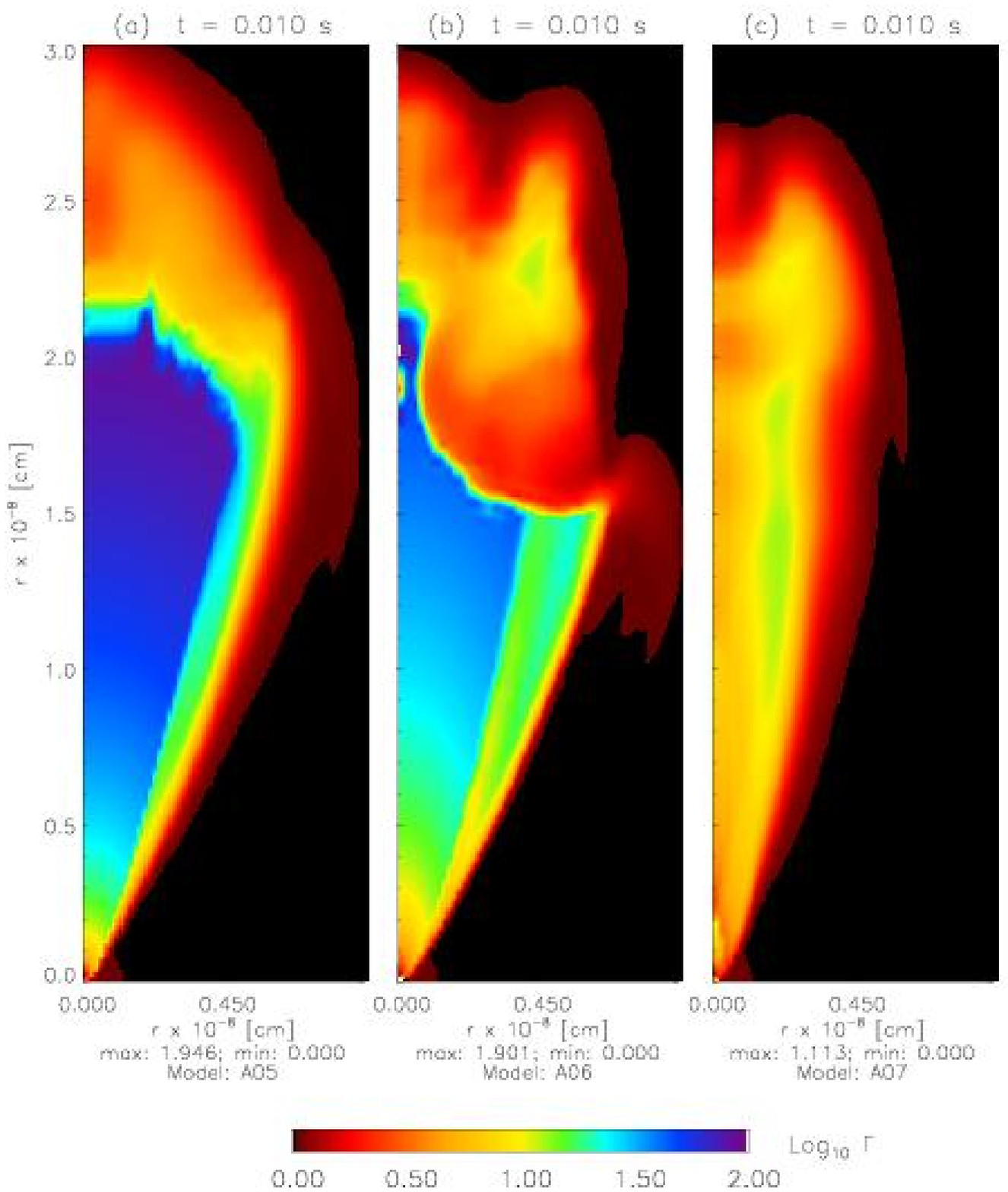}
\end{center}
\caption{Logarithm of the Lorentz factor for models A05 (panel a), A06
(panel b) and A07 (panel c) after 10\,ms of energy deposition. The
color scale is limited in order to enhance the details of the outflow
region. The maximum and minimum Lorentz factor values (in logarithmic
scale) of each model are given below the panels.}
\label{fig:lor-Amodels-isoangle2}
\end{figure*}

 We have performed a series of runs where we compared for {\em fixed}
energy deposition rates the influence of an increasing angular width
of the region of energy release, $\theta_0$ from $30\gra$ to
$75\gra$. The models considered in this study are divided into two
groups. The first group consists of A02, A03 and A04
(Table~\ref{tab:models}, Fig.~\ref{fig:lor-Amodels-isoangle1}) which
have a moderate energy deposition rate of $\dot{E} = 2\e{50}\,$erg/s
(group M hereafter).  The second group possesses a higher deposition
power of $10^{51}\,$erg/s, and consists of models A05, A06 and A07
(Table~\ref{tab:models}, Fig.~\ref{fig:lor-Amodels-isoangle2}; group H
hereafter).

For a fixed, moderate energy deposition rate the opening angle of the
outflow is initially quite insensitive to the increase of the
deposition angle $\theta_0$: for the M-group, while $\theta_0$ grows
from $30\gra$ to $75\gra$, $\theta_w$ only decreases from $11.3\gra$
to $8.5\gra$ (we define $\theta_w$ as the angle of the outflow having
$\Gamma > 10$) in 10\,ms. After longer evolutionary times all the
models of group M tend to form relativistic jets with opening angles
that depend on the interaction between the beam, the cocoon and the
external medium, but not on $\theta_0$. For models of group H,
depending on the value of $\theta_0$, we obtain outflows with opening
angles after 100\,ms of evolution that exhibit a rather small range of
variation between $15\gra$ (A05; $\theta_0=30\gra$) and $8.13\gra$
(A07; $\theta_0=75\gra$). Hence, the trend for group H is a slight
non-monotonic decrease of the outflow opening angle as we deposit the
energy in a wider region.

These results confirm, on the one hand, that the opening angle of the
deposition region is not directly related to the final opening angle
of the outflow and, on the other hand, that the mechanism of
collimation of this series of models is significantly different from
the mechanism discussed by \cite{LE00}. These authors propose
that a baryon-rich wind emerging from the torus is able to collimate
the baryon poor jet moving along the symmetry axis of the system. For
type-A models the role of such a baryon-rich wind is played by an
almost static cloud that is formed during the initial relaxation phase
of the torus. This cloud has a rather low velocity ($\simeq 0.027c$)
and a density of $\simeq 10^6\,$g/cm$^{-3}$ at a distance of
$10^7\,$cm. The density does not fall off as $r^{-2}$ (as in case of
the baryon-rich wind of \citealp{LE00}), but with a much smaller
power. Therefore, the main effect of the environment is, in our case,
to provide sufficient inertial confinement for preventing excessive
sideways expansion of the axial ultrarelativistic flow.

The insensitivity of $\theta_w$ to $\theta_0$, arises because the
opening angle of the relativistic outflow is mainly determined by the
inclination angle of the torus walls around the rotation axis. As
discussed in \S~\ref{sec:depwithedot}, the torus does not have a sharp
surface that separates it from the halo. Instead, a gradient of
density and pressure connects it with the surrounding medium. We have
argued that the region where the pressure of the torus and of the
emerging fireball match each other determines the initial
$\theta_w$. In the considered set of models, the energy deposition
rate per unit of volume decreases with growing $\theta_0$. This fact
implies a smaller pressure (at any given radial distance). Thus,
pressure equilibrium between the fireball and the torus occurs at a
smaller polar angle, which in turn reduces the value of $\theta_w$ as
can be seen in Figs.~\ref{fig:lor-Amodels-isoangle1} and
\ref{fig:lor-Amodels-isoangle2}.
\begin{figure*}
\begin{center}
\begin{tabular}{cc}
\includegraphics[width=0.95\columnwidth]{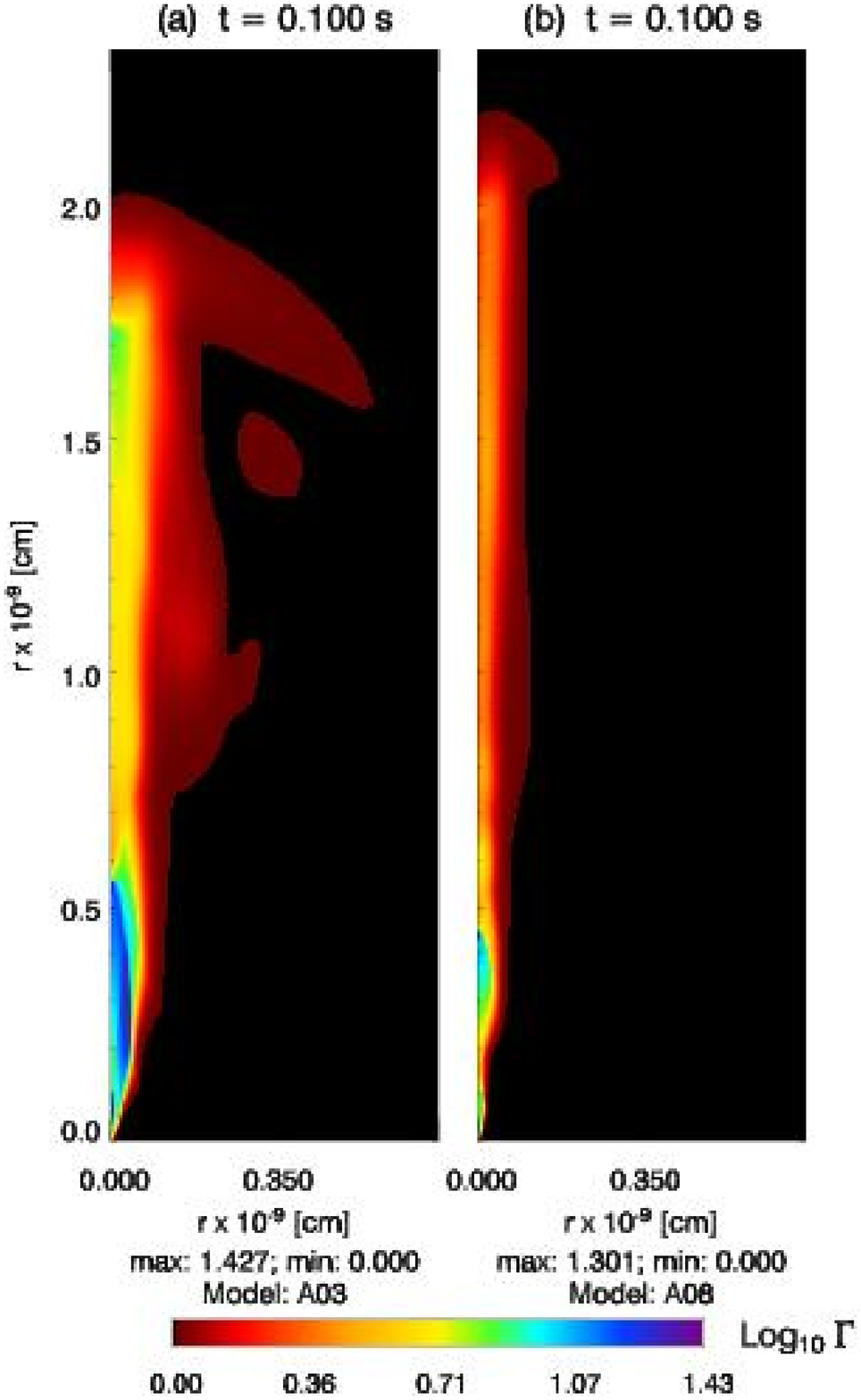} & 
\includegraphics[width=0.95\columnwidth]{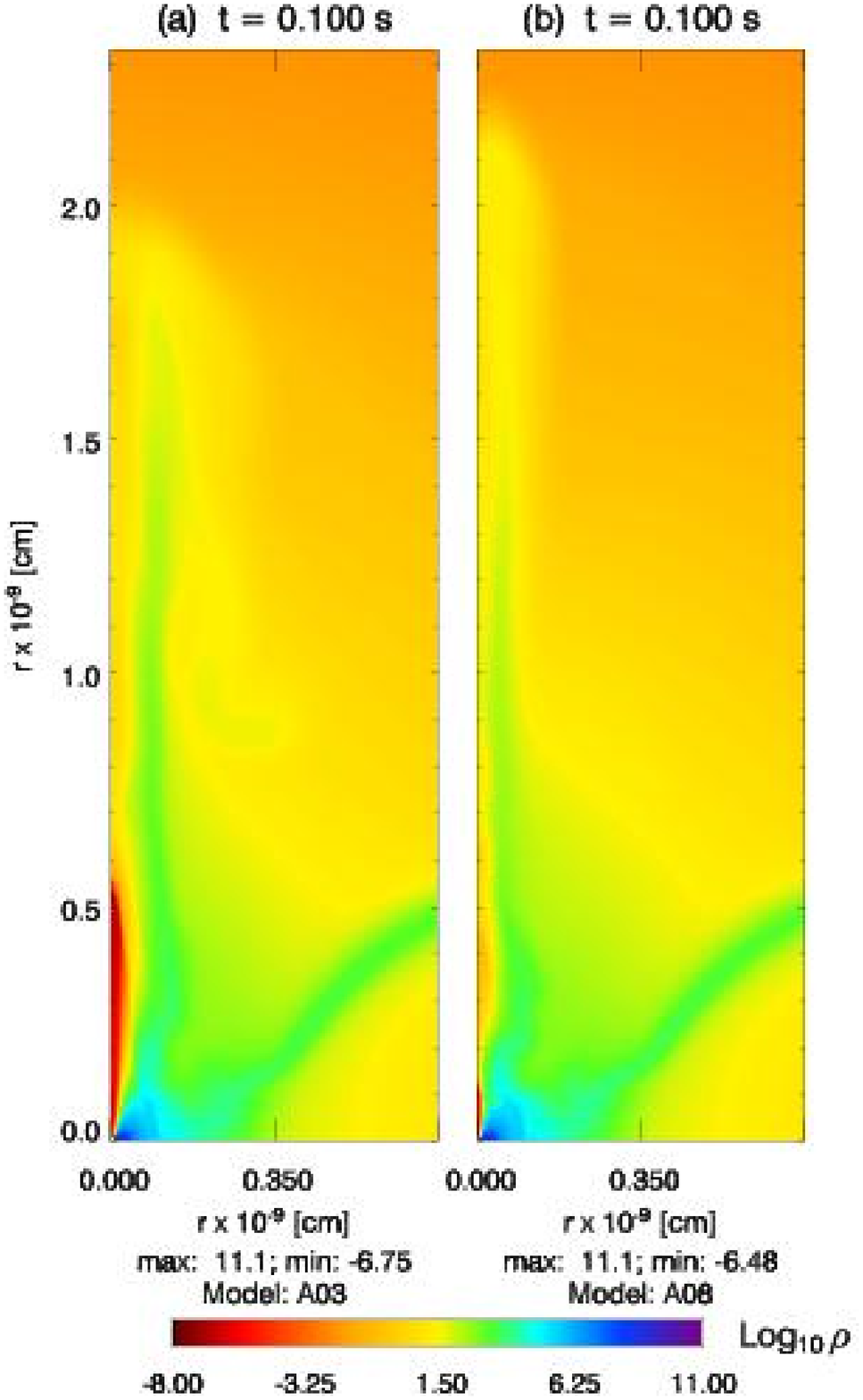} 
\end{tabular}
\end{center}
\caption{Logarithm of the rest mass density (right two panels) and of
the Lorentz factor (left two panels) for models A03 (panels a), and
A08 (panels b) after $100\,$ms of energy deposition. The color scales
are limited in order to enhance the details of the outflow region. The
maximum and minimum values of the respective variable are given below
the panels. Note the strong pinching of the jet of model A08 at
$r\approx10^8$\,cm which is due to the shedding of the most external
shells of the accretion torus.}
\label{fig:rho-A03-A08}
\end{figure*}

Another consequence of increasing $\theta_0$ is the transition from a
sheared, ultrarelativistic wind to a knotty, moderately relativistic
jet. For models of group M and $\theta_0 \geq 45\gra$, this transition
takes place very soon after the birth of the fireball ($\Delta t_{\rm
trans} \simlt 10\,$ms; Fig.~\ref{fig:lor-Amodels-isoangle1}) at small
radial distances (models A03 and A04). In the case of the largest
$\theta_0$ considered (model A04, $\theta_0 = 75\gra$), the maximum
Lorentz factor is only $6.97$, and the jet is prone to large amplitude
Kelvin-Helmholtz (KH) instabilities that lead to a wrinkled surface
and an inhomogeneous beam. Increasing the energy deposition rate
(models of group H), we observe the same qualitative behavior, but the
wind-to-jet transition happens at a larger value of $\theta_0$
($\approx 75\gra$).

Model A04 produces an inhomogeneous, low-velocity jet because the
imposed energy deposition rate per unit volume for the chosen value of
$\theta_0$ is only slightly above the threshold value of $\dot E$ that
is necessary to launch an outflow at all. We have confirmed this by
performing another simulation with the same energy deposition rate but
with $\theta_0 = 90\gra$, which did not produce any outflow.  For a
given threshold value for the energy deposition rate per unit of
volume, the minimum total energy deposition rate to produce an outflow
increases with larger $\theta_0$. The formation of jets or winds
depends on the energy released per unit of time and of volume. The
larger the energy release, the larger is the chance for wind
formation.
 
\subsubsection{Dependence on the deposition volume $V_{\rm dep}$ and 
on the total energy deposition rate $\dot E$ for ${\dot E}/V_{\rm
dep}$=constant.}
\label{sec:depwithedens}

In this section we will study the dependence on the deposition volume
if the rate per unit volume at which energy is released in the system
is fixed.  The two models that we consider here are A03 (having a
large deposition rate and a large deposition angle) and A08 (with
smaller deposition rate and deposition angle but the same energy
released per unit of time and volume as A03).  Both models develop a
mildly relativistic jet and display very similar properties at the
beginning of the evolution (Fig.~\ref{fig:rho-A03-A08}). However, they
later evolution differs. For example, the propagation velocity of the
jet head is smaller for A03 than for A08, because model A03
accumulates more mass in the cocoon, and a part of this mass enters
the beam thus leading to a more massive jet ($M_f^{A03} =
4.5\e{24}\,$g while $M_f^{A08} = 1.4\e{22}\,$g).
\begin{figure*}
\begin{center}
\includegraphics[width=0.95\textwidth]{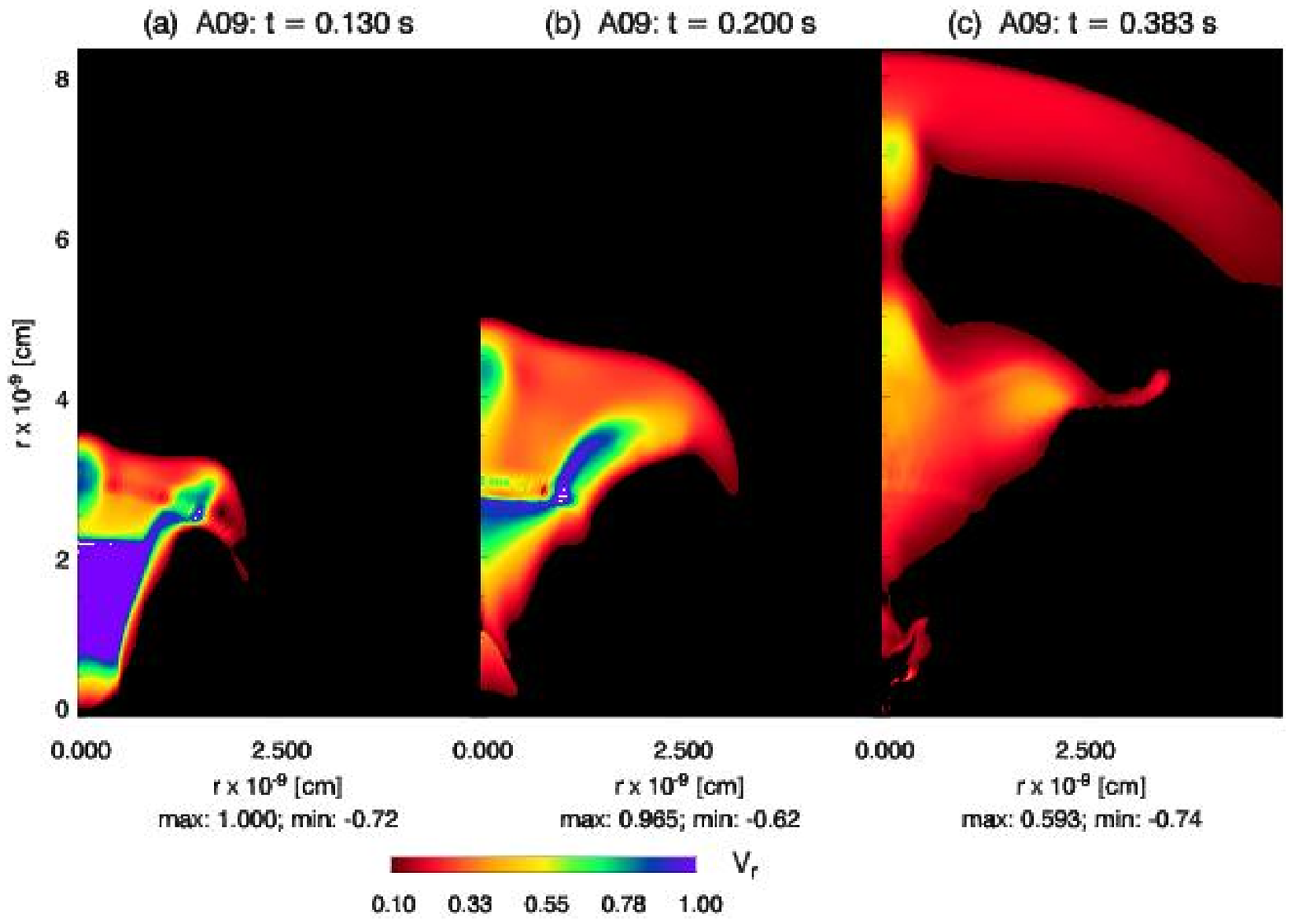}
\end{center}
\caption{Snapshots of the radial velocity of model A09 after the
shutdown of the energy release (the snapshot times are provided above
each panel). The color scale is limited in order to enhance the
details of the outflow region. 
%Note that the scale has been adjusted
%in the two left panels in order to enlarge the details of the earlier
%phases of the evolution. 
The maximum and minimum radial velocity
values of each model are given below the panels. Although the outflow
has reached Lorentz factors above 700 after 100\,ms of evolution, and
continues accelerating (panel a) until the rear end of the outflow
catches up with the reverse shock (panel b), after 500\,ms only a
subrelativistic outflow has survived and propagates with radial
velocities of $\sim 1.25\e{10}\,$cm\,s$^{-1}$.}
\label{fig:vlx-A09-evol}
\end{figure*}
\begin{figure*}
\begin{center}
\includegraphics[width=0.95\textwidth]{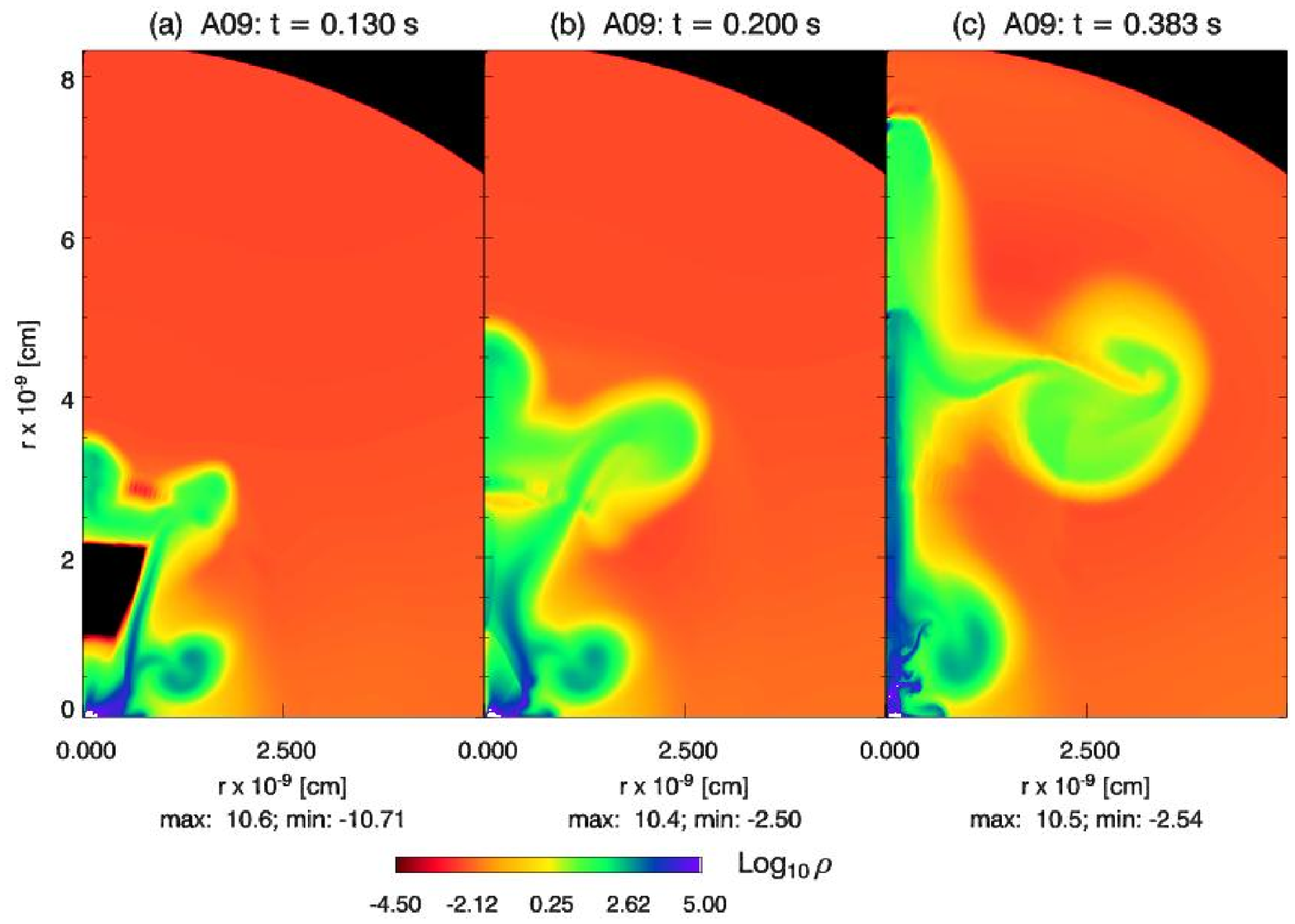}
\end{center}
\caption{Snapshots of the logarithm of the rest mass density of model
A09 before the shutdown of the energy release (the snapshot times are
provided above each panel). The color scale is limited in order to
enhance the details of the outflow region. 
%Note that the scale has
%been adjusted in the two left panels in order to enlarge the details
%of the earlier phases of the evolution. 
The maximum and minimum rest
mass density values (in logarithmic scale) of each model are given
below the panels.}
\label{fig:rho-A09-evol}
\end{figure*}

This can be understood from the fact that in model A03 the cone where
energy is deposited has a larger half-opening angle $\theta_0$ and
thus overlaps more with the outer layers of the torus than it does in
model A08. The energy release in this overlap region causes mass
shedding from the outer layers of the torus where the densities (for
$31.4\gra < \theta < 45\gra$) are 2 to 4 orders of magnitude higher
than those in the solid angle corresponding to the deposition cone of
model A08. Hence, the outflow in model A03 becomes
baryon-enriched. The outflow velocities are correspondingly lower,
although the rate of energy deposition per unit volume is the same as
in model A08.

Independent of the differences in mass, velocity and maximum Lorentz
factors reached in the two models, the final opening angle becomes
less than $4\gra$ in both cases (Table\ref{tab:models}), because a
series of recollimation shocks in the beam prevents a large sideways
expansion of the forming jet. Nonetheless, the cocoon of the jet of
model A03 is twice as thick ($\approx 3\e{8}\,$cm when the jet has
propagated for $100\,$ms) as that of model A08. These differences in
the cocoon can be understood from the different angular extensions of
the deposition regions, too.

\subsubsection{Evolution after the switch off of the central energy source}
\label{sec:switchoff-A}

We have done simulations where, starting from model A09 (which has the
largest $\dot E$ of all type-A models), the energy deposition was
switched off after a source activity time $t_{\rm sa} = 100\,$ms. In
this case the ultrarelativistic wind solution was destroyed within
less than $2t_{\rm sa}$, because the fireball sweeps up mass when it
moves through the high density halo in models of type-A
(Figs.~\ref{fig:vlx-A09-evol},~\ref{fig:rho-A09-evol}). The swept up
mass prevents the terminal shock of the ultrarelativistic wind from
moving at ultrarelativistic speeds (it actually propagates at a speed
of $\approx 0.97c$ at 100\,ms). Therefore the ultrarelativistic wind
is decelerated when it catches up with the reverse shock
(Figs.~\ref{fig:vlx-A09-evol}b,~\ref{fig:rho-A09-evol}b). In case of
type-A models that form jets, a similar argument can be used to
explain the mildly relativistic speeds at which the heads of the jets
propagate ($\approx 0.67c$). Hence, models of type-A will, in general,
not lead to successful GRBs, because they accumulate too much mass in
the fireball and decelerate well before the outflow becomes optically
thin.

Note that a necessary condition for a fireball to produce a successful
GRB is that the rest mass carried by the fireball multiplied by $c^2$
should be about two orders of magnitude smaller than the energy of the
fireball by the time at which transparency sets in. If the mass of the
fireball grows too much before the fireball reaches the photosphere,
the Lorentz factor corresponding to the terminal shock of the wind,
and as a consequence, the fluid Lorentz factor, will be reduced below
the critical limit (namely, $\Gamma \approx 100$) to account for the
properties of observed GRBs. An estimate of the maximum Lorentz factor
the fireball could ultimately reach (if all the internal energy could
be converted into kinetic energy) is given by the ratio $\eta \equiv
E/Mc^2$, where $E$ is the total energy of the fireball at any given
time after the energy deposition is switched off. $M$ is the mass of
the fireball computed according to the definition given in the caption
of Table~\ref{tab:models}. In model A09 $\eta$ decreases with time
once the energy deposition is switched off, because of the increasing
mass of the fireball. Thus, this model will not meet the criterion
$\eta \approx 100$ which is required for producing an observable GRB.

Although type-A models are not expected to produce GRBs, we can
estimate the kind of observational signature that this kind of events
might generate. We assume that after some hundred milliseconds of
evolution the bipolar outflow can be roughly approximated by two
spherical, homogeneous clouds of hot radiation--dominated plasma, each
with a mass $M \sim 10^{-5}\ms$, radius $R_0 \sim 10^9\,$cm,
temperature $T_0 \sim 1.5\e{9}\,$K and internal energy $U_0 \sim
2\e{50}\,$erg
(Figs.~\ref{fig:vlx-A09-evol}c,~\ref{fig:rho-A09-evol}c). Assuming
that the subsequent evolution of such a cloud is adiabatic, its
temperature and internal energy will decrease as the cloud
expands. Transparency is reached when the radius of the plasma cloud
becomes
\begin{equation}
R_{\rm t} \approx 3\e{13}\: {\rm cm} \:
		     \left( \frac{\kappa}{\kappa_e}  \right)^{1/2}
		     \left( \frac{M}{10^{-5}\: {\rm \ms} }
		     \right)^{1/2}		     
\end{equation}
\noindent
on a time scale (assuming that the cloud expands with about the light
speed)
\begin{equation}
t_{\rm t} \approx R_{\rm t} / c \approx 10^{3}\: {\rm s} \:
		     \left( \frac{\kappa}{\kappa_e}  \right)^{1/2}
		     \left( \frac{M}{10^{-5}\: {\rm \ms} }
		     \right)^{1/2}
		     \:,
\end{equation}
where $\kappa \approx \kappa_e \approx 0.2\,$cm$^2$g$^{-1}$ is the
mean opacity (with $\kappa_e$ being the opacity caused by electron
scattering for an electron--to--baryon ratio of 0.5).

When transparency sets in, both the temperature and the internal
energy of the cloud have decreased to the following values

\begin{eqnarray}
T_{\rm t} \approx  5\e{4}\; {\rm K}\: & {\displaystyle
                     \left( \frac{T_0}{1.5\e{9}\: {\rm K}} \right) 
		     \left( \frac{\kappa}{\kappa_e}  \right)^{-1/2}}
                     \: \times \nonumber \\ 		
		  &  {\displaystyle
		     \left( \frac{M}{10^{-5}\: {\rm \ms} } \right)^{-1/2}
		     \left( \frac{R_0}{10^9\: {\rm cm}} \right)}\: ,
\end{eqnarray}
\begin{eqnarray}
U_{\rm t} \approx 7\e{45}\; {\rm erg\, cm}^{-3} \: & {\displaystyle
                     \left( \frac{T_0}{1.5\e{9}\: {\rm K}} \right)^4
		     \left( \frac{\kappa}{\kappa_e} \right)^{-1/2}}
                     \: \times \nonumber \\ 		
		  &  {\displaystyle
		     \left( \frac{M}{10^{-5}\: {\rm \ms} } \right)^{-1/2}
		     \left( \frac{R_0}{10^9\: {\rm cm}} \right)^4}\: .
\end{eqnarray}

After optically thin conditions are reached, most of the internal
energy of the cloud will be radiated over a time scale of the order of
$t_{\rm t}$ and, therefore, the peak luminosity $L_{\rm m}\approx
U_{\rm t} / t_{\rm t}$ will be roughly

\begin{eqnarray}
L_{\rm m} \approx 7\e{42}\: {\rm erg\,
                     s}^{-1} \;  & {\displaystyle
                     \left( \frac{T_0}{1.5\e{9}\: {\rm K}} \right)^4
		     \left( \frac{\kappa}{\kappa_e} \right)^{-1}}
                     \: \times \nonumber \\ 		
		  &  {\displaystyle
		     \left( \frac{M}{10^{-5}\: {\rm \ms} } \right)^{-1}
		     \left( \frac{R_0}{10^9\: {\rm cm}} \right)^4}\: .
\end{eqnarray}

 Therefore, we expect a low--luminosity, soft UV-flash to be emitted
as a result of the nonrelativistic outflow expanding from BH-tori
systems in high--density merger halos. Due to the small value of $L_m$
only galactic events might be detectable.

\subsection{The case of low-density halo}
\label{sec:type-B}
\begin{figure}
\begin{center}
\includegraphics[width=0.95\columnwidth]{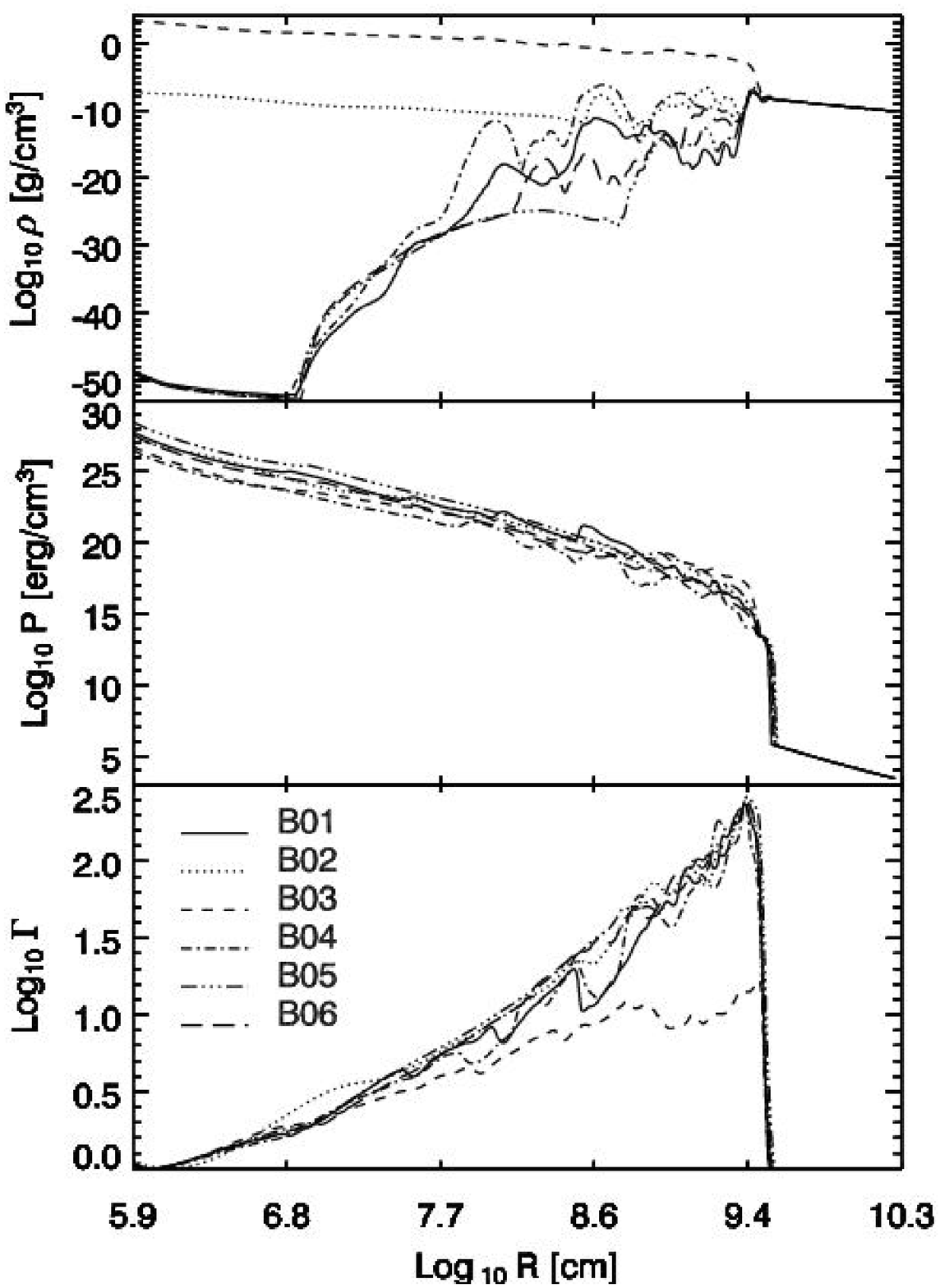}
\end{center}
\caption{Logarithm of the rest-mass density (top panel; in \grcm3), of
the pressure (middle panel) and of the fluid Lorentz factor (bottom
panel) vs radius along the symmetry axis for the models B01 to B06
after 100\,ms.}
\label{fig:r-p-W-axis-B}
\end{figure}

In case of a type-B progenitor, our parameter study comprises the same
three different aspects as for type-A models. We first study the
dependence on the energy deposition rate which is varied from
$10^{49}\,$erg\,s$^{-1}$, to $2 \cdot 10^{50}\,$erg\,s$^{-1}$ keeping
the value of the half-opening angle of the deposition cone fixed to
$\theta_0 = 45^{\circ}$ and the power-law index of the vertical
decrease of the energy deposition rate set to $n=5$ (see
\S~\ref{sec:type-A}). The corresponding models are B01, B04 and B05
(Table~\ref{tab:models}). The second aspect is the dependence of the
results on $\theta_0$ for $\dot E = 2 \cdot 10^{50}\,$erg\,s$^{-1}$
and $n=5$ (models B01, B02 and B03; Table~\ref{tab:models}). Thirdly we
compare models where the energy deposition rate per unit volume is
fixed, but $\theta_0$ was varied (models B02 and B06).

We have further computed two models where the energy deposition rate
was assumed to vary with time (models B07 and B08). These models were
set up according to the following prescription for the energy
deposition rate (guided by the results obtained by \citealp{SRJ04})
\begin{eqnarray}
\dot{E}_{\rm B07} = \left\{ \begin{array}{ll}
	   \dot{E}_0 \displaystyle{\frac{t}{0.01}}  & \mbox{for $t
	   \leq 0.01\,$s, } \\
	   \dot{E}_0                 & \mbox{for $0.01\,\mbox{s} < t
	   \leq 0.03\,$s, }  \\
	   \dot{E}_0 \displaystyle{\left(\frac{t}{0.03}\right)^{-3/2}}  & \mbox{for $t >
	   0.03\,$s,}
	          \end{array}
          \right.
\label{eq:B07}
\end{eqnarray}
where $\dot{E}_0=2.35\cdot10^{50}$\ergsec. With this value of
$\dot{E}_0$ the same amount of energy is released in model B07 until
time infinity as in model B01 during its on-time of 100\,ms. We point
out that due to the rather steep power-law decrease of the energy
deposition rate with time in B07, more than 60\% of the total energy
is released within the first 100\,ms. In model B08, the energy
deposition rate follows the same average time dependence as in B07 but
is sinusoidally modulated with a time period $t_{\rm var} = 10$\,ms
according to
\begin{equation}
\dot{E}_{\rm B08} = \dot{E}_{\rm B07} \cdot ( 0.5 \sin( 2 \pi  t/t_{\rm var}
) + 1 )\, .
\label{eq:B08}
\end{equation}

In all other models of type-B we have initiated the energy deposition
by raising the energy deposition rate linearly from zero to its
terminal value within 10\,ms. This timescale leads to a more modest
onset of the hydrodynamic evolution. It can be motivated by the fact
that in neutron star mergers the neutrino luminosities and, therefore,
the energy deposition by $\nu\bar{\nu}$ annihilation increase within a
few milliseconds in response to the viscous shear heating of the
merger remnant \citep[see, \eg][]{Jetal99,RL03}.

\begin{table*}
 \caption{Type-B progenitor models. The first column gives the model
 name, and the second one the total energy $E_d$ released in two
 hemispheres during the simulated time. Columns 3 to 6 display the
 ratio of the total energy (internal plus kinetic) in the region of
 the outflow that has a Lorentz factor larger than a certain threshold
 value (given by the subscript of the energy in each column) relative
 to the energy that is released (in one hemisphere). The seventh column
 provides the fraction of the deposited energy (per hemisphere) that
 is converted into kinetic energy of the outflow with Lorentz factor
 larger than 100. Columns 8 to 10 show the maximum Lorentz factor
 $\Gamma_{\rm max}$, the half-opening angle $\theta_w$ and the mass of
 the fireball $M_f$, respectively. The half-opening angle of the
 fireball is computed as the maximum $\theta$--coordinate of all
 computational cells where the fluid has a Lorentz factor above 10 and
 positive velocities. The mass of the fireball is computed by adding
 up the mass in all computational cells that match the same criterion
 as the one used to compute $\theta_w$.  The last column displays an
 estimate of the asymptotic value of the Lorentz factor of the outflow
 computed as $\Gamma_{\infty} = E_{\Gamma>10}/(M_f c^2)$, \ie assuming
 that all internal energy is eventually converted into kinetic energy,
 and no mass is subsequently swept up from the environment. All the
 data in the table correspond to a time of evolution of 0.5\,s.}
 \label{tab:B-models}
 \centering
 \begin{tabular}{@{}lcccccccccc}
 \hline
 \vspace{-0.4cm}
\\
Model & $E_d$ [erg] &
 $\displaystyle{\phantom{\Biggl(} \frac{E_{\Gamma>100}}{E_d}\phantom{\Biggr)} }$ & 
$\displaystyle{\frac{E_{\Gamma>50}}{E_d}}$ & 
$\displaystyle{\frac{E_{\Gamma>10}}{E_d}}$ & 
$\displaystyle{\frac{E_{\Gamma>2}}{E_d}}$ & 
$\displaystyle{\frac{E_{k,\Gamma>100}}{E_d}}$ &
$\Gamma_{\rm max}$ & $\theta_w$ & $M_f$ [g] & $\Gamma_{\infty}$
%\vspace{0.1cm}
\\
                                                               \hline
B01  & $2\e{49}$ &0.29 &0.36 &   $\phantom{\int^{10}}$ 0.59   $\phantom{\int^{10}}$   &0.70 &$4.4\e{-3}$& 859
&$24\gra$ & $7.4\e{24}$ & 1765 \\ 
B02  &  $2\e{49}$ &0.09 &0.15 &   0.35   &0.48 &$4.8\e{-3}$& 687
&$15\gra$ & $6.4\e{24}$ & 1217 \\ 
B03  &  $2\e{49}$ &0.00 &0.00 &$1.2\e{-4}$&0.04 &0.00      & 16  &
$3\gra$ & $1.9\e{22}$ & 142 \\ 	
B04  &  $10^{48}$ &0.17 &0.26 &   0.52   &0.65 &$5.1\e{-3}$& 492
&$15\gra$ & $3.6\e{23}$ & 1601 \\  
B05  &  $10^{50}$ &0.30 &0.37 &   0.60   &0.71 &$8.3\e{-3}$& 979
&$25\gra$ & $3.6\e{25}$ & 1848 \\    
B06  &  $10^{49}$ &0.19 &0.29 &   0.54   &0.66 &$3.7\e{-3}$& 717
&$18\gra$ & $3.4\e{24}$ & 1761 \\   
B07  &$1.65\e{49}$&0.14 &0.25 &   0.59   &0.72 &$3.4\e{-3}$& 839
&$25\gra$ & $7.6\e{24}$ & 1429 \\
B08  &$1.67\e{49}$&0.18 &0.25 &   0.49   &0.60 &$4.5\e{-3}$& 789
&$21\gra$ & $5.6\e{24}$ & 1607 \\
 \hline
 \end{tabular}
\end{table*}
\begin{figure*}
\begin{center}
\includegraphics[width=0.95\textwidth]{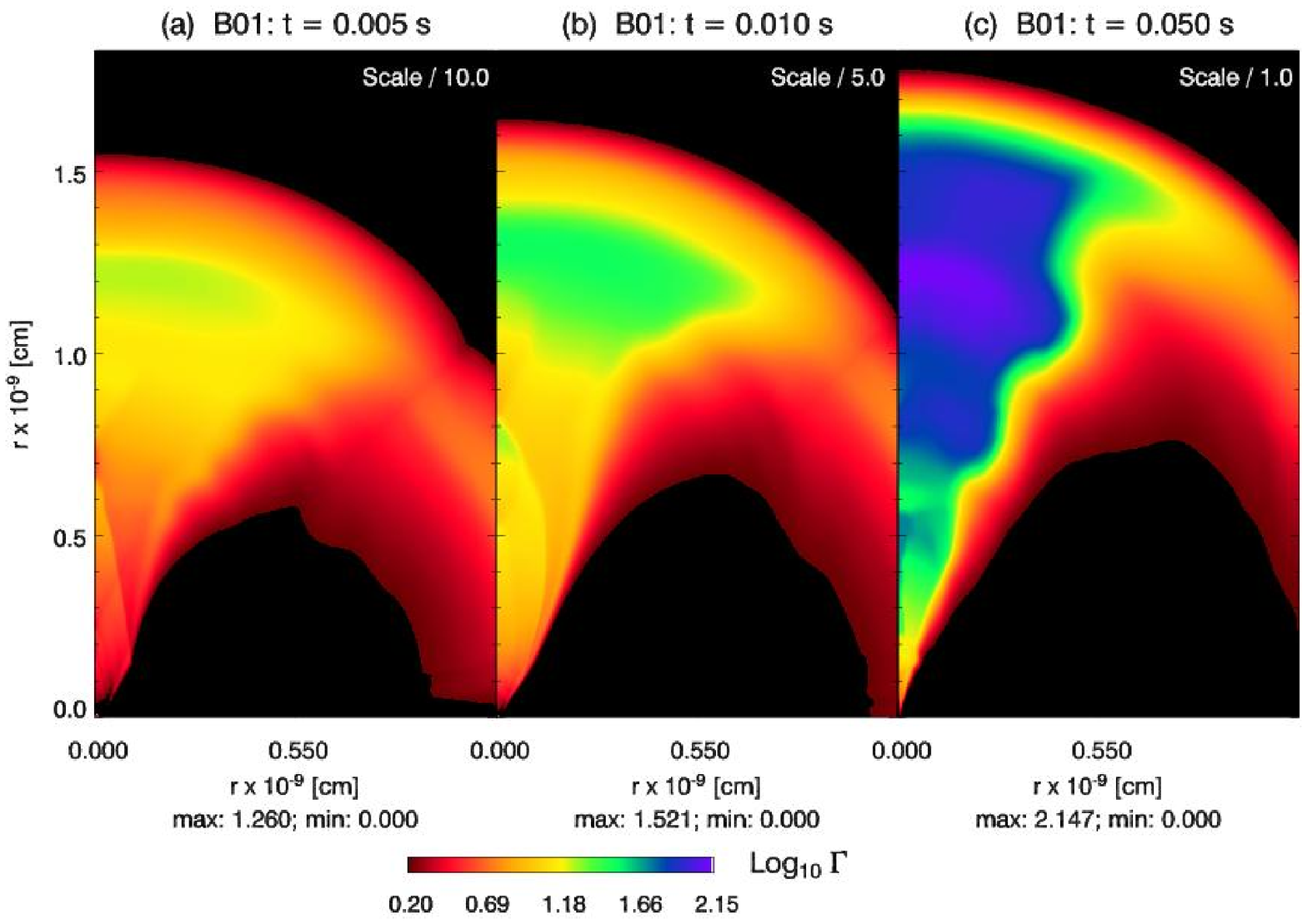}
\end{center}
\caption{Snapshots of the logarithm of the Lorentz factor of model B01
before the shut off of the energy release (the snapshot times are
provided above each panel). The color scale is limited in order to
enhance the details of the outflow region. Note that the distance
scale has been adjusted in the two left panels in order to enlarge the
details of the earlier phases of the evolution. The maximum and
minimum Lorentz factor values (in logarithmic scale) of each model are
given below the panels.}
\label{fig:lor-B01-evol}
\end{figure*}

In type-A models, a relativistic or ultrarelativistic outflow (\ie
fireball) is obtained when the energy deposition rate is larger than a
certain threshold value (\S~\ref{sec:type-A}).  The threshold for
type-B models is much smaller than that for models of type-A in case
the opening angle of the deposition cone is chosen to be the same (for
$\theta_0 = 30\gra$ the threshold is well below
$10^{48}\,$erg$\,$s$^{-1}$).  This can be understood by the lower ram
pressure that the low-density gas exerts on the fireball close to its
site of initiation in type-B models.
\begin{figure*}
\begin{center}
\includegraphics[width=0.95\textwidth]{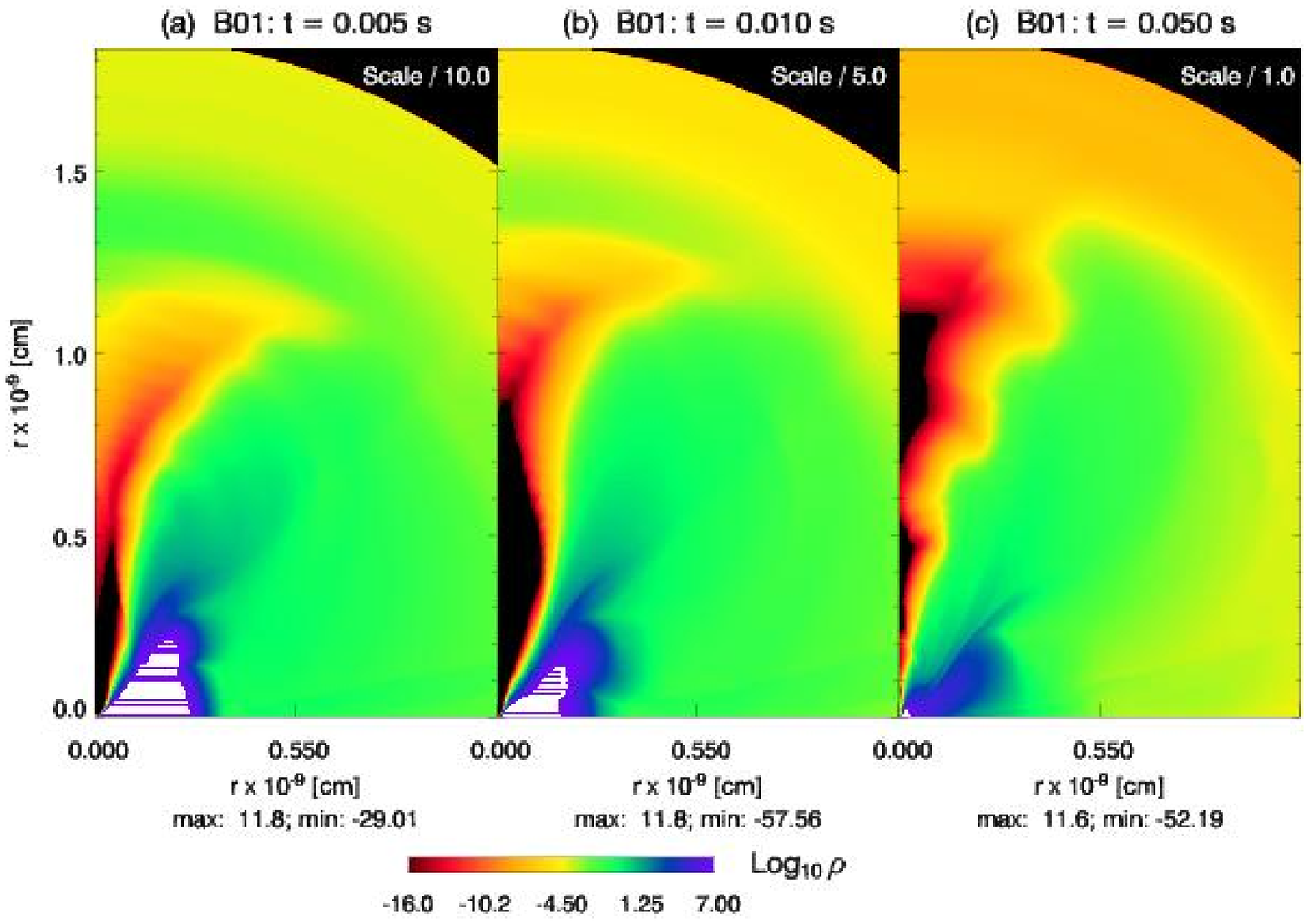}
\end{center}
\caption{Snapshots of the logarithm of the rest mass density of model
B01 before the shut off of the energy release (the snapshot times are
provided above each panel). The color scale is limited in order to
enhance the details of the outflow region. Note that the distance
scale has been adjusted in the two left panels in order to enlarge the
details of the earlier phases of the evolution. The maximum and
minimum density values (in logarithmic scale) of each model are given
below the panels.}
\label{fig:rho-B01-evol}
\end{figure*}

\begin{figure*}
\begin{center}
\includegraphics[width=0.80\textwidth]{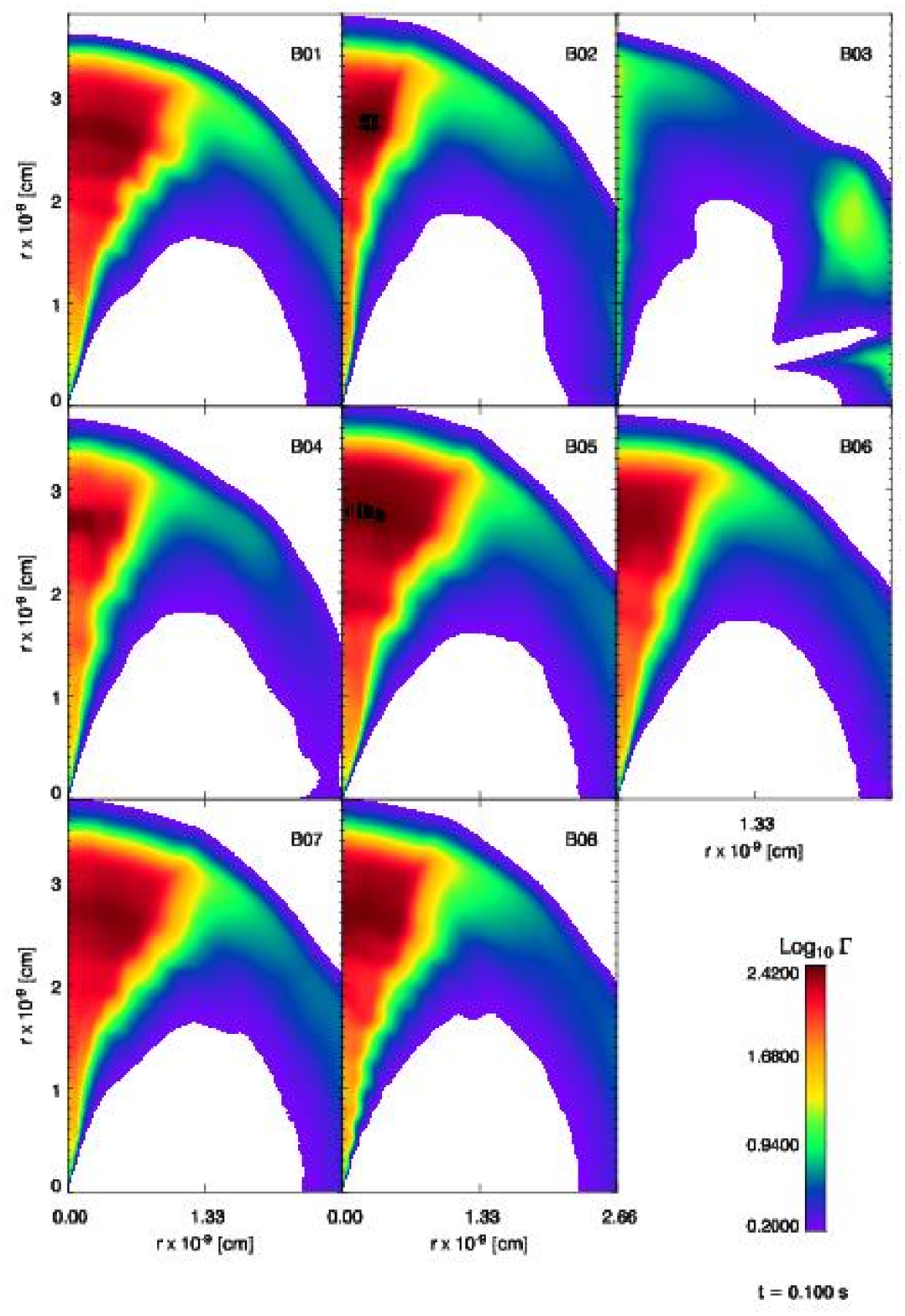}
\end{center}
\caption{Logarithm of the Lorentz factor of the models of type-B after
0.1\,s of their evolution. The color scale is limited in order to
enhance the details of the outflow region.}
\label{fig:lor-Bmodels}
\end{figure*}

\begin{figure*}
\begin{center}
\includegraphics[width=0.80\textwidth]{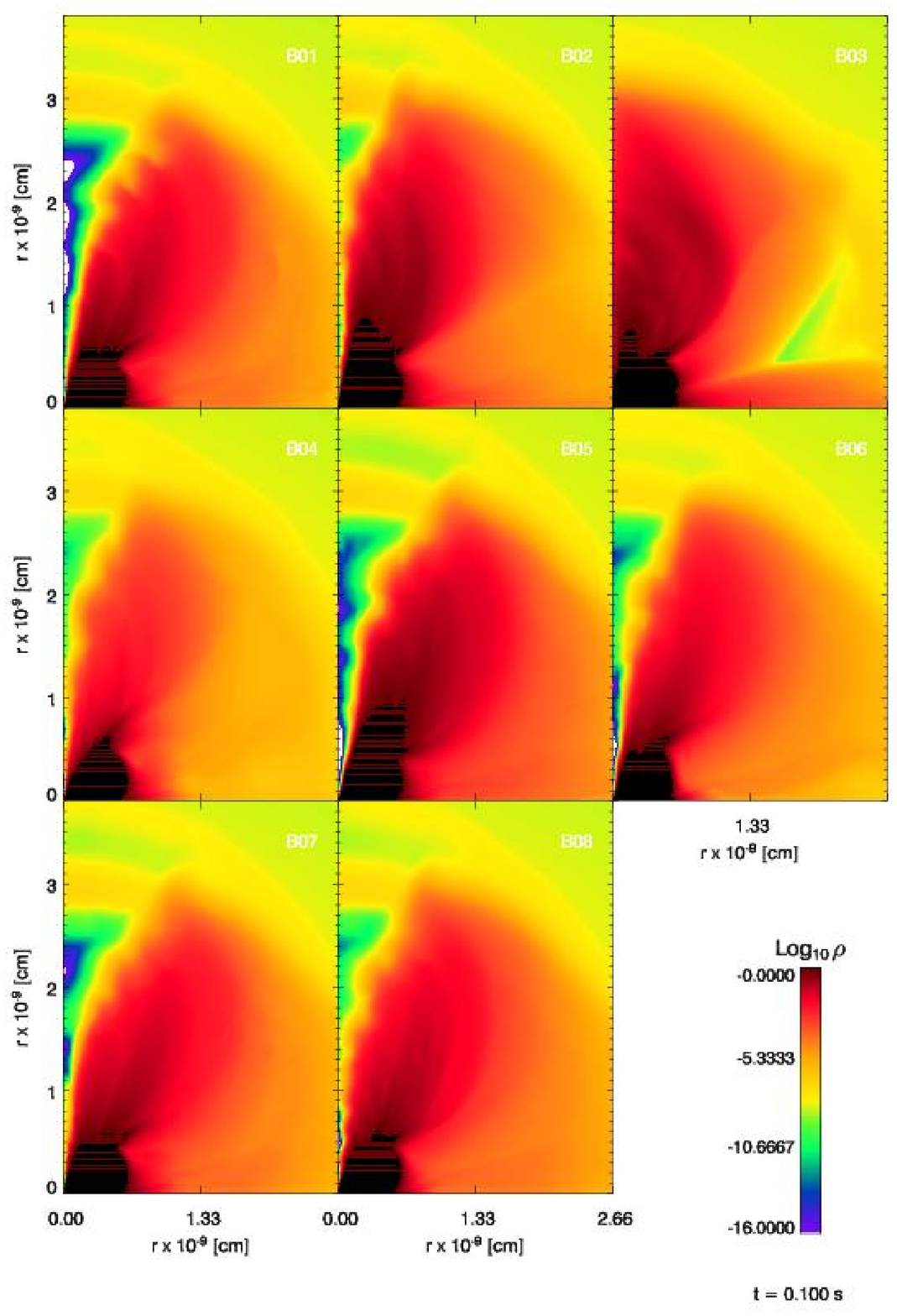}
\end{center}
\caption{Logarithm of the rest-mass density (in \grcm3) of the models
of type-B after 0.1\,s of their evolution. The color scale is limited
in order to enhance the details of the outflow region.}
\label{fig:rho-Bmodels}
\end{figure*}

The shape and the internal structure of the fireballs in models of
type-B (\eg Figs.~\ref{fig:lor-Bmodels} and \ref{fig:rho-Bmodels})
differ substantially from those of type-A models (\eg
Fig.~\ref{fig:lor-Amodels}). Fireballs in type-B models that reach
ultrarelativistic speeds (all the B-cases shown in
Table~\ref{tab:models} except B03) display a conical core with an
opening angle of $\approx 5\gra$ to $\approx 20\gra$ which spreads
into a $26\gra$ to $35\gra$ wide conical region at larger radii. In
addition, the full structure is surrounded by an approximately
spherical shell of matter that expands with moderate Lorentz factors
($< 10$). In this shell the variation of the Lorentz factor with polar
angle depends strongly on the model. This structure is produced by the
deposition of energy in the surface layers of the torus.  Different
from models of type-A, the unshocked region of the outflow (i.e., the
region of the outflow between the fiducial point and the first
recollimation shock) extends to small radial distances (after
$100\,$ms, this region covers \simlt$10^7\,$cm;
Fig.~\ref{fig:r-p-W-axis-B}, except for models B02 and B03) and there
is no simple stratification in the angular direction. In models of
type-B KH instabilities originating in the walls of the torus are
stronger. They produce variations of Lorentz factor
(Fig.~\ref{fig:lor-B01-evol}) and density
(Fig.~\ref{fig:rho-B01-evol}) in the outflow, in particular close to
the angular edge of the central conical structure. In the radial
direction the KH instabilities cause a modulation of the on--average
increasing Lorentz factor, and local fluctuations of the density and
the pressure.

The growth of KH modes from the interface between two relativistic
fluids depends, among other parameters, on the contrast of density
($\chi$) and of specific enthalpy between the two fluids. Roughly, the
larger the density contrast is, the faster the growth of the KH
modes. Thus, the faster growth of KH instabilities close to the
accretion torus in most of the models of type-B can be attributed to
the larger density contrast between the torus and the outflow ($\chi
\equiv \rho_{\rm torus} / \rho_{\rm outflow} \simeq 10^{14}$ ) as
compared to models of type-A ($\chi \simeq 10^{12}$).  As a
consequence of the growth of KH instabilities, mass is entrained in
the relativistic outflow which modifies the local speed of the
fluid. The total mass entrained is comparable to that found for type-A
models. However, in contrast to those models, the amount of matter
piled up in the front part of the outflow is much smaller now (because
the halo is less dense). This allows a highly relativistic propagation
velocity of the fireballs of type-B models
(Table~\ref{tab:models}). Indeed, the propagation speed is so large
that models of type-B are viable to produce GRBs as will be further
discussed in \S~\ref{sec:switchoff}.

The mixing of non-relativistic baryons with the relativistic fluid of
the outflow leads to a change of the adiabatic index in the fireball
which occurs mainly along the boundaries of the outflow.  In a cone of
about $20\gra$ the outflow is still strongly radiation dominated
because the average temperatures and densities are $3\e{7}\,$K and
$10^{-12}\,$\grcm3 (except in case of models B02 and B03 where the
density is much larger, although they are still radiation dominated;
Fig.~\ref{fig:r-p-W-axis-B}), respectively. Thus the adiabatic index
in this region is very close to $4/3$. However, the fit of the
physical variables by power laws as a function of the distance along
the rotational axis produces results which are significantly different
from those obtained for type-A models (\S~\ref{sec:depwithedot}), and
which are not in agreement with the analytic estimates of
\cite{LE00}. This disagreement has two reasons. On the one hand, the
interaction between the external wind medium and the relativistic
outflow is strongly affected by KH instabilities and not restricted to
a thin layer as assumed by \citealp{LE00}). On the other hand, the
accretion torus in our models is thick (i.e., vertically extended) and
not an infinitesimally thin ring as in the analytic treatment by
\cite{LE00}. This prevents the outflow from being sufficiently smooth
to be well fitted by a single power law.

\subsubsection{Dependence on the total energy deposition rate $\dot E$}
\label{sec:depwithedot-B}

To study the dependence of the results for models of type-B on the
assumed total energy deposition rate we have fixed the half-opening
angle of the deposition cone to $45\gra$. The models involved in this
study are B01, B04 and B05 (Table~\ref{tab:models}) with energy
deposition rates of $2\e{50}\,$\ergsec, $10^{49}\,$\ergsec and
$10^{51}\,$\ergsec, respectively.

Increasing the energy deposition rate in models of type-B leads, on
the one hand, to a progressive widening of the ultrarelativistic
outflow and, on the other hand, to an increase of the average Lorentz
factor of the fireball at all times (Fig.~\ref{fig:lor-Bmodels}). From
Table~\ref{tab:models} we learn that the opening angle of the wind is
smaller than the deposition angle $\theta_0$. Thus, as for models of
type-A, the opening angle is not set by our choice of $\theta_0$, but
is constrained by the presence of the thick torus.

Within an outflow crossing time of the torus in the axial direction
($\approx 0.5\,$ms; note that the torus is more compact than in models
of type-A, \ie the crossing time is smaller), the fluid accelerates up
to Lorentz factors of $\approx 3$, \ie the Lorentz factor exceeds the
inverse of the release angle ($\theta_0^{-1} = 1.27$). Hence, the
sideways expansion of the outflow is drastically reduced and the
initially imprinted opening angle remains almost unchanged during the
later propagation of the fireball. Indeed, if the outflow is not
affected by strong KH instabilities (as in the case of model B04), the
fluid quickly enters an ultrarelativistic regime where, after
$10\,$ms, the maximum Lorentz factor can be larger than 30
(Table~\ref{tab:models}).

The amount of mass in the fireball (Table~\ref{tab:models}) increases
linearly with $\dot E$. A linear fit (above the critical threshold)
yields $\log M_f = 0.97 \log \dot{E} - 24.2$, where $\dot E$ and $M_f$
are given in cgs units. The mass mainly accumulatess in the cocoon
which surrounds the outflowing fireball, and near the leading edge of
the outflow. In contrast to the wind solutions found for type-A
models, the density varies by up to 10 orders of magnitude over radial
scales of $\approx 10^8$\,cm~$- 10^9$\,cm within the central cone of
the fireball (Fig.~\ref{fig:r-p-W-axis-B}).

\subsubsection{Dependence on the half-opening angle of the energy
deposition cone}
\label{sec:depwithangle-B}

 For a fixed, moderate energy deposition rate of $2\e{50}\,$erg/s, we
study the influence of an increasing opening half-angle of the
deposition cone from $45\gra$ to $75\gra$ (models B01, B02 and B03;
Table~\ref{tab:models}; Fig.~\ref{fig:lor-Bmodels}). We point out that
these models also differ in the energy deposition rate per unit of
volume (${\dot E}/V_{\rm dep}$).

Depending on the opening angle, either a highly relativistic outflow
is produced (models B01 and B02 display $\Gamma_{\rm max}$\simgt 240),
or only a moderately relativistic jet is generated (model B03,
$\Gamma\,$\simgt 10; Fig.~\ref{fig:lor-Bmodels}). Therefore, the value
of ${\dot E}/V_{\rm dep}$ determines whether a collimated jet (for
small values) or a wide-angle jet (for larger values) is produced.

 Type-B models display a moderately relativistic ($\Gamma\,$\simgt 10)
shell surrounding the central ultrarelativistic cone of the outflow,
which makes it difficult to define an opening angle of the outflow. We
define $\theta_w$ as the angle of the cone around the symmetry axis
where $\Gamma > \Gamma_{\rm op} \equiv 10$. Using values of
$\Gamma_{\rm op}$ in the interval $[10-25]$ does not lead to
substantially different values of $\theta_w$ (except for model B03
where the maximum value of $\Gamma$ is $\Gamma_{\rm max}=17.22$). With
this definition, and for a fixed, moderate total energy deposition
rate the opening angle of the outflow $\theta_w$ is sensitive to an
increase of the deposition angle $\theta_0$ (Table~\ref{tab:models})
because the layer of the outflow between $\approx 17\gra$ and $\approx
35\gra$ enclosing the central relativistic or ultrarelativistic cone
is mostly filled with matter which was pulled up and accelerated from
that part of the torus overlapping with the deposition region. In the
three models under consideration the deposition region extends beyond
$35\gra$. The energy per unit of mass released in the above mentioned
layer is sufficient to speed up the fluid beyond $\Gamma \simgt 100$,
however only in models B01 and B02.

After 100\,ms, $\theta_w$ ranges from $2.3\gra$ (model B03) to
$30\gra$ (model B01), the values decreasing with time
(Table~\ref{tab:models}). In models B01 and B02 this decrease is
caused by the change from a relativistic to an ultrarelativistic flow
due to the continuing deposition of energy. When the fluid speed in
the radial direction approaches the speed of light the transverse
velocity component can only be very small, \ie the sideways expansion
is largely suppressed. In case of B03, a highly collimated jet forms
very early (in less than 20\,ms), and after this wind--to--jet
transition the opening angle of the outflow remains almost constant.
Different from type-A models the transition is not caused by the
inertial confinement due to the external medium, but by the growth of
KH instabilities which tend to pinch the conical channel. These KH
instabilities are of much larger amplitude for models with large
$\theta_0$, because the over-pressure generated by the release of
energy is smaller, \ie more torus matter is entrained into the
outflow. Indeed, the energy in the ultrarelativistic outflow and the
mass of the fireball $M_f$ tends to decrease with increasing
$\theta_0$. Since the entrained mass is not homogeneously distributed
but localized towards the lateral edges of the fireball, in these
regions the Lorentz factor of the outflow decreases below
10. Therefore, they are not included in our computation of the
fireball mass (Table~\ref{tab:models}). The process of mass
entrainment is extreme in the case of model B03, where for most of the
outflow mass $\Gamma < 10$ holds.

Type-B models produce radiation dominated, wide-angle, shocked jets
(see the profiles of the density, pressure and Lorentz factor in
Fig.~\ref{fig:r-p-W-axis-B}). Since they are radiation dominated the
pressure variations effectively decouple from the density fluctuations
(Aloy \etal 2002).  While the radiation pressure displays rather small
variability across and along the fireball and falls almost
monotonically about 15 orders of magnitude in the radial direction,
the density variation in the same direction is more than 50 orders of
magnitude and clearly non-monotonic (Fig.~\ref{fig:r-p-W-axis-B}). The
temperature of the fireball behaves similar as the pressure but with
even less variability. The Lorentz factor, which also shows a
non-monotonic increase with distance, is modulated by a number of
shocks whose shape is either biconical (up to $\approx 10^9$\,cm) or
radial (at larger distances). As a result of the non-monotonic
behaviour, the outflow of type-B models cannot be properly fitted by
power laws as in case of type-A models. The cause of this difference
is easy to understand. The outflow of type-B models is much less dense
than that of type-A models. Thereby, the entrainment of high density
matter from the torus produces larger relative density perturbations
than in type-A models.

The differences between models of type-B and type-A (and also the
theoretical models of \citealp{LE00}) are particularly large in the
radially outer part of the fireball, while in the unshocked region the
two types of models differ much less. This is expected because, on the
one hand, the shocked region is neither included in simplified
theoretical considerations nor can it be fitted by power-laws. On the
other hand, the exact scaling laws of the outflow depend strongly on
the details of the interaction between the surface of the fireball and
the external medium. While this interaction is assumed by \cite{LE00}
to be confined to a thin layer, we find that it takes place in an
angularly extended region where large-amplitude KH instabilities
occur.
\begin{figure}
\begin{center}
\includegraphics[width=0.95\columnwidth]{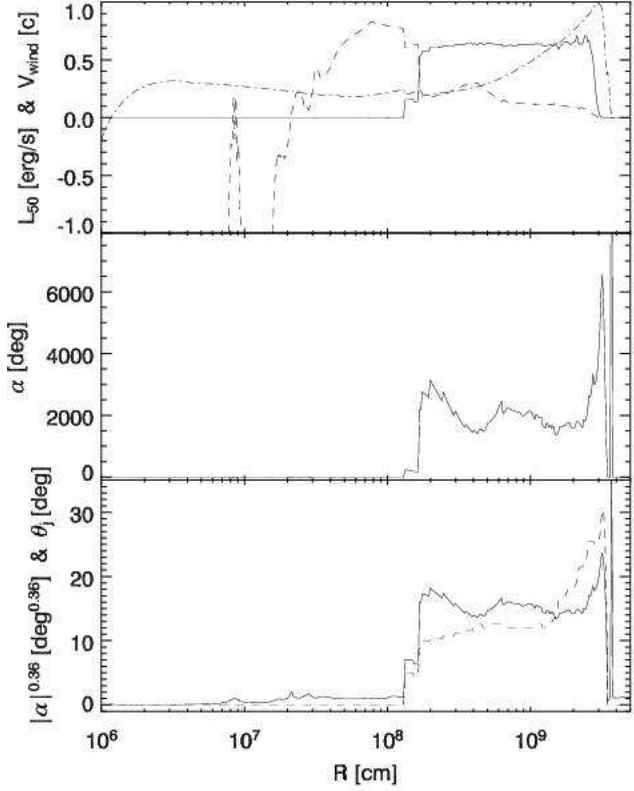}
\end{center}
\caption{The top panel shows the luminosity of the ultrarelativistic
fireball (solid line) and of the surrounding moderately relativistic
wind (dashed line) in units of $10^{50}$\ergsec, as well as the
angularly-averaged, radial velocity of the wind $v_w/c$ (dashed-dotted
line) as a function of the radial distance for model B01 at
$t=0.1\,$s. The middle panel shows the expected fireball half-opening
angle (in degrees) as predicted by the analytic model of \cite{LE00},
i.e., $\alpha = \pi \beta_w^{-1} \Gamma_{\rm fo}^{-2}(L_f/L_w)$, where
$\beta_w=v_w/c$ is the velocity of the surrounding wind medium ($v_w$)
in units of the speed of light ($c$), $\Gamma_{\rm fo}$ is the Lorentz
factor at the base of the jet (which is of order unity in all our
simulations), and $L_f$ and $L_w$ are the luminosities of the fireball
and of the external wind, as defined in the text. For the calculation
of $\alpha$, the angularly-averaged, radial velocity $v_w$ of the
external wind medium (see top panel) is computed at all radii
$R$. Finally, the bottom panel shows $|\alpha|^{0.36}$ as a function
of the radius (solid line) compared with the local half-opening angle
of the fireball $\theta_j$ as a function of the radius (dashed
line). The value of $\theta_j$ is computed as the maximum polar angle
where, at a distance $R$, the Lorentz factor of the fluid is larger
than 10. Note that the fireball half-opening angle $\theta_w$ (given
in Tables~\ref{tab:models} and \ref{tab:B-models}) is defined as the
maximum of all the local half-opening angles $\theta_j$.}
\label{fig:LE-angle}
\end{figure}

As in case of type-A models (see Sect.~\ref{sec:depwithangle-A}) the
theory of \cite{LE00} is also not suitable to describe the collimation
of the ultrarelativistic outflow obtained in our type-B models, where
the presence of the torus is responsible for the jet collimation,
too. This fact is supported by the results shown in
Fig.~\ref{fig:LE-angle} where the values of the half-opening angle
($\alpha$) of the ultrarelativistic outflow are displayed calculated
according to the simplified analytic model of \cite{LE00}, which
assumes that an external, baryonic wind is responsible for the
collimation of the jet: $\alpha \approx \pi \beta_w^{-1} \Gamma_{\rm
fo}^{-2}(L_f/L_w)$, where $\beta_w = v_w/c$, $v_w$ is the velocity of
the wind, $\Gamma_{\rm fo}$ is the Lorentz factor at the base of the
jet ($\Gamma_{\rm fo} \approx 1$ in all our simulations), and $L_f$
and $L_w$ are the luminosities of the fireball and of the external
wind, respectively. These luminosities are evaluated as $L_f=2\pi R^2
c^3 (1-2R_g/R)^{3/2} \int_{0}^{\theta_w(R)} d\theta \sin^2\theta \rho
\Gamma \beta_r (h\Gamma-1)$ and $L_w=2\pi R^2 c^3 (1-2R_g/R)^{3/2}
\int_{\theta_w(R)}^{\pi/2} d\theta \sin^2\theta \rho \Gamma \beta_r
(h\Gamma-1)$, respectively. In these expressions, $h$ is the specific
enthalpy and $\beta_r=v_r/c$ with $v_r$ being the radial velocity. For
this quantitative comparison we identify the moderately relativistic
outflow with $\Gamma < 10$ around the fireball with the baryonic wind
which produces the jet collimation in Levinson \& Eichler's model. We
further assume that the velocity of the wind $v_w$ can be approximated
at each radial distance $R$ by the angularly-averaged, radial velocity
of the external medium.  The latter velocity is computed as the
arithmetic mean of the radial velocities of all cells at a distance
$R$ in the interval between $\theta_w(R)$ (the polar angle at which
the Lorentz factor becomes smaller than 10 at a radial distance $R$)
and $\pi/2$.

 Comparing the values of the fireball half-opening angle shown in
Table~\ref{tab:models} with the huge (meaningless) values shown in the
bottom panel of Fig.~\ref{fig:LE-angle}, we conclude that the
non-relativistic wind is clearly not responsible for the fireball
collimation. We point out that the value of $\alpha$ is not unique. It
changes by a factor of $\sim 3$ depending on the radial distance at
which it is computed. A similar variability is also found for the
local half-opening angle of the fireball, $\theta_j$, calculated as
the maximum polar angle where, at a distance $R$, the Lorentz factor
of the fluid is larger than 10 (Fig.~\ref{fig:LE-angle} bottom
panel). We have found that the function $|\alpha|^{0.36}$ reasonably
fits $\theta_j$.

\subsubsection{Dependence on the deposition volume $V_{\rm dep}$ and 
on the total energy deposition rate $\dot E$ for ${\dot E}/V_{\rm
dep}$=constant.}
\label{sec:depwithedens-B}

In two models of type-B we have varied the half-opening angle of the
energy deposition cone simultaneously with the total energy deposition
rate $\dot E$, while keeping the rate of energy deposition per unit of
volume ${\dot E}/V_{\rm dep}$ constant. The two corresponding models
are B02 (with a large energy deposition rate and deposition angle
$\theta_0 = 60\gra$) and B06 (with a smaller deposition rate and
$\theta_0 = 41.4\gra$).

Both models develop a wide-angle ultrarelativistic jet because the
lateral edges of the deposition region and, further downstream, also
the lateral boundaries of the outflow are causally connected, \ie KH
modes can grow modulating the variation of the physical variables
inside the outflow. Both models show similar global properties
(propagation speed of the fireball, opening angle of the wind, mass of
the fireball) during the time when energy is deposited
(Table~\ref{tab:models}). A more detailed comparison reveals minor
differences, for example, in the maximum Lorentz factor (larger in
model B02), the propagation velocity (larger in B02), and the fireball
opening angle after a canonical time of 100\,ms (smaller in B02 than
in B06, although in both cases $\theta_w \approx 22\gra$; see
Table\ref{tab:models}). The model with the smaller deposition angle
(B06) sweeps up less mass from the external medium, and drags less
mass off the torus (the deposition region overlaps with the torus by
$\approx 16\gra$) than the model with the larger deposition angle and
deposition rate (the energy deposition cone overlaps with the torus by
$\approx 35\gra$ in B02). This explains why the outflow in the latter
model is more massive ($M_f^{B02} = 2.5\e{24}\,$g) than in model B06
($M_f^{B06} = 1.4\e{24}\,$g).

Looking at the density distribution (Fig.~\ref{fig:rho-Bmodels}) we
see that the ultrarelativistic region of model B02 is denser (on
average) and displays much smaller density fluctuations than model B06
(Fig.~\ref{fig:r-p-W-axis-B} top panel). Furthermore, model B06
develops a more extended unshocked region
(Fig.~\ref{fig:r-p-W-axis-B}), where the Lorentz factor is larger than
that of model B02 after an evolutionary time of 100\,ms
(Fig.~\ref{fig:r-p-W-axis-B} bottom panel).  The propagation velocity,
although very similar, is systematically smaller in model B06. The
differences between the two models arise from the larger angular
extent of the deposition region in model B02. Since the deposition
region overlaps more with the torus in model B02 than in model B06
there is more mass pulled up from the torus in the former model. This
explains the larger density and, as a result, the smaller fluid
Lorentz factor in the shocked region of the fireball.

\subsubsection{Dependence on the time variability of the energy source}
\label{sec:depwithvariability-B}

In models B07 and B08 we have varied the energy deposition rate
according to the prescriptions given in Eqs.~(\ref{eq:B07}) and
(\ref{eq:B08}). The energy released in models B07 and B08 until time
infinity is the same as in model B01 during 100\,ms. The width of the
deposition region is also equal to that of model B01. With both models
we try to determine the consequences of an intrinsic source
variability as compared to the variability caused by the interaction
with the accretion torus and the merger halo.

After 100\,ms of continuous deposition of energy, the decay of the
energy deposition rate of model B07 for $t > 0.03\,$s has not produced
any obvious signature in the Lorentz factor of the outflow, which in
fact looks similar as in model B01 (Fig.~\ref{fig:lor-Bmodels}). Even
the variations of Lorentz factor in the outflow are almost
indistinguishable between the model with constant energy release and
the models with a variable, {\em burst-like} deposition of
energy. However, the ultrarelativistic central core of the outflow of
model B07 is (on average) $\sim 50\%$ denser than that of model B01
(Fig.~\ref{fig:rho-Bmodels}), which is a consequence of a larger mass
entrainment resulting from the smaller ${\dot E}/V_{\rm dep}$ near the
system axis after $\sim 35\,$ms relative to model B01.

The periodic variation of the energy deposition rate yields a
modulated growth of the Lorentz factor of model B08 the periodicity
following that of the energy release (in particular up to distances of
$\sim 2\e{9}\,$cm, where the grid resolution starts to become too
coarse, and along the boundaries of the ultrarelativistic core of the
fireball; Fig.~\ref{fig:lor-Bmodels}). Model B08 also has a larger
mass entrainment than model B01 for the reason mentioned above. It is
slightly larger than in model B07 because of the sinusoidal variation
of the deposition rate: ${\dot E}/V_{\rm dep}$ of model B08 is $50\%$
smaller than that of model B07 during the epochs of minimum energy
release (\ie when $\sin( 2 \pi t/t_{\rm var})=-1$), when a larger
amount of mass is entrained. Once that mass is inside of the fireball,
it is never expelled again.

\section{Evolution after the switch-off of the central energy source}
\label{sec:switchoff}

In the previous sections we have discussed the evolution of the
relativistic outflow that forms by the release of thermal energy
around a black hole-accretion torus system. The simulations were
carried on for a canonical time of 100\,ms during which the source was
assumed to release energy either at a constant rate (models B01 to
B06) or with a rate decaying with time (models B07 and B08) as
$t^{-3/2}$ (see \S~\ref{sec:type-B}). After 100\,ms, the different
fireballs have reached a distance of $\simeq 3\e{9}\,$cm. However,
according to the standard model, the GRB phenomenon can be produced
only at radial distances of $\simgt 10^{13}\,$cm from the energy
source, \ie the outflow must still propagate and persist over another
4 orders of magnitude in radius before an observable signal
emerges. In order to investigate longer timescales than the epoch of
energy release, we have followed the evolution of the type-B models
for another 400\,ms after the central energy supply had either been
shut off (B01 to B06) or during which it continued at a decreasing
rate (models B07 and B08).

In the following sections we will concentrate on the evolution of the
prototype model B01 (\S~\ref{B01-switchoff}), and afterwards we will
outline the particularities of the other type-B models
(\S~\ref{B02-B08-switchoff}).

\subsection{Model B01}
\label{B01-switchoff}

\begin{figure*}
\begin{center}
\includegraphics[width=0.95\textwidth]{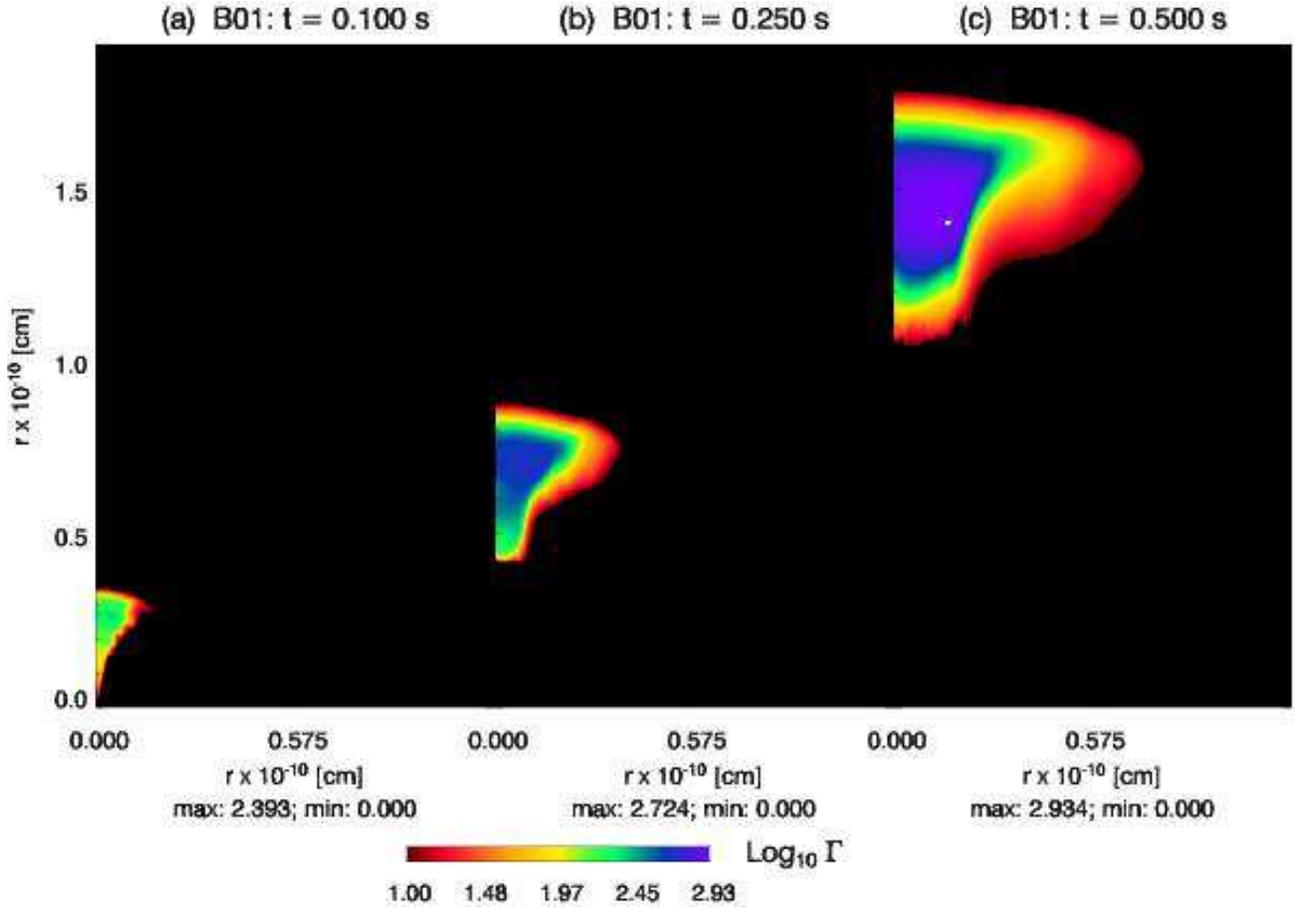}
\end{center}
\caption{Evolution of logarithm of the Lorentz factor of model B01
after the shut down of the central energy supply.  Time is measured
from the moment when the energy deposition was started.}
\label{fig:lor_B01n}
\end{figure*}

\begin{figure}
\begin{center}
\includegraphics[width=0.95\columnwidth]{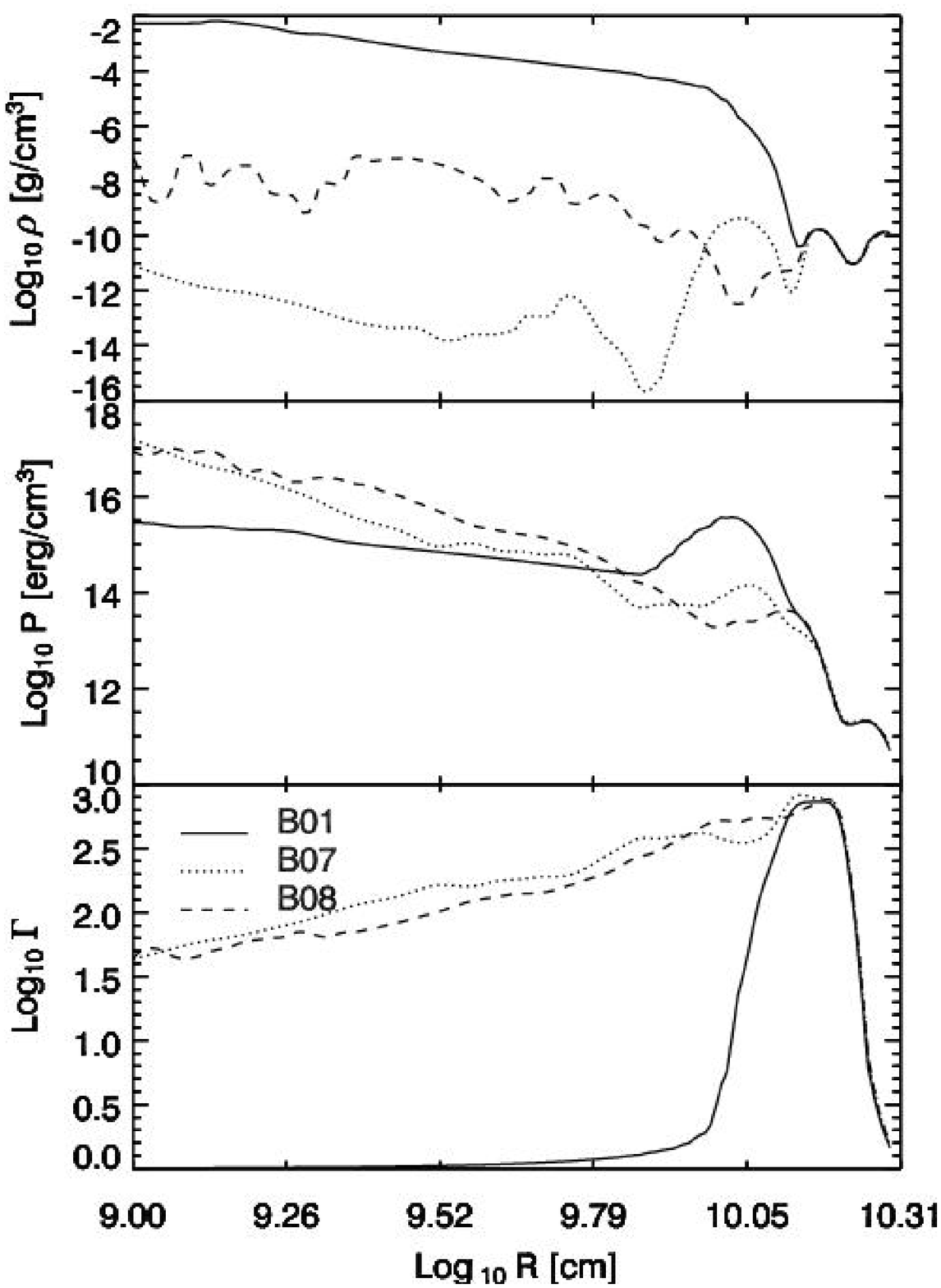}
\end{center}
\caption{Logarithm of the rest-mass density (top panel), of the
pressure (middle panel) and of the fluid Lorentz factor (bottom panel)
vs radius along the symmetry axis for models B01, B07 and B08 (see
legend in the plot) after 500\,ms.}
\label{fig:r-p-W-axis-B01-B07-B08}
\end{figure}

\begin{figure}
\begin{center}
\includegraphics[width=0.95\columnwidth]{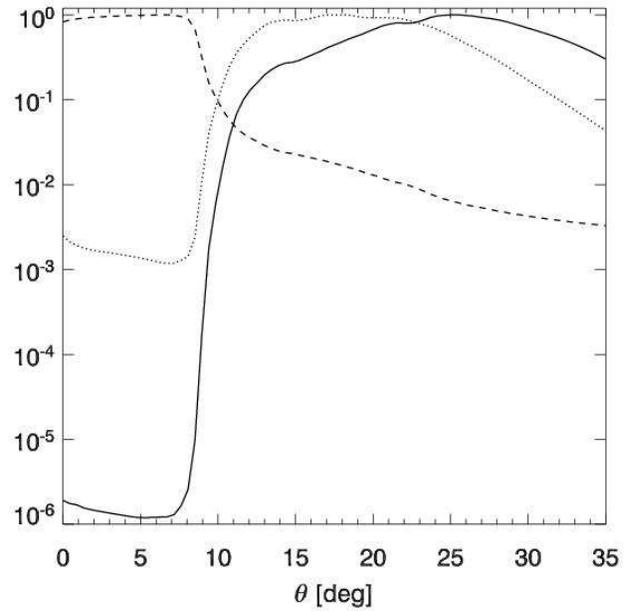}
\end{center}
\caption{Lorentz factor (dashed line), pressure (dotted line) and
rest-mass density (solid line) in polar direction at a radial distance
of $1.4\e{10}\,$cm after 500\,ms of evolution of model B01. The
quantities are normalized to their respective maxima in the displayed
angular interval. Note the rarefaction visible between the center and
the edges of the fireball (at $\approx 8\gra$).}
\label{fig:r-p-W-angle-B01}
\end{figure}

After the epoch of 100\,ms of a constant rate of energy release has
ended (except for the first 10\,ms during which the energy deposition
rate is linearly increased), the high-Lorentz factor outflow possesses
a conical shape which decouples from the energy deposition region, and
moves away from the black hole (Fig.~\ref{fig:lor_B01n}).  The shape
is mainly a consequence of the continuing radial acceleration of the
matter in the outflow. The angular structure of the adiabatically
expanding fireball is non-uniform because of a larger baryon loading
at larger polar angles. The Lorentz factor reaches values between 20
and 50 (approximately, between $\approx 15\gra$ and $30\gra$) where
the fireball is loaded with baryons extracted from the outermost
layers of the torus. These interaction layers have still mildly
relativistic or subrelativistic velocities at the time the energy
release is switched off. In contrast, the ultrarelativistic core has
swept up only matter from the dilute halo. One, therefore, finds
variations of the Lorentz factor (Fig.~\ref{fig:lor_B01n}) which
depend on the lateral angle (in addition to the radial variations of
$\Gamma$). The maximum Lorentz factor ($\Gamma_{\rm max} \simeq 850$)
is attained close to the lateral edges of the fireball core (at polar
angles around $8\gra$) and in its rear part, while in the region
around the axis ($\theta < 3\gra$) the fluid Lorentz factor is about
$20\%$ smaller than its maximum value
(Fig.~\ref{fig:r-p-W-angle-B01}). For small angles, the Lorentz factor
exhibits a fast rise from the backward end of the fireball to its
center where values beyond 800 are found
(Fig.\ref{fig:r-p-W-axis-B01-B07-B08}). From this local maximum the
Lorentz factor decreases slightly towards the terminal shock of the
wind (see the violet shade in the right panel of
Fig.~\ref{fig:lor_B01n}). For larger angles ($\theta > 15\gra$) the
radial profile of the Lorentz factor is similar although the values
here are much smaller ($\Gamma \simeq 30$). This radial profile of the
Lorentz factor is the result of the beginning of a coasting phase,
that starts soon after the energy release ceases.

\begin{figure*}
\begin{center}
\includegraphics[width=0.95\textwidth]{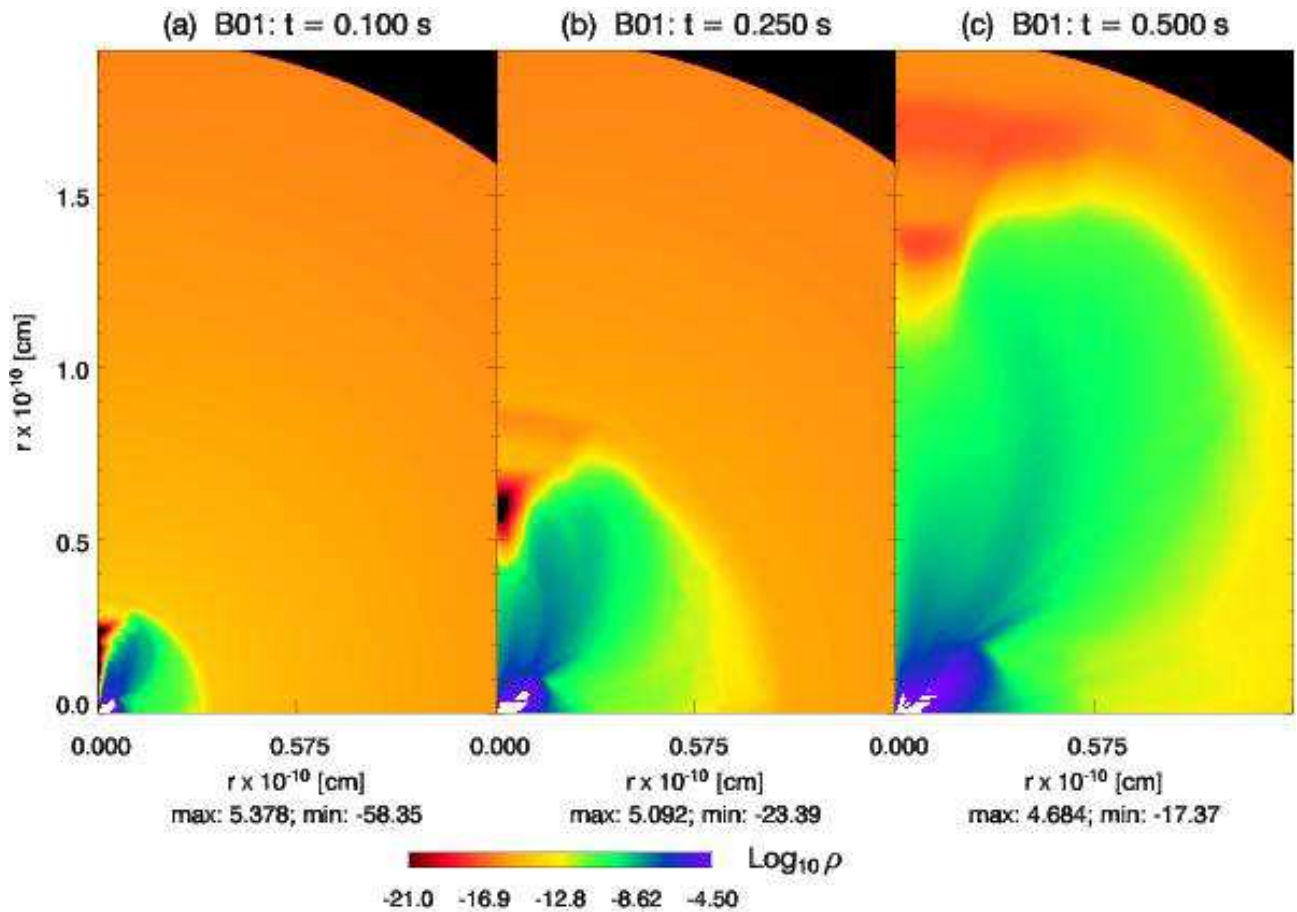}
\end{center}
\caption{Evolution of the logarithm of the rest-mass density (in
\grcm3) of model B01 after the shutdown of the central energy supply.
Time is measured from the moment when the energy deposition was
started.}
\label{fig:rho_B01n}
\end{figure*}

The rest-mass density (Fig.~\ref{fig:rho_B01n}) in the fireball does
not preserve the irregular structure imprinted by the KH instabilities
that had developed near the lateral edges during the earlier
evolution. After the shutdown of the energy deposition, two {\em
voids} (density minima) show up in the central part of the fireball,
one close to its rear part and another one close to the front of the
fireball (solid line in Fig.\ref{fig:r-p-W-axis-B01-B07-B08}). The two
voids are radially separated by a $20$ times denser ``wall'',
corresponding to a reverse shock that sweeps backwards into the
fireball.

The pressure in the fireball decreases with increasing radial distance
and increasing polar angle. Across the fireball, in the polar
direction, the pressure is lowest close to the side edges of the
fireball core and is highest around the symmetry axis (where the
density is about a factor of 2 larger than at the sides;
Fig.~\ref{fig:r-p-W-angle-B01}). The pressure variation in the radial
direction can be as large as three orders of magnitude
(Fig.\ref{fig:r-p-W-axis-B01-B07-B08}). However, due to the fact that
the fireball is radiation dominated, the variation of pressure (or
temperature) is much smaller than that of the rest-mass density.
The structure of the outflow can be understood by the combination of
three main dynamical effects. First, there is a trend to adopt a
radially stratified wind structure due to the expansion of the outflow
as it moves radially outwards. This explains the decrease of the
pressure, temperature and density (and the increase of the Lorentz
factor) in the radial direction. Second, there are local variations of
the physical variables caused by the KH instabilities during the early
interaction of the fireball with the external medium and,
particularly, with the torus. The largest of these instabilities
survive beyond the shutdown of the energy deposition. Most of the
small scale variations are (numerically) suppressed because of the
logarithmic coarsening of the grid in the radial direction. However,
KH modes with wavelengths smaller than the width of the thick,
expanding layer surrounding the central ultrarelativistic cone are
also physically damped \citep{Bir91}. Third, there is a genuinely
relativistic effect \citep{Po00,RZ02} that leads to the generation of
a rarefaction wave and a shock (instead of two shocks) for collisions
between fluids in case of a small impact angle (an example of this
phenomenon in the context of parsec scale jets can be found in
\citealp{Alo03}). The effect arises in the presence of a non-vanishing
tangential velocity component at a relativistic shock. The generalized
jump conditions require that the quantity $h\Gamma v_t$ is equal
(where $h$ and $v_t$ are the specific enthalpy and the velocity
component tangential to the shock front, respectively) on both sides
of the shock.  The rarefaction propagates from the collision region
into the medium that has smaller density and higher specific
enthalpy. The shock moves in the opposite direction into the fluid
that has larger density and smaller specific enthalpy.

This described situation applies to our models when the fireball
expands sideways (i.e., in the lateral direction) and interacts with
the external medium and the torus. Tangential velocity components are
present due to the radial fluid motion in the fireball, the moderately
relativistic rotation of the torus, and the expansion of the
nonrelativistic ejecta that are {\it blown off} the torus surface due
to the energy deposition. The interaction leads, on the one hand, to
the formation of a rarefaction wave that starts from the lateral edges
of the fireball, propagates very slowly towards the axis, and leaves
behind a region with negative angular ($\theta-$)velocities (\ie
velocities directed to the axis). On the other hand, the interaction
generates a shocked layer that encloses the fireball's side edges.
Because of the rarefaction wave the pressure near the edges of the
fireball core (which is a region of low density and high specific
enthalpy) is smaller than near the axis (in
Fig.~\ref{fig:r-p-W-angle-B01}, the rarefaction extends from $\approx
7\gra$ to $\approx 14\gra$). Analytic modeling of the collision of the
fireball with the external medium in two dimensions (assuming planar
symmetry) shows that a rarefaction wave forms instead of a shock when
the impact angle is smaller than $\sim 35^{\circ}$.

We point out that the existence of such a rarefaction wave was
included neither in analytic models ---\cite{LE00} assumed that a thin
shocked layer that results from the propagation of two oblique
shocks--- nor in phenomenological fireball-jet models assumed to be
either homogeneous \citep[e.g.,][]{Piran99,Piran00} or structured
\citep[\eg][]{KG03,Salmonson03}.
\begin{figure}
\begin{center}
\includegraphics[width=0.95\columnwidth]{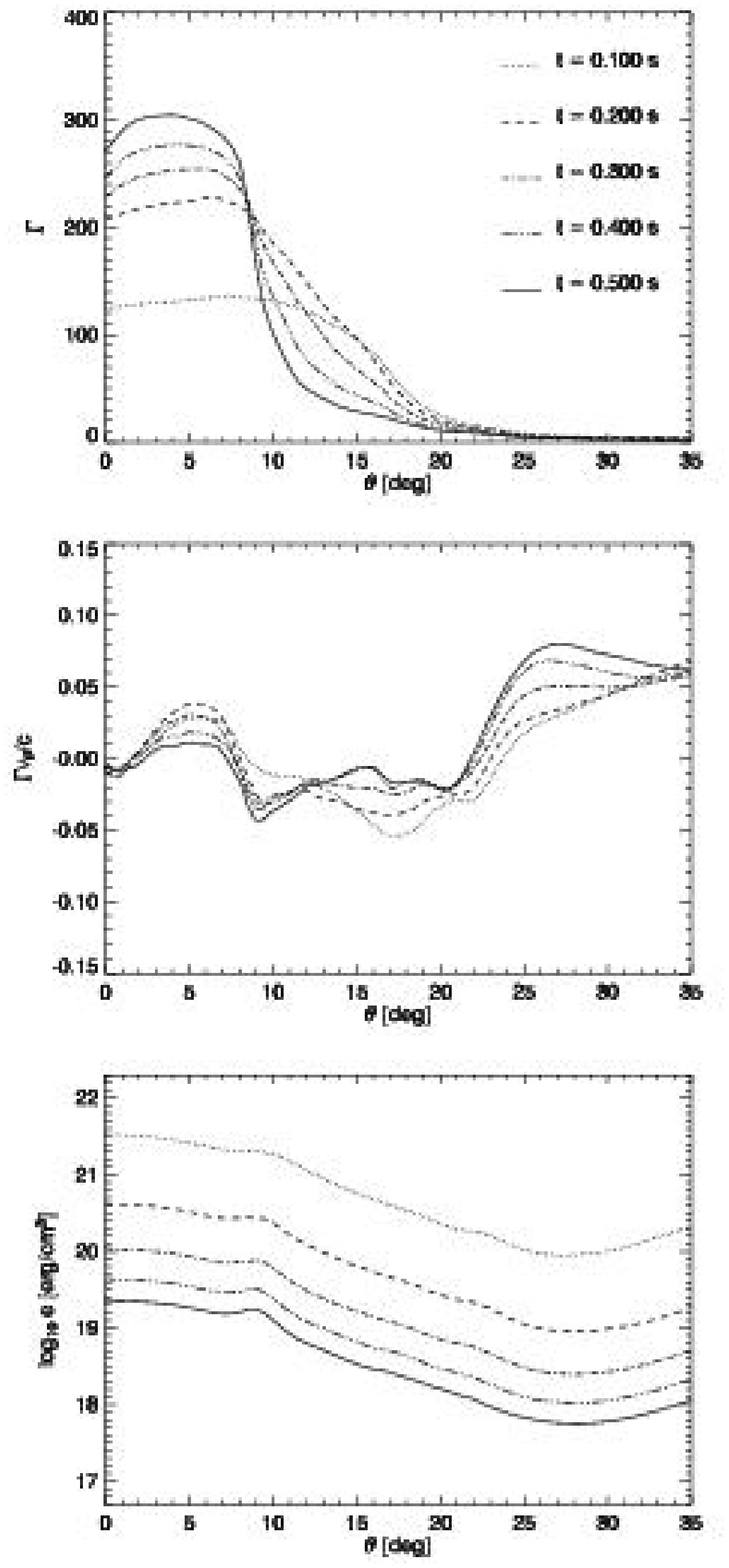}
\end{center}
\caption{Radially averaged values of the Lorentz factor (top panel),
angular component of the four-velocity (intermediate panel) and energy
density (bottom panel) as a function of the polar angle for model
B01. The different lines correspond to different moments in time
spaced by 100\,ms (the times are listed in the top panel).}
\label{fig:raverages_B01n}
\end{figure}

From the above discussion it is clear that the ultrarelativistic
outflow is laterally inhomogeneous and not bounded by {\em sharp
edges} (as assumed in the so-called homogeneous or {\it top-hat} jet
models).  The outflow profile can also not be fitted by a monotonic
function (a power-law or a Gaussian function as assumed in the {\it
structured} jet models). Let us now consider the one-dimensional
angular outflow structure by calculating the radial average of the
quantities displayed in Fig.\ref{fig:raverages_B01n}. We apply the
formula $\bar{A} = (\Delta V_r)^{-1} \int^{R_{max}}_{R_{min}} r^2 A
dr$, with $\Delta V_r = (R_{max}^3 - R_{min}^3) / 3$, $R_{max}$ and
$R_{min}$ being the radii of the front and of the rear end of the
relativistic outflow, respectively. We find that the evolution of the
angular profiles is rather different from the behaviour discussed by
\cite{KG03}. The angular wedge with high Lorentz factors becomes
narrower with time and the angular location of the peak Lorentz factor
moves to smaller angles. At the same time the peak value increases and
becomes larger than the value on the symmetry axis
(Fig.~\ref{fig:raverages_B01n}). A good fit for the radially averaged
Lorentz factor ($\bar{\Gamma}$) as a function of the polar angle in
the interval $[0\gra,90\gra]$ is given by the expression
\begin{eqnarray}
\bar{\Gamma} = 1 + \frac{a_0 + a_1x + a_2x^2 + a_3x^3}
                        {a_4 + a_5x + \exp{(a_6x)}}\: . 
\label{eq:lorfit}
\end{eqnarray}

The values of the seven coefficients $a_i$ and the $\chi^2$ of the fit
are listed in Table~\ref{tab:fits} for the type-B models at a time of
0.4\,s after the shutdown of the central energy source. Attempts to
fit simpler functions in a narrower interval (namely,
$[0\gra,20\gra]$) were unsuccessful. In particular, a Gaussian profile
of the form $ae^{bx^2}+1$, or a quadratic ansatz of the form
$a/(b+cx^2)$ had to be rejected because the resulting $\chi^2$ values
are 176 and 860, respectively.

Figure~\ref{fig:raverages_B01n} shows that during the last 0.4\,s the
increase of the averaged Lorentz factor per unit of time in a $10\gra$
cone around the axis is much smaller than during the first
0.1\,s. However, we also see that the averaged Lorentz factor has not
yet converged to an asymptotic value after 0.5\,s, \ie acceleration
continues. Table~\ref{tab:B-models} gives a rough estimate of the
maximum bulk Lorentz factor (defined in the table caption) that can be
asymptotically reached by the fireball when its whole internal energy
is converted into kinetic energy (provided there is no further
acceleration or deceleration due to the presence of a pressure
gradient in the ambient medium of swept-up matter, and that the medium
at larger distances from the fireball origin is uniform and very
diluted).  The values of $\Gamma_{\infty}$ are still $\sim 2$ times
larger than the corresponding values of $\Gamma_{\rm max}$ except for
model B03 (Table~\ref{tab:B-models}). This is consistent with our
hypothesis according to which the value of $\Gamma_{\infty}$ sets an
upper limit to the maximum bulk Lorentz factor of the
outflow. Furthermore, the fact that $\Gamma_{\rm max} <
\Gamma_{\infty}$ suggests that there is still room for additional
acceleration of the fireball fluid.

\begin{table*}
 \caption{Fit coefficients of the lateral profiles
 (Eq.~\ref{eq:lorfit}) of the radially averaged Lorentz factor 0.4\,s
 after of the cease of the energy release. Note that model B03 cannot
 be matched with the functional form given in Eq.~(\ref{eq:lorfit})
 and. $\chi^2$ provides a measure of the quality of the fit.}
 \label{tab:fits}
 \centering
 \begin{tabular}{@{}lcccccccc}
 \hline

Model& $a_0$ & $a_1$ & $a_2$ & $a_3$ & $a_4$ & $a_5$ & $a_6$ &
$\displaystyle{\phantom{\Bigl(} \chi^2 \phantom{\Bigr)}}$\\
\hline 
B01$\phantom{\int^{10}}$  & $1.2\e{5}$ & $-5.3\e{5}$ & $-4.9\e{6}$ & $2.5\e{7}$ & 
       $4.4\e{2}$ & $-3.1\e{3}$ & 31  & 5.3 \\
B02  & $5.9\e{2}$ & $-1.5\e{4}$ & $1.6\e{5}$ & $-2.2\e{5}$ &
       $1.5\e{0}$ & $-9.1\e{2}$ & 28 & 1.6 \\
B04  & $1.5\e{3}$ & $-2.2\e{4}$ & $8.3\e{4}$ & $3.4\e{5}$ & 
       $8.1\e{0}$ & $-1.6\e{2}$ & 28 & 0.4 \\
B05  & $2.1\e{5}$ & $-1.2\e{6}$ & $-5.5\e{6}$& $3.3\e{7}$ & 
       $6.2\e{2}$ & $-4.4\e{3}$ & 31 & 2.8 \\
B06  & $4.5\e{4}$ & $-3.5\e{5}$ & $-1.0\e{6}$& $1.3\e{7}$ &
       $1.9\e{2}$ & $-1.7\e{3}$ & 35 & 1.0 \\
B07  & $7.1\e{3}$ & $2.3\e{4}$  & $-7.1\e{5}$& $2.7\e{6}$ & 
       $1.8\e{1}$ & $-1.9\e{1}$ & 24 & 3.2 \\
B08  & $1.3\e{3}$ & $-1.4\e{4}$ & $9.7\e{5}$ & $-7.9\e{4}$ & 
       $2.5\e{0}$ & $-5.8\e{1}$ & 20 & 0.9 \\
 \hline
 \end{tabular}
\end{table*}
Profiles of the radially averaged energy density ($\bar{e}$) can 
be fitted as a function of the polar angle, in the interval
$[0\gra,90\gra]$, by polynomials of the form
\begin{eqnarray}
\bar{e} = \sum_{i=0}^{4} b_ix^i\:. 
\label{eq:enefit}
\end{eqnarray}
 
The values of the five fit coefficients $b_i$ 0.4\,s after the shut
down of the energy deposition are listed in
Table~\ref{tab:fits-ener}. Simple second-order polynomials provide good
fits to $\bar{e}$ in the interval $[0\gra,20\gra]$.

\begin{table}
 \caption{Fit coefficients of the lateral profiles of the radially
 averaged energy density (Eq.~\ref{eq:enefit}) at $t=0.5$\,s.}
 \label{tab:fits-ener}
 \centering
 \begin{tabular}{@{}lcccccc}
 \hline

Model& $b_0$ & $b_1$ & $b_2$ & $b_3$ & $b_4$ & $\displaystyle{\phantom{\Bigl(} \chi^2 \phantom{\Bigr)}}$ \\
\hline 
B01 $\phantom{\int^{10}}$  &   19  &  4.1  &  -44  &   81  &  -41  &   0.003  \\ 
B02  &   20  & -0.26 &  -37  &   88  &  -53  &   0.004  \\
B03  &   19  & -4.2  & -2.6  &   22  &  -15  &   0.002 \\
B04  &   18  & -0.72 &  -40  &  110  &  -79  &   0.010 \\
B05  &   20  &  4.3  &  -33  &   44  &  -13  &   0.004 \\
B06  &   19  &  8.6e &  -73  &  140  &  -82  &   0.003 \\
B07  &   19  & -0.99 &  -21  &   44  &  -21  &   0.005 \\
B08  &   19  &  2.5  &  -45  &   94  &  -53  &   0.002 \\
 \hline
 \end{tabular}
\end{table}

Radial averages of the angular component of the four-velocity
($\overline{\Gamma v_{\theta}}$) display profiles with negative values
in some regions of the body of the fireball. This is a consequence of
the above mentioned rarefaction wave that results from the interaction
between the external medium and the fireball. These negative
velocities tend to collimate the outflow.  Hence, as $\overline{\Gamma
v_{\theta}}$ is a measure of the lateral expansion velocity in the
comoving frame, it turns out that the part of the fireball which has
average Lorentz factors 10\simlt$\Gamma$\simlt100 is contracting while
matter with relatively large Lorentz factors (\simgt$100$) expands
with velocities that are considerably subsonic
(Fig.~\ref{fig:raverages_B01n}). In this point we qualitatively agree
with \cite{KG03}: The lateral expansion is slower than that assumed by
the simplest homogeneous models (which consider expansion velocities
close to the sound speed or even larger). We also point out that the
absolute values of $\overline{\Gamma v_{\theta}}$ tend to decrease
with time.

 An upper bound for the duration of a GRB event might be set by the
time interval between the moments when the forward and the rear ends
of the expanding fireball reach transparency at large radii. This time
difference is linked to the difference in speed between the forward
and rear ends of the fireball in the radial direction. It turns out
that the velocity of propagation of the forward edge corresponds to
that of the terminal shock of the wind or the head of the jet in our
models.  This velocity is determined by the amount of mass swept up by
the bow shock of the jet. Since the halo density in models of type-B
is low, the propagation speed of the fireball is ultrarelativistic
(Table~\ref{tab:models}). On the other hand, the velocity of the rear
part of the wind is set by the speed of the contact discontinuity that
separates the fireball from the external medium after the energy
release has stopped. Behind this wave, matter fills the wake of the
fireball, \ie the ultrarelativistic wind structure is lost. The
contact discontinuity propagates at a speed very close to the speed of
light because the relativistic wind is very rarefied. In addition, the
speed of this wave is not constant, because the plasma in the fireball
experiences ongoing acceleration during the transformation of its
internal energy into kinetic energy. In order estimate this velocity
we have approximated the hydrodynamic states at the rear, radial edge
of the fireball by two constant states corresponding to the fireball
(right) and to the external medium (left). The two states are (in cgs
units): $(p_{\rm L},\rho_{\rm L},\Gamma_{\rm L}) \approx
(10^{14},4\e{-4},2)$ and $(p_{\rm R},\rho_{\rm R},\Gamma_{\rm R})
\approx (6\e{14},10^{-7},\Gamma_{\rm R})$. The generic solution of the
Riemann problem (see above) consists of a contact discontinuity
flanked by two rarefaction waves (one advancing into the fireball and
the other one extending into the external medium) where the fluid
progressively accelerates from $\Gamma_{\rm L}$ to $\Gamma_{\rm R}$.
We have computed the exact solution of the Riemann problem set by this
series of states as a function of $\Gamma_{\rm R}$ with $80 \leq
\Gamma_{\rm R} \leq 220$, which are typical values of the Lorentz
factor near the rear edge of the fireball as it accelerates after the
energy deposition is stopped.  The Lorentz factor associated with the
contact discontinuity ($\Gamma_{\rm CD}$) is a monotonic function of
$\Gamma_{\rm R}$ that can be fitted by the linear relation
\begin{equation}
\Gamma_{\rm CD} \approx 6.42\e{-2}\Gamma_{\rm R} + 12.60 \: ,
\label{eq:rearvelo}
\end{equation}

\noindent
\ie $\Gamma_{\rm CD} \leq \Gamma_{\rm R}$ for ultrarelativistic
fireballs. 
\begin{figure}
\begin{center}
\includegraphics[width=0.95\columnwidth]{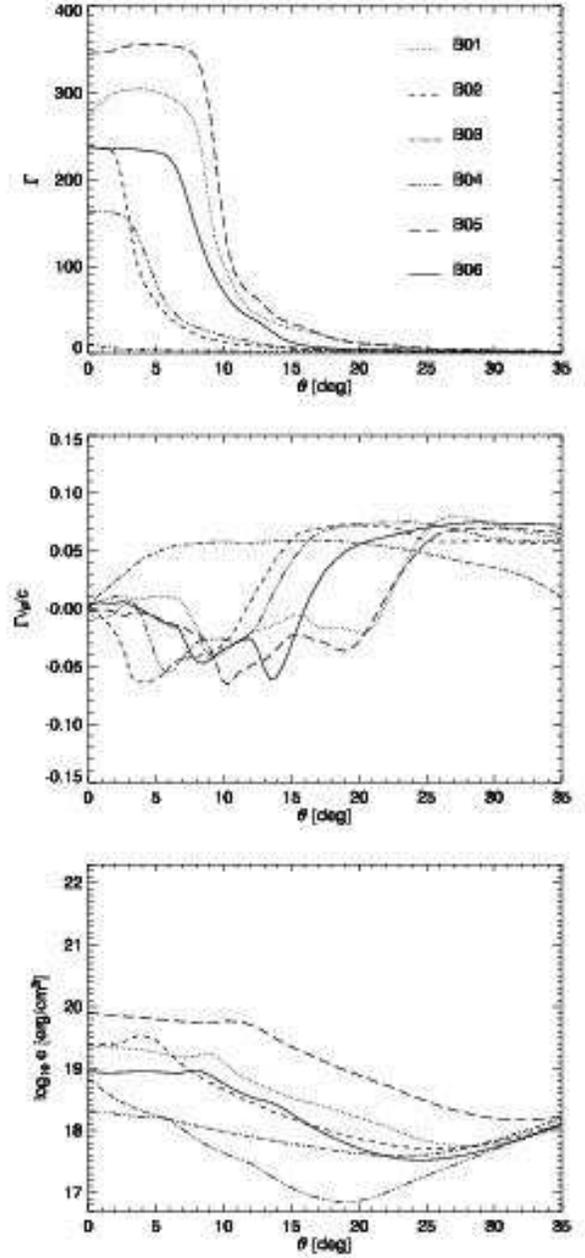}
\end{center}
\caption{Radially averaged values of the Lorentz factor (top panel),
angular component of the four-velocity (intermediate panel) and energy
density (bottom panel) as functions of the polar angle after 0.5\,s of
evolution. The different lines correspond to different models from B01
to B06.}
\label{fig:raverages_B01n-to-B06n}
\end{figure}
\begin{figure}
\begin{center}
\includegraphics[width=0.999\columnwidth]{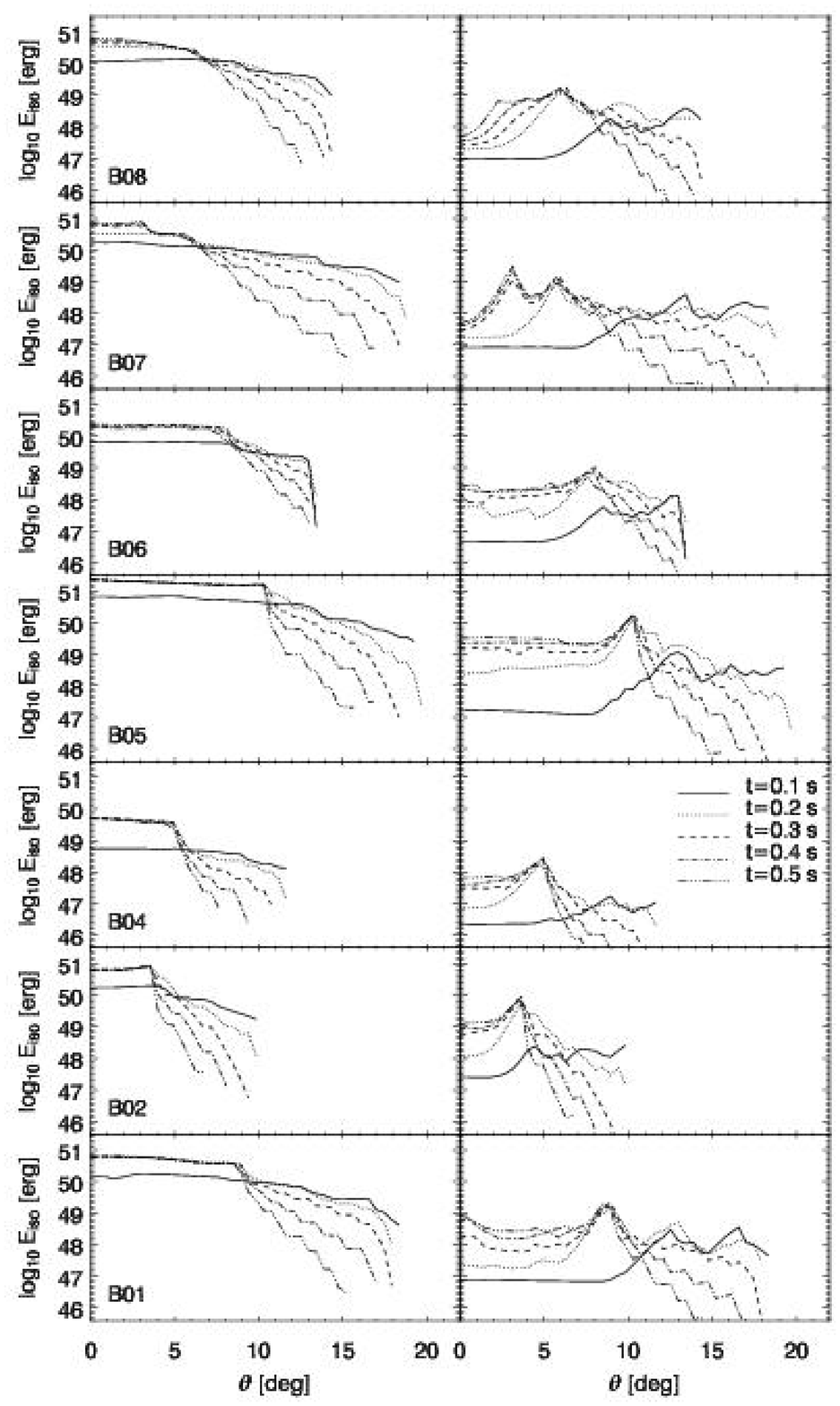}
\end{center}
\caption{Isotropic equivalent total (left) and kinetic (right)
energies of matter with $\Gamma >100$ for all B-type models at
different evolutionary times. Model B03 is omitted in this figure
because it does not develop an outflow with $\Gamma > 100$. Note that
the central plateau of the isotropic equivalent total energy increases
substantially between 0.1\,s and 0.2\,s when the source is already
off. This is due to the continuous acceleration of matter to Lorentz
factors above 100 as a result of the conversion of internal to kinetic
energy in the fireball (even after the shut-off of the energy
release).}
\label{fig:isoenergy_W>100}
\end{figure}

$\Gamma_{\rm R}$ is, in turn, smaller than the Lorentz factor
associated with the leading radial edge of the fireball during the
acceleration phase of the outflow, which leads to a progressive
stretching of the fireball in the radial direction. A possible
important consequence of this stretching is that the duration of the
GRB event might be considerably longer than the time interval of the
activity of the central engine. If the fireball stretches radially on
its way to the optically thin regime, the on-time of the energy source
would not necessary be a suitable measure of the GRB duration. In
order to derive an order of magnitude estimate of the stretching
effect it is necessary to extrapolate from our computed evolution to
much later times. For this purpose we have calculated from our
hydrodynamic models the speed of the fireball tail and compared it
with the analytic estimate of Eq.~(\ref{eq:rearvelo}). It turns out
that the analytic result agrees reasonably well with the numerical
computation. The contact discontinuity, which roughly corresponds to
the tail of the fireball, moves with an average velocity of $v_{\rm
CD} = 0.99^{+0.009}_{-0.04}$. Hence, if the burst is produced
after the fireball has traveled a distance $R$, an upper bound to its
duration will approximately be given by
\begin{eqnarray}
\Delta t_b \simlt \left(R - \displaystyle{\frac{v_{\rm CD}}{v_p}} 
                  (R - t_{\rm sa} c)
\right) c^{-1} \, ,
\label{eq:duration}
\end{eqnarray}
where $v_p$ is the propagation velocity of the fireball in the
radial direction and $t_{\rm sa}=0.1\,$s is the time of source
activity.

Taking a typical distance $R=10^{13}\,$cm, Eq.~(\ref{eq:duration})
yields a maximum burst duration of $\Delta t_b \simeq
4.3^{+10.3}_{-3.0}\,$s, which is considerably longer than the time
interval of source activity ($t_{\rm sa}=0.1\,$s). It is emphasized
that Eq.~(\ref{eq:duration}) and the arguments given in this context
do not allow one to make quantitatively meaningful predictions for
observable GRB durations. The estimate of Eq.~(\ref{eq:duration})
assumes that no deceleration of the fireball head occurs by the
accumulation of matter that is swept up from the ambient medium until
the fireball reaches the distance $R$. A detailed model of the
long-term evolution and structure of the fireball also requires a
realistic, time-dependent model for the torus accretion and the
associated energy release (see the subsequent discussion of models B07
and B08). Moreover, the velocities of the head and the tail of the
fireball cannot be extracted to very high accuracy from our
relativistic hydrodynamics results (mainly because of the limited
resolution at large radial distances), but are nevertheless used to
extrapolate the simulations by several orders of magnitude in
radius. The velocities also do not need to be decisive for the actual
duration of an observable GRB, in particular, because the tail
velocity was deduced from the motion of the rear contact discontinuity
of the fireball, which might not be the backward boundary of highly
relativistic matter with $\Gamma > 100$ that forms the $\gamma$-ray
emitting part of the GRB fireball. Equation~(\ref{eq:duration}) is
mainly meant to manifest the fact that the GRB duration, in particular
for short bursts, is not necessary given by the on-time of the energy
source, but can be significantly longer due to the discussed fireball
stretching. Indeed, the stretching may account for the major part of
the GRB duration. The numerical estimates obtained from
Eq.~(\ref{eq:duration}) at best are rough upper limits to this
effect. In contrast, in the collapsar model for long bursts, for
example, the accretion time and thus the period of source activity
clearly dominates the observed GRB duration.

\subsection{Models B02 to B06}
\label{B02-B08-switchoff}
\begin{figure*}[hbt!]
\begin{center}
\includegraphics[width=0.80\textwidth]{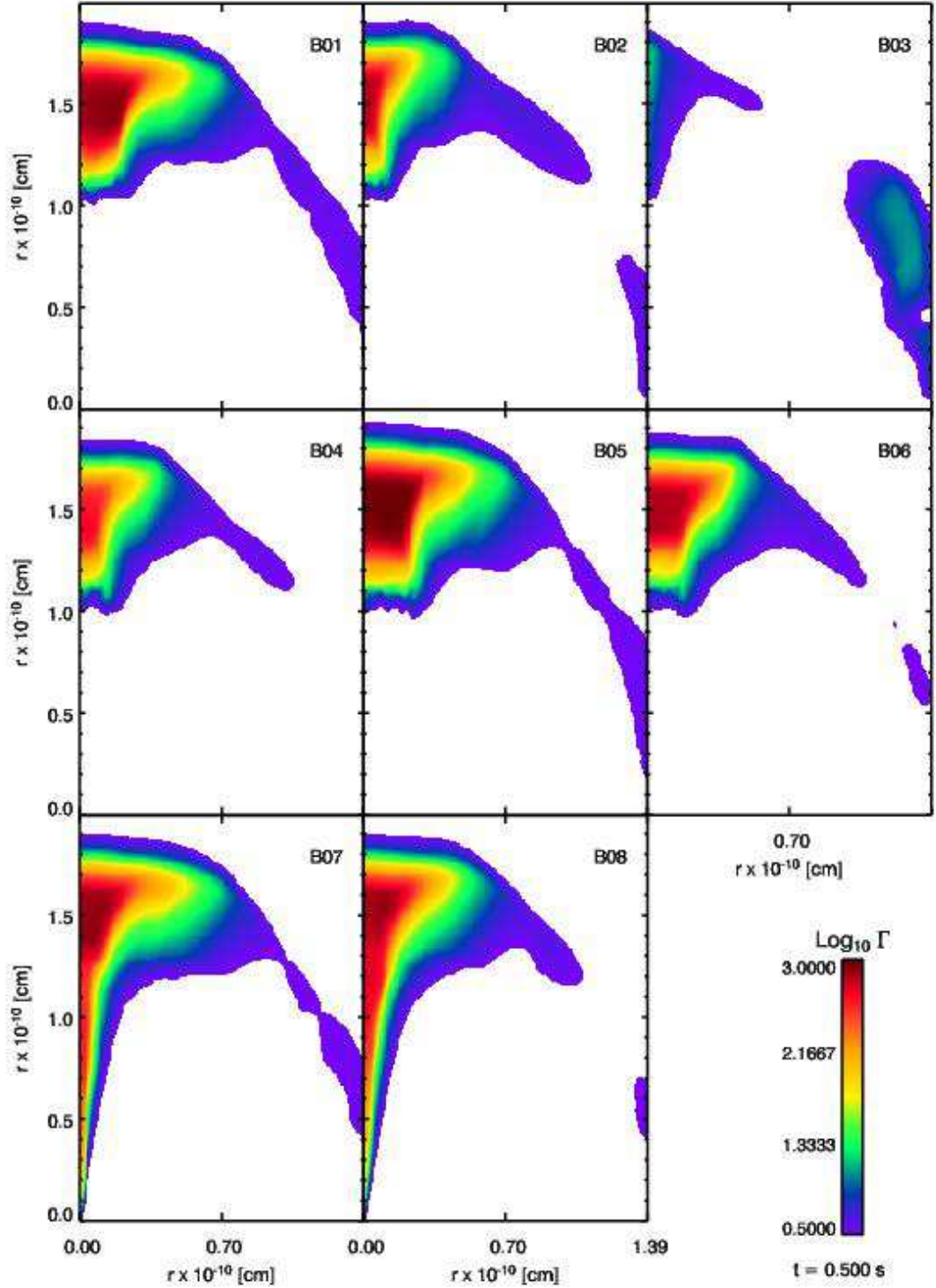}
\end{center}
\caption{Logarithm of the Lorentz factor of the type-B models after
0.5\,s of evolution time. The color scale is limited in order to
enhance the details of the outflow region.}
\label{fig:lor-Bmodels_t=0.5}
\end{figure*}
\begin{figure*}[hbt]
\begin{center}
\includegraphics[width=0.80\textwidth]{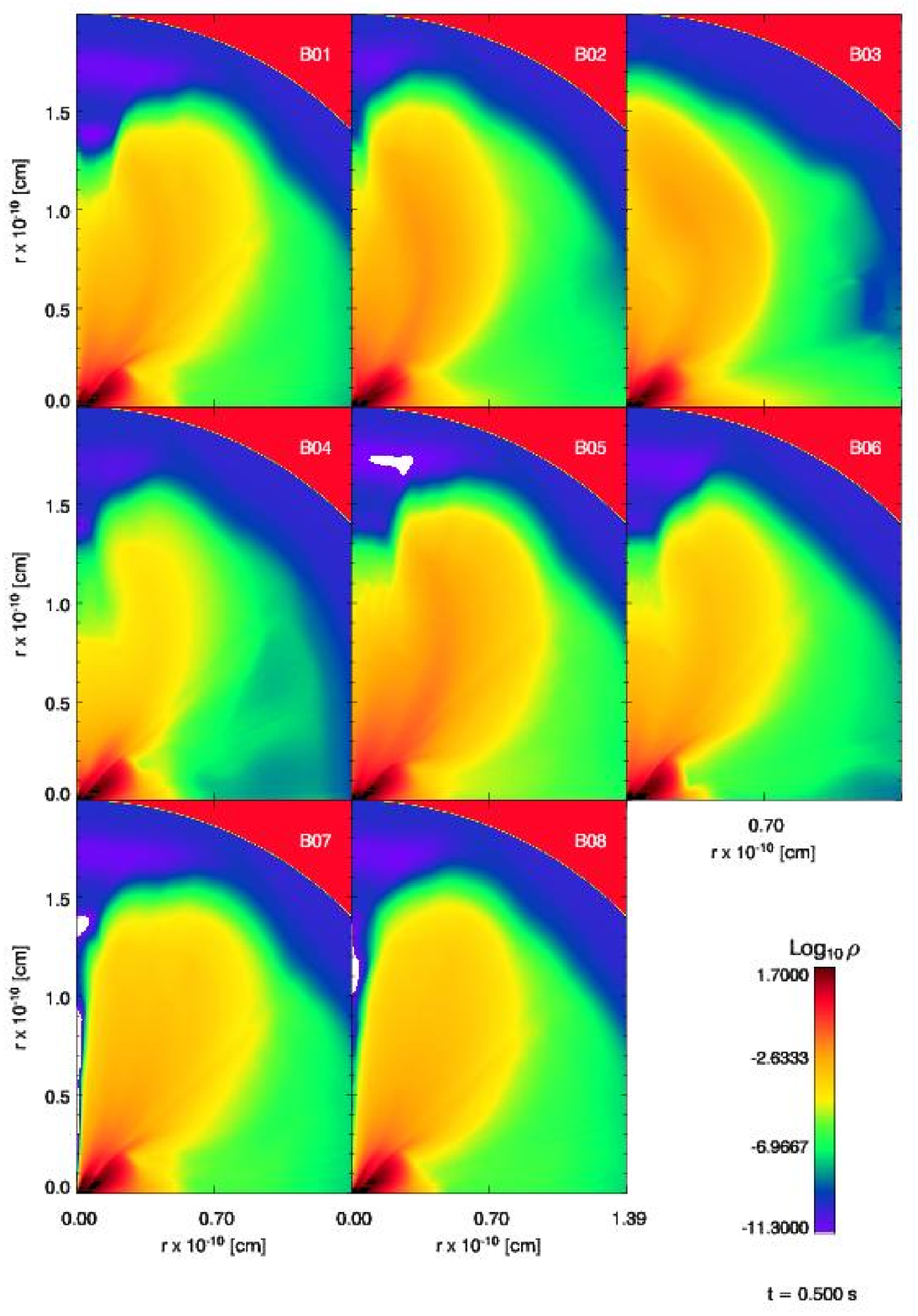}
\end{center}
\caption{Logarithm of the rest-mass density (in \grcm3) of the
B-models after 0.5\,s of evolution time. The color scale is limited
in order to enhance the details of the outflow region.}
\label{fig:rho-Bmodels_t=0.5}
\end{figure*}

This series of models has in common with B01 that the source activity
time is limited to 0.1\,s. In this section we shall discuss which are
the similarities and differences of their evolution after the end of
the energy release relative to model B01. In
Fig.~\ref{fig:lor-Bmodels_t=0.5} the fluid Lorentz factor of all
type-B models is displayed at the end of our computed evolution
(0.5\,s). The high Lorentz factors suggest that all models, except
B03, may be suitable for producing cosmic GRBs (at least they fulfill
the canonical requirement of $\Gamma$\simgt100). All models have
outflow opening angles which are smaller than the opening angle of the
deposition region (see also Table~\ref{tab:B-models}), confirming once
again that the opening angle of the outflow is set by its interaction
with the side {\em walls} of the torus. The diversity of opening
angles (from $3\gra$ to $25\gra$) is caused by the differences in the
energy that is released around the BH per unit of time and unit of
volume, which also depends on the angular width of the deposition
region. The trend is that models with a larger ${\dot E}/V_{\rm dep}$
develop larger outflow opening angles. Model B03 is special only by
the fact that due to the large opening angle of the deposition cone
and a modest total rate of energy release, most of the energy is
deposited in the densest part of the torus and hence can only trigger
a moderately relativistic outflow. In case of models B02 and B04 the
final opening angle of the ultrarelativistic outflow, might be smaller
than at 0.5\,seconds because both models show substantially negative
values of $\overline{\Gamma v_{\theta}}$ ($\approx -0.5c$;
Fig.~\ref{fig:raverages_B01n-to-B06n}).

Stretching of the fireballs in the radial direction should occur in
all models discussed in this section, and is indeed found up to
0.5\,s. The radial extension increases similarly in all models. In
agreement with our numerical results and, according to our estimate of
Eq.~(\ref{eq:rearvelo}), the small differences result from somewhat
different velocities of the tail of the fireballs (the forward radial
front of the fireballs moves in all the cases except B03 at $v_p
\simeq c$).

The maximum Lorentz factor found within the fireball at $t=0.5\,$s
varies substantially from model to model and lies in the range from
$\sim 16$ to $\sim 1000$ (Table~\ref{tab:B-models}), following the
trend that a larger ${\dot E}/V_{\rm dep}$ leads to a larger maximum
Lorentz factor. Figure~\ref{fig:isoenergy_W>100} shows that soon after
the end of energy release the total (internal plus kinetic) isotropic
equivalent energy of the fireball core reaches a maximum and
saturates, while there is an ongoing transformation of internal into
kinetic energy (which is constantly rising). During this conversion of
energies the Lorentz factors within the fireball grow. They have not
saturated until the end of the simulations.

A comparison of the estimated terminal Lorentz factors
$\Gamma_{\infty}$ of the different models (Table~\ref{tab:B-models})
reveals that models B01, B04, B05 and B06 are somewhat more efficient
than models B02 and B03, the main reason being the larger opening
angle of the energy deposition cone in the latter two cases in
combination with a smaller energy deposition rate per volume. This
causes an increase of the baryon loading of the fireball relative to
the energy stored in the pair-photon plasma and is a crucial
difference between models B03 and, \eg B04. The latter develops an
energetic ultrarelativistic jet although the energy deposition per
solid angle is smaller by a factor of $\sim 10$ than in B03. In model
B03 the estimate of the asymptotic Lorentz factor is not meaningful
because the model develops a collimated jet with only a very narrow
and light relativistic core (where $\Gamma$\simgt10;
Fig.~\ref{fig:raverages_B01n-to-B06n}). As the estimate
$\Gamma_{\infty}$ only considers the mass of the outflow with $\Gamma
> 10$, the amount of mass that fulfills this criterion in model B03 is
dynamically negligible compared to the mass of the whole outflow,
which would have to be used for calculating the asymptotic value of
$\Gamma$ in this case.
\begin{figure}
\begin{center}
\includegraphics[width=0.95\columnwidth]{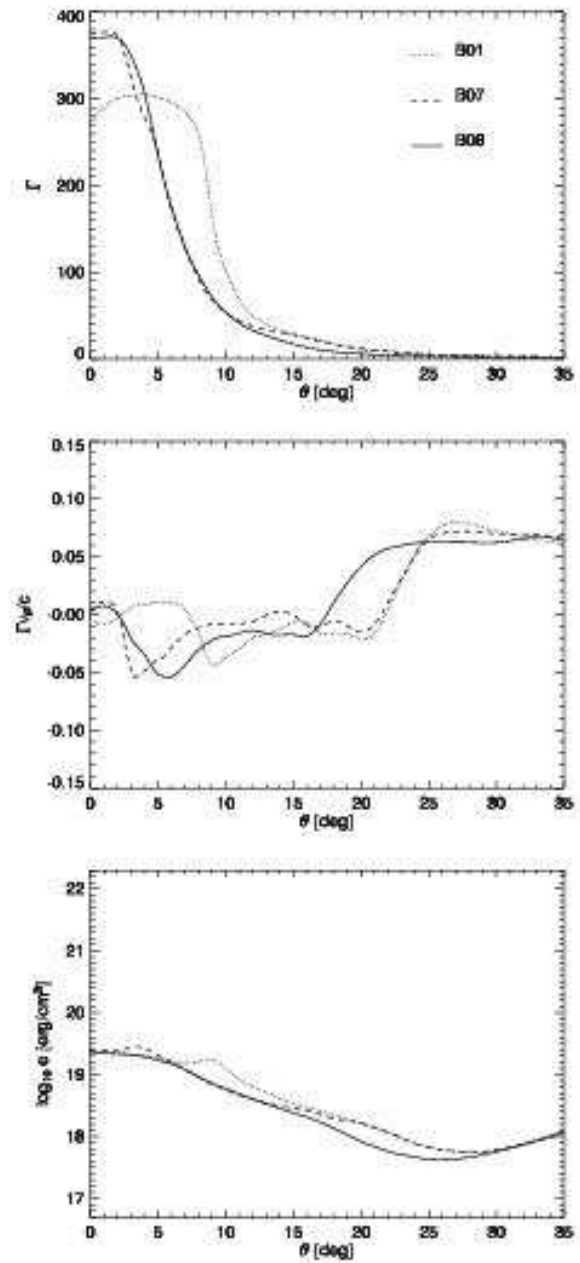}
\end{center}
\caption{Radially averaged values of the Lorentz factor (top panel),
angular component of the four-velocity (intermediate panel) and energy
density (bottom panel) as functions of the polar angle after 0.5\,s of
evolution. The different lines correspond to models B01, B07 and B08,
respectively.}
\label{fig:raverages_B01n-B07n-B08n}
\end{figure}

All models (except B03) display a similar structure of the radially
averaged energy density (Fig.~\ref{fig:raverages_B01n-to-B06n}). Close
to the symmetry axis and up to a few degrees off axis it is nearly
constant. At larger angles $\theta$ a hump appears where the energy
density is about a factor of three higher than near the axis. From
this local maximum the energy density declines gently towards the
lateral edges of the fireball where it starts rising again due to the
interaction with the external medium. The energy hump within the
fireball coincides with a region of large compression of the fluid
where the comoving lateral velocity reaches a local minimum that, in
some models (e.g., in B02, B04 and B05), is an absolute minimum;
(Fig.~\ref{fig:raverages_B01n-to-B06n}, middle panel).

The isotropic equivalent total energy as a function of polar angle
$\theta$ for matter with Lorentz factors larger than 50
(Fig.~\ref{fig:isoenergy_W>100}) consists of two parts: a central core
of $3\gra$ (model B02) to $9\gra$ (model B01) where it is nearly
uniform and maximal, surrounded by an extended layer where it drops
roughly like $\theta^{-2}$. Considering that the equivalent isotropic
total energy is proportional to the energy per unit of solid angle,
our energetic profiles are in very good agreement with those assumed
in the {\it structured} jet models for the afterglows of long GRBs
\citep[\eg][]{Salmonson03}. However, we point out that our results are
supposed to apply for the GRB event itself and for short bursts. The
generation of the energy profile of the burst seems neither to depend
on the width of the deposition region (it is similar for a wide range
of $\theta_0$ values) nor on the absolute value of the energy
deposition rate (or its time dependence). Nevertheless, we are not in
a condition yet to claim that this is a {\it universal} profile,
because it might be linked to the particular spatial dependence of the
assumed energy deposition law or to the properties of the BH-torus
system. Future simulations will have to investigate these aspects.

\subsubsection{Models B07 and B08}
\label{B07-switchoff}

\begin{figure*}
\begin{center}
\includegraphics[width=0.95\textwidth]{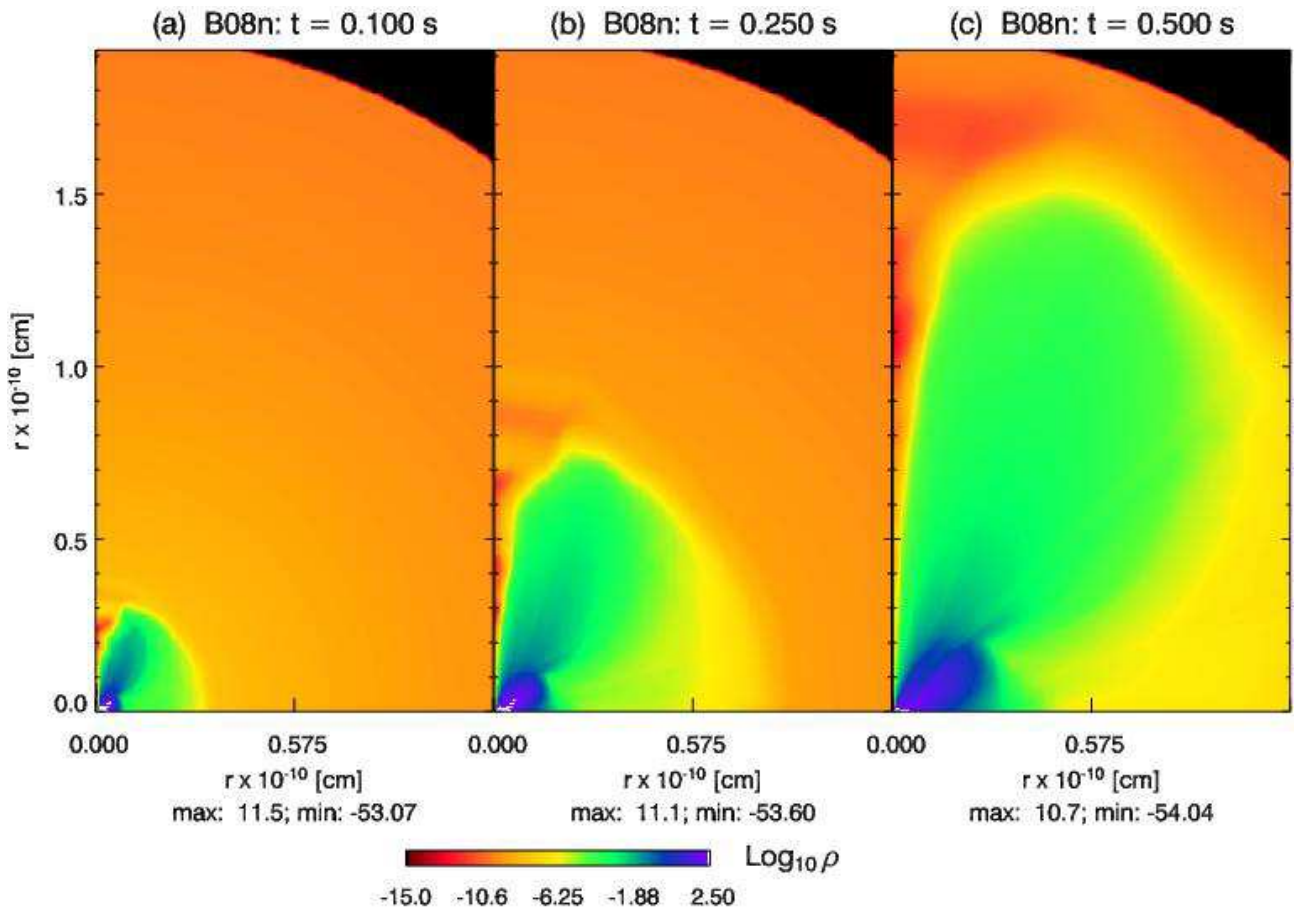}
\end{center}
\caption{Evolution of the logarithm of the rest-mass density (\grcm3)
of model B08 after 100\,ms.  Time is measured from the moment when the
energy deposition was started.}
\label{fig:rho_B08n}
\end{figure*}

\begin{figure*}
\begin{center}
\includegraphics[width=0.95\textwidth]{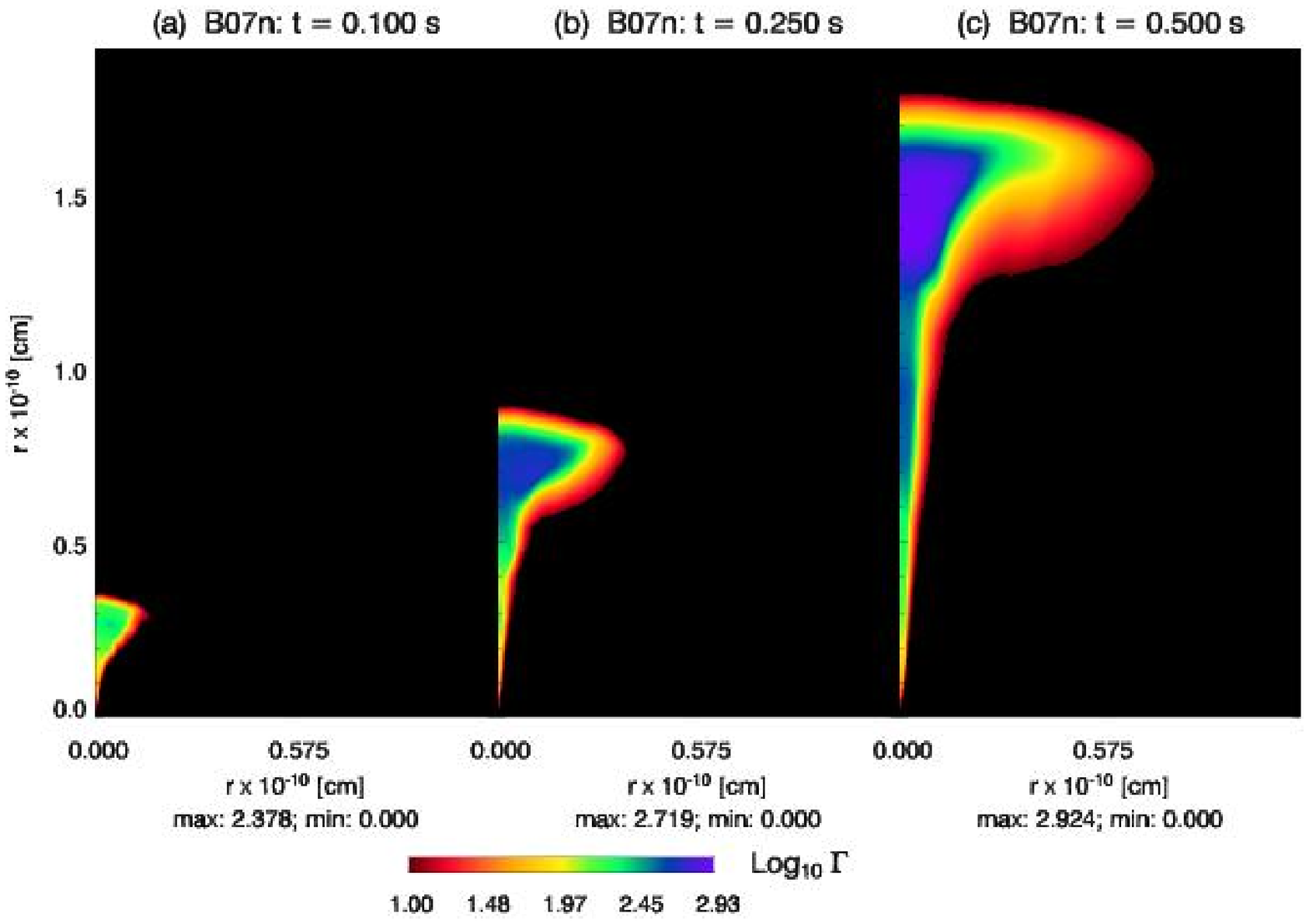}
\end{center}
\caption{Evolution of the logarithm of the Lorentz factor of model B07
after 100\,ms.  Time is measured from the moment when the energy
deposition was started.}
\label{fig:lor_B07n}
\end{figure*}

In these models, the energy is released with a time-variable
deposition rate (Eq.\ref{eq:B07}) and continues for longer that
100\,ms. Note that due to the assumed power-law decay the energy
deposition rate is still at a level of $3.5\e{48}$\ergsec after
500\,ms. This is the reason why the high-Lorentz factor outflow that
has formed an almost conical fireball does not detach from the energy
deposition region and, thus, the axial funnel of low density and high
Lorentz factor stays open until the end of the computed evolution
(Figs.~\ref{fig:lor-Bmodels_t=0.5}, \ref{fig:rho-Bmodels_t=0.5},
\ref{fig:lor_B07n},\ref{fig:rho_B08n}). This axial {\em spine}
decreases in width with time and, if we continued to follow the
evolution of B07 or B08 for a longer period, the energy supply at some
later time would not be sufficient any longer to keep the funnel open.

The structure of the fireball is non-uniform
(Figs.~\ref{fig:lor_B07n},~\ref{fig:lor-Bmodels_t=0.5}) and its
morphology does not differ substantially from that of model B01
(except for the long trailing tail which results from the continuous
energy deposition). The plateau of the maximal, radially averaged
Lorentz factor is more narrow ($\approx 2\gra$ in both model B07 and
B08) than in model B01 (Fig.~\ref{fig:raverages_B01n-B07n-B08n}),
because of the progressive contraction of the fireball fluid (which is
larger in model B07 than in B08;
Fig.~\ref{fig:raverages_B01n-B07n-B08n}). Furthermore, the radially
averaged maximum Lorentz factor after 0.5\,s is larger for these two
models (with values above 375) than for B01. However, the absolute
maximum Lorentz factor ($\Gamma_{\rm max}$) at the end of the
simulations corresponds to model B01 (Table~\ref{tab:B-models}). The
reason for these differences is that B01 is more efficient than B07 or
B08 in using the released energy to accelerate matter to Lorentz
factors above 100 (see the values of $E_{\Gamma>100}/E_d$ in
Table~\ref{tab:B-models}).

The estimates for the asymptotic Lorentz factor $\Gamma_{\infty}$ for
models B07 and B08 in Table~\ref{tab:B-models} may not be meaningful
because energy is released even beyond the simulated epoch of
evolution.

The time variability of the energy deposition rate of model B08
produces a wrinkled boundary of the fireball with a modulation period
of 10\,ms as the energy release (Fig.~\ref{fig:rho_B08n}, left
panel). A similar wrinkled surface occurs in model B05
(Fig.~\ref{fig:rho-Bmodels}), but different from model B08 the
modulation of the density in the surface is not exactly periodic, and
the values of the density in the fireball core up to $\sim 2\e{9}$\,cm
are much higher ($\sim 10^{-2}$\grcm3 very near the axis;
Fig.~\ref{fig:rho-Bmodels}). After 0.25\,s models B07 and B08 display
a fairly similar structure. This is, to a large extent, due to the
grid coarsening in radial direction. Beyond a distance of $\approx
10^{10}\,$cm the grid spacing exceeds 10 light-milliseconds and,
therefore, most of the variability on scales less than $\approx
15\,$ms is damped or erased. For this reason, we cannot draw firm
conclusions on the short-time variability from our models, and in
particular we cannot answer the question how it depends on the
intrinsic variability of the source or the interaction with the
external medium.
\begin{figure*}
\begin{center}
\includegraphics[width=0.65\textwidth]{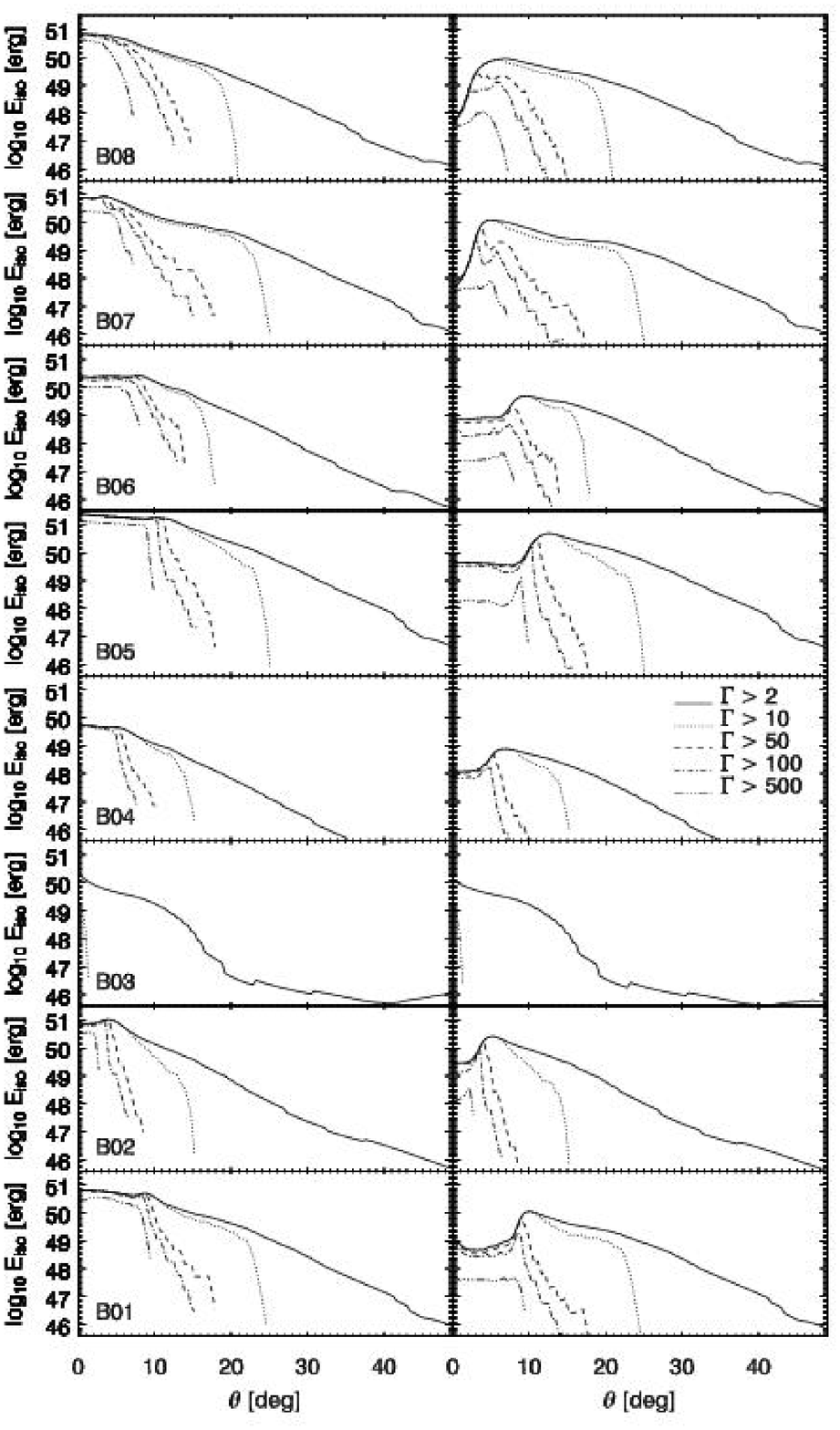}
\end{center}
\caption{Isotropic equivalent total energy (left panels) and
isotropic equivalent kinetic energy (right panels) corresponding to
outflows with Lorentz factors above certain thresholds at 500\,ms of
computed evolution. Resuls for different thresholds are plotted with
different line styles. Note that in the panels corresponding to model
B03 there are no lines for Lorentz factor thresholds above 50, because
there is no outflow with such a large Lorentz factor.}
\label{fig:isoenergy_final}
\end{figure*}

The radially averaged energy density of models B01, B07 and B08 varies
somewhat differently with the polar angle
(Fig.~\ref{fig:raverages_B01n-B07n-B08n}). The central, most energetic
core of the fireball of models B07 and B08 is more narrow. Model B08
does not possess a clear plateau around the symmetry axis and also
has, on average, a smaller energy content compared to the other two
models. The kinetic energy in the central core of the
ultrarelativistic outflow is smaller than in the surrounding layer in
all B-models at $t=0.5\,$s (Fig.~\ref{fig:isoenergy_final}) because
the higher Lorentz factors in the cores are compensated by larger
densities outside. The efficiency for converting thermal to kinetic
energy until $\sim 0.5\,$s is $\sim 25\%$ smaller in case of model B07
than in model B01 or B08 (Table~\ref{tab:B-models}). However, in all
three cases and all other B-models the kinetic energy in the fireball
relative to the total energy of matter with $\Gamma > 100$ is still at
most a few per cent at $t=0.5\,$s (Table~\ref{tab:B-models}), \ie there
is ample room for further fluid
acceleration. Figure~\ref{fig:isoenergy_final} shows that only model
B03 is an exception, as it has essentially reached its terminal
state. Figure~\ref{fig:isoenergy_final} also shows that the
contribution to the kinetic energy of the parts of the outflow having
bulk Lorentz factors larger than 50 is small while most of the kinetic
energy is transported by fluid with Lorentz factors below 10. It is
also evident that the fraction of kinetic to internal energy at
$0.5\,$s is closer to unity for matter with low Lorentz factors
farther off axis, \ie this material will not reach much higher
asymptotic velocities during its adiabatic expansion. The
ultrarelativistic core, however, can be expected to continue
accelerating, when its internal energy is converted to kinetic
energy. 

\begin{figure*}
%\begin{center}
\begin{tabular}{ll}
\includegraphics[width=0.95\columnwidth]{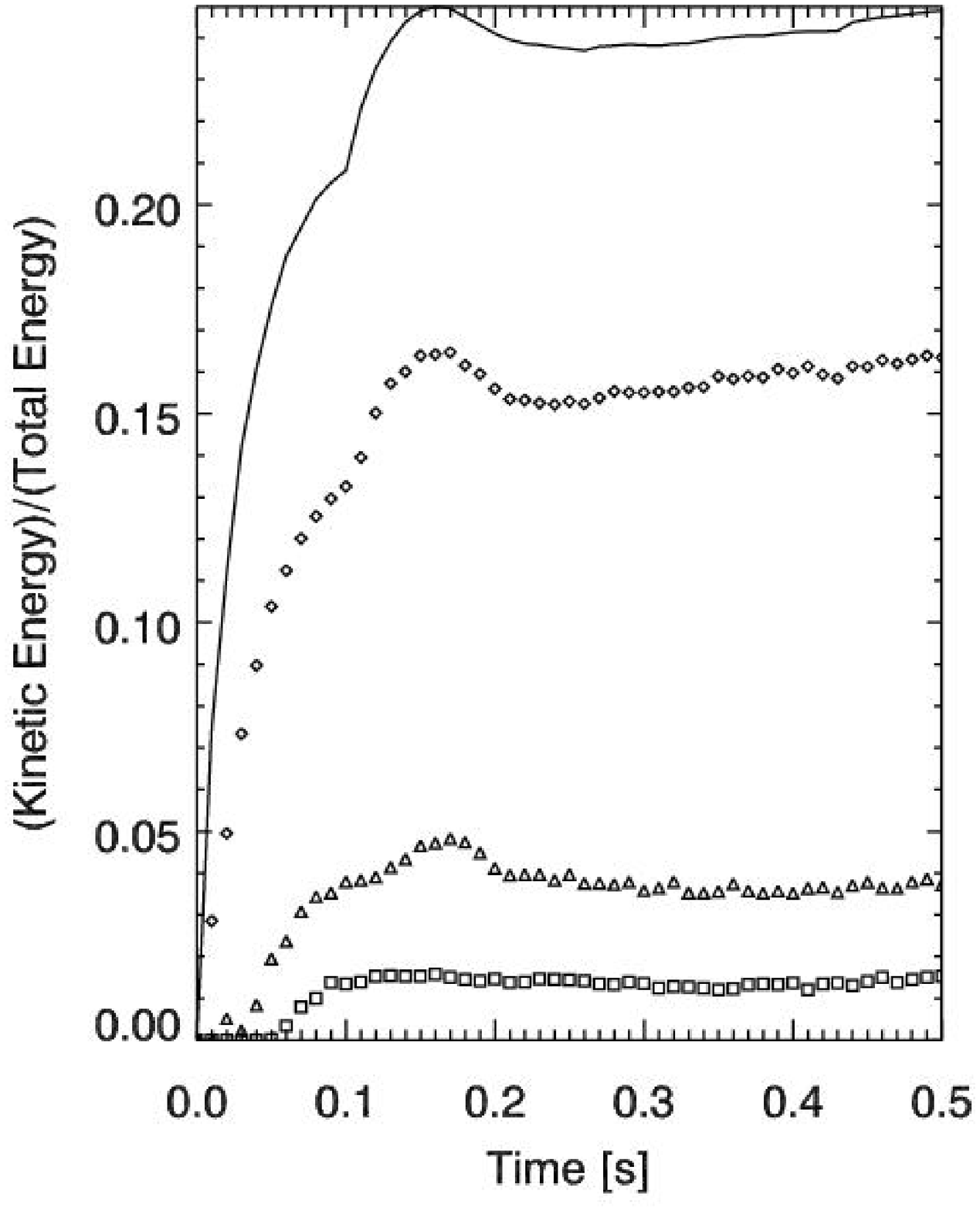} &
\includegraphics[width=0.95\columnwidth]{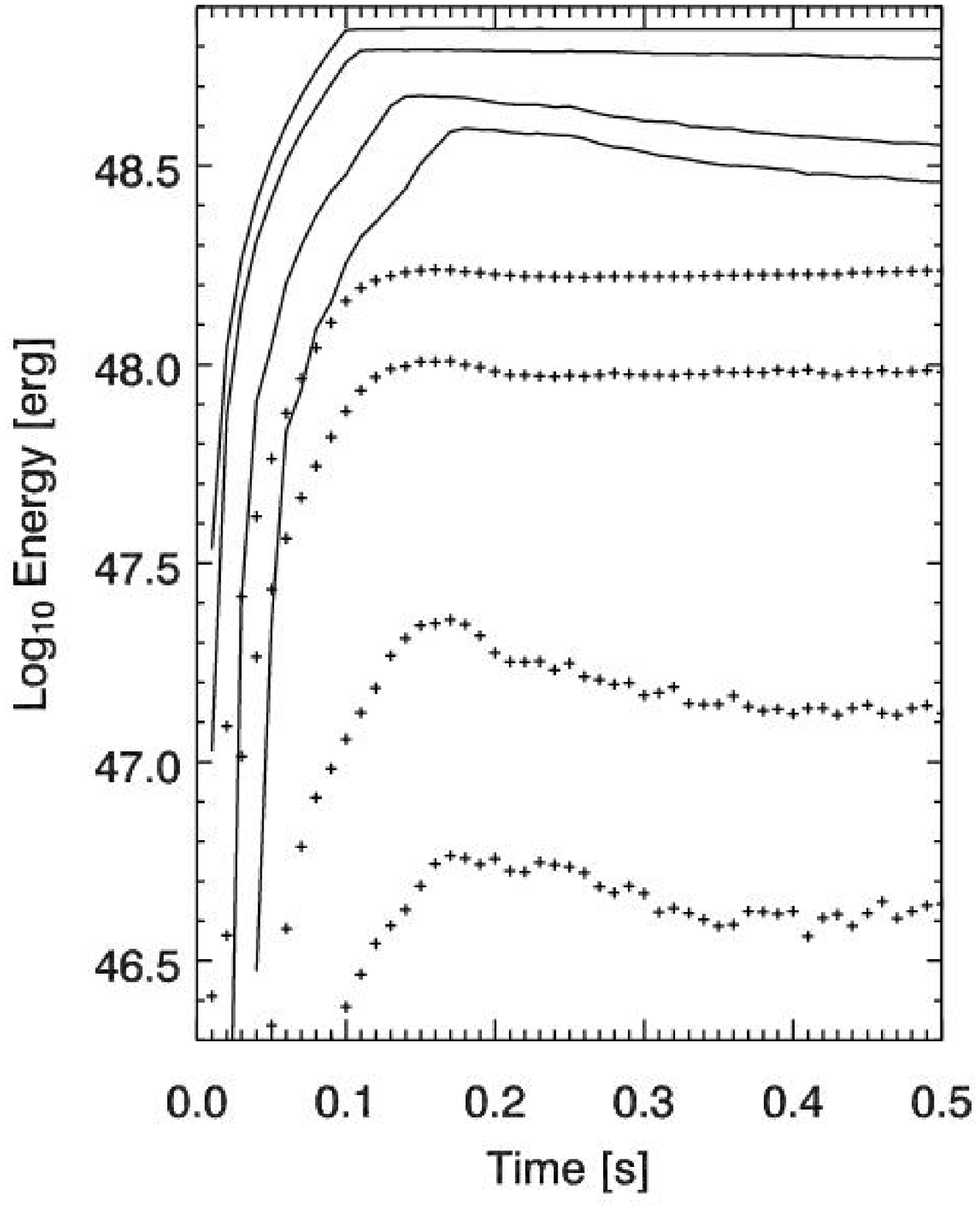} 
\end{tabular}
%\end{center}
\caption{Time evolution of the ratio of kinetic to total (internal
plus kinetic) energy (left panel) of relativistic outflow in model B01
with Lorentz factor above 2, 10, 50 and 100, from top to bottom,
respectively. The right panel shows the corresponding evolution of the
total energy (solid lines) and of the kinetic energy (crosses) in the
same order from top to bottom.}
\label{fig:kinetictototal}
\end{figure*}

The evolution of the ratio of kinetic energy to total (internal plus
kinetic) energy in the outflow of the prototype model B01 is shown in
Fig.~\ref{fig:kinetictototal}. In this figure the kinetic and total
energies of the relativistic outflow are computed as the integral
values for all cells with $\Gamma > \Gamma_{\rm min}$ for different
lower limits $\Gamma_{\rm min}$ of the Lorentz factor. One can notice
a slight, non-monotonic increase of the fraction of kinetic energy
with time. Several conclusions can be drawn from
Fig.~\ref{fig:kinetictototal}. First, if the rate of increase can be
extrapolated until the time when transparency sets in, it must be
expected that most of the internal energy will be converted to kinetic
energy. Second, after 0.5\,s of evolution about half of the kinetic
energy of the outflow is in fluid with $\Gamma \geq 10$. However, it
is impossible to reliably predict the final distribution of kinetic
energy versus Lorentz factor (after, \eg $\sim 1000$\,s), because the
trends shown for fluid elements with different values of $\Gamma_{\rm
min}$ are slightly different and, in some cases it is hardly possible
to deduce an accurate number for the net increase of the ratio of
kinetic to total energy (e.g., in case of $\Gamma_{\rm min} \geq 100$;
Fig.~\ref{fig:kinetictototal}).

\section{Summary and conclusions}
\label{sec:summary}

We have performed general relativistic, axisymmetric hydrodynamic
simulations for studying the creation and evolution of relativistic
outflow in response to the deposition of thermal energy above the
poles of a black hole--accretion torus system. The latter is
considered to be a necessary condition for obtaining baryon-poor
fireballs, in particular if the energy is provided by
neutrino-antineutrino annihilation. Above the poles of the black hole
a favorable environment is encountered for the creation of an
ultrarelativistic e$^\pm$-pair-photon plasma by the mentioned process,
different from the situation in case of neutron stars, where neutrino
energy deposition in the surface-near layers produces a
nonrelativistic baryonic wind \citep{DWS86,Wo93,QW96,TBM01}.

Being guided by numerical models of \cite{RJ99} and \cite{Jetal99},
the BH-torus systems for our studies were set up to represent the
remnants of NS+NS or NS+BH mergers. We chose Schwarzschild BHs of
$2.44\,M_{\odot}$ or $3\,\,M_{\odot}$, being girded by a thick torus
of $0.13\,M_{\odot}$ or $0.17\,M_{\odot}$, respectively. The accretion
torus was constructed either by relaxation of an initial toroidal
cloud of matter or by creating a near-equilibrium configuration with
the technique of \cite{FD02}. In the first case the BH-torus system is
surrounded by a fairly dense ($\sim 100\,$g$\,$cm$^{-3}$) gas halo
with a radius of about $10^9\,$cm (containing a mass of roughly
$10^{-4}\,M_{\odot}$), e.g., considered to be the consequence of a
neutrino--driven wind from a transient, metastable NS which underwent
collapse to a BH--torus system only after some time delay.  In the
second case the density of the surrounding gas is much lower and
decreases radially approximately as $r^{-3.4}$. Correspondingly, we
discriminate between high-density type-A models and low-density type-B
models.  Both are considered as possible results of a violent,
preceding merger history. Thermal energy was deposited within a cone
with chosen opening angle above the poles of the black hole, starting
at the inner grid boundary at $4GM_{\mathrm{BH}}/c^2$ or
$2GM_{\mathrm{BH}}/c^2$ for type-A or type-B models, respectively.
Cases with constant energy deposition rate were studied or with a
burst-like initial phase, followed by a long-time, gradual decay.  The
energy deposition rate was assumed to depend only on the vertical
height $z$ measured along the rotation axis and to drop like $z^{-5}$.
This prescription is an approximate representation of deposition maps
obtained by detailed calculations of neutrino-antineutrino
annihilation in the vicinity of the BH-torus system
\citep{RJ99,Jetal99}. Disregarding the modest lateral bending of the
computed surfaces of constant $\nu\bar\nu$-energy deposition rate
turned out not to be essential for the discussion in this work, but
the deviation of these surfaces from spheres had important
consequences.

We calculated sequences of type-A and type-B models, systematically
varying the total rate of energy deposition ($\dot E$) while keeping
the lateral width of the deposition region unchanged.  Alternatively,
we changed the opening angle of the energy-deposition cone for a fixed
integral rate or for a fixed energy deposition rate per unit of volume
(${\dot E}/V_{\rm dep}$), respectively. In summary, our studies
revealed the following dynamical behavior of the evolving outflow:

\begin{itemize}
\item 
Relativistic outflow develops in form of narrow, knotty jets or wide,
ultrarelativistic winds or jets. The generation of relativistic
outflow requires a high energy deposition rate or low environmental
density. In case of the high-density type-A models the energy
deposition rate has to exceed a certain threshold value (around
$10^{49}\,$erg$\,$s$^{-1}$) in order to produce any outflow. Slightly
above this threshold an inhomogeneous jet is formed rather than an
ultrarelativistic wind.

\item
In case of type-A models the collimation of the outflow is mainly
determined by the interaction with the high-density ambient
medium. The opening angle of the polar outflow increases with ${\dot
E}/V_{\rm dep}$ and varies from a few degrees for the low-density
(baryon-poor) jets that develop just above the ejection threshold to
more than 25$^{\mathrm{o}}$ for the smooth, low-density, wide-angle
winds in case of the highest considered ${\dot E}/V_{\rm dep}$.  The
interaction between outflow and ambient medium resembles the
hydrodynamic collimation mechanism discussed by \cite{LE00},
and the unshocked part of the baryon-poor outflow can be well fitted
by power laws with indices close to the ones predicted by the analytic
model of these authors.  Because of the large mass that is swept up,
the propagation speed of the shocked, forward part of the outflow
becomes only mildly relativistic in all models, not allowing for
favorable conditions to produce GRB emission. Instead, a
low--luminosity (at most $\sim 10^{43}$\ergsec) soft UV--flash at a
temperature of $\sim 5\e{4}\,$K with a duration of $\sim 1000$\,s can
be expected as the observable signature of merger events surrounded by
high-density halos. Due to their small luminosity, only galactic
events might be detected.
\item
In case of type-B models all employed energy deposition rates (above
$\sim 10^{49}\,$erg$\,$s$^{-1}$) led to non-uniform, wide-angle,
radially and laterally structured ultrarelativistic jets provided the
rate of energy deposition per unit of volume was sufficiently
high. The propagation velocities are found to be very close to the
speed of light, corresponding to Lorentz factors of several hundred,
also in the forward, shocked part of the flow. This suggests favorable
conditions for the production of GRBs. The simple power-law fits of
the analytic model of \cite{LE00} do not work well. The latter
authors assumed that a baryon-poor jet is confined by the ram pressure
of a baryonic (nonrelativistic) wind which originates from a toroidal
ring of negligible thickness.  They further assumed that the
baryon-rich wind collides with the relativistic jet and gets deflected
into a very thin, shocked layer that envelopes the relativistic flow.
Our hydrodynamic simulations do not support this picture.  We found
that the collimation of the baryon-poor flow in type-B models is
caused by the interaction of the accelerating fluid with the dense
accretion torus, whose vertical extension cannot be ignored. This
interaction produces a thick shell of gas which surrounds the
ultrarelativistic core and which shows large inhomogeneities due to
Kelvin-Helmholtz instabilities.  The half-opening angle of the
ultrarelativistic flow is set within the first millisecond of its
expansion on a distance that corresponds to the vertical thickness of
the accretion torus. It is determined by the curvature of the torus
walls around the poles of the BH and was found to be around
20$^{\mathrm{o}}$ in the computed type-B models.

\item
Type-A and type-B models also showed a different behavior after the
deposition of energy was switched off or had decayed.  In case of a
high-density halo around the BH-torus system (type-A) the conical,
relativistic wind structure is destroyed within a few source activity
timescales ($t_{\rm sa}$), because the tail of the outflow moves into
a cleared funnel and is much faster. Therefore, it catches up with the
leading edge of the flow, which gets decelerated by swept-up
matter. In contrast, in type-B models with their steeply decreasing
halo density, the front part of the outflow is steadily accelerated
while its rear end moves somewhat slower because of the lack of
support by further strong energy release at the origin. Therefore the
fireball cone is preserved instead of being destroyed. However,
preservation is a necessary condition for GRB production. The fireball
can in fact be radially stretched before it reaches optically thin
conditions. Assuming that the GRB duration, $\Delta t_{\mathrm{b}}$,
is defined by the time difference between the front and rear ends of
the fireball reaching the ``transparency radius'', e.g.,
$10^{13}\,$cm, we can estimate, by extrapolating the results of our
computed models, a burst duration $\Delta t_{\mathrm{b}}$ that might
be significantly longer than the on-time ($\sim 0.1\,$s) of the
source: $\Delta t_{\mathrm{b}} \cong 4.3^{+10.3}_{-3.0}\,$s. Note that
this stretching can be the dominant contribution to $\Delta
t_{\mathrm{b}}$ in case of source activity times of significantly less
than one second as expected for the accretion timescale of post-merger
disks. The latter timescale is expected to vary with the torus
mass. It is set by the viscous transport within the compact accretion
torus, in contrast to collapsars where the accretion disk is fed by an
external reservoir of several solar masses of stellar matter and the
accretion timescale is therefore determined by the collapse timescale
of the massive, rotating star.  Even if the accretion phase of the
remnant BHs of NS+NS or NS+BH mergers lasts only fractions of a
second, our simulations suggest that such events can account for the
measured durations of short GRBs.  Of course, our estimation of
$\Delta t_{\mathrm{b}}$ can only be considered as an exercise for
demonstrating a fundamental possibility. It has no predictive power
for GRB durations, because we need to extrapolate our hydrodynamic
results over several orders of magnitude in radius. This is rather
uncertain due to difficulties in extracting accurate values for the
fluid velocities from the computations in case of very high Lorentz
factors and due to the geometrical dilution of the spatial grid
resolution with increasing radial distance. Extrapolation of our
results from about 0.5$\,$s ($\sim 0.4\,$s) after the onset (shutdown)
of the energy release to more than 300$\,$s later also ignores how the
fireball properties continue to change and how the long-time
propagation and expansion of the fireball may depend on the structure
of the ambient medium of the merger site. Moreover, the GRB emission
might be shorter than our estimate if it is produced in a region that
is smaller than the whole fireball. Nonthermal emission of radiation
requires the dominant energy of the flow to be kinetic energy of
relativistic baryons but not internal energy. Unfortunately our
simulations had to be stopped before definite statements about the
terminal fireball structure and the final ratio of its kinetic to
internal energy were possible.

\item
The total energy (internal plus kinetic) of the relativistic fireball
scales roughly linearly with the deposited energy and saturates after
the energy release has ended.  A burst-like, short initial phase of
energy deposition, followed by a long-time, gradual decay turned out
to channel somewhat less energy into the relativistic outflow and to
be slightly less efficient in converting internal energy to kinetic
energy than a constant rate of energy release with a sudden end.  In
both cases a minor part (a few per cent at most) of the energy was in
form of kinetic energy after 0.5 seconds of computed evolution, with
the tendency to continue rising.

\item
The relativistic flow in type-B models reveals a highly inhomogeneous
structure in both the radial and lateral directions.  A small negative
radial gradient of the radiation-dominated pressure corresponds to a
radial increase of the Lorentz factor also in the shocked region of
the flow, where the average behavior of different quantities can
roughly be described by power laws of the radius (though different
from those of the analytic model of \citealp{LE00}).  While pressure
and Lorentz factor reveal only moderate variations (factors of a few),
the density shows huge fluctuations up to three orders of
magnitude. These inhomogeneities are a relic of Kelvin-Helmholtz
instabilities which are triggered by the interaction of the fireball
with the torus medium.  In lateral direction the structure cannot be
fitted by simple top-hat, power-law or Gaussian profiles but can
posses off-axis maxima of the Lorentz factor. This feature results
from a genuinely relativistic effect that occurs in the presence of
tangential velocity components at a propagating shock. It is favored
by the fact that the surfaces of constant energy deposition deviate
from spheres leading to relatively high energy deposition rates at
large radii off the symmetry axis. A relativistic shock with large
tangential velocities causes the formation of a rarefaction wave that
moves slowly from the lateral edges towards the axis of the polar
outflow and allows the Lorentz factor in some of our B-models to
become about $20\%$ larger in the rarefied region than near the
symmetry axis of the radial flow.  At the same time the
ultrarelativistic core, where Lorentz factors of several hundred are
reached, is kept collimated by a subsonic negative lateral velocity
component while only the surrounding, mildly relativistic ejecta
exhibit a very slow sideways expansion. Thus we expect the structure
of the fireball to be essentially preserved during the subsequent
evolution.

\item The lateral profile of the radially averaged energy of type-B
models shows a central hot core (of $\approx 3\gra - 9\gra$ width) of
nearly constant value, surrounded by a layer where the radially
averaged energy drops proportional to $\theta^{-2}$. This profile
arises naturally in all our models, and agrees well with that of the
so-called {\it structured} jet model that has been used to explain the
phenomenological correlation between the GRB peak luminosity and pulse
lags \citep{NMB00} of long GRBs \citep[\eg][]{Salmonson00} or the
nearly constant energy inferred from observations by \cite{Frail01}
and from theoretical models by \cite{RLR02} and \cite{ZM02}. The
lateral profile of the Lorentz factor displays a more complex
structure whose functional form cannot be properly fitted by a
$\theta^{-2}$ or Gaussian decay (which is required in the standard
structured jet model) mainly because of off-axis maxima of the Lorentz
factor. In case these profiles are able to survive until the fireball
reaches transparency, the angular energetic profiles of short bursts
may be very similar to those of the long bursts, but the Lorentz
factor distribution might be more complex.

\end{itemize}

Despite the coarsening of the computational grid with increasing
radius, the fireball could be well followed from the shutdown of the
energy source at $t_{\rm sa}$ (or during the long-time, gradual decay
of the energy release, respectively) to the end of our simulations (at
0.5$\,$s or $5\,t_{\rm sa}$) when the fireball reaches a radial
extension of about $1.5\times 10^{10}\,$cm.  For discussing gross
observational features we therefore feel safe to assume that global
fireball properties can be sufficiently reliably extrapolated to much
larger distances.  Inhomogeneities on angular scales
$\theta_{\mathrm{ih}} > \Gamma^{-1}$ should persist and are expected
to be smoothed out only after the Lorentz factor has dropped below
$\theta_{\mathrm{ih}}^{-1}$ \citep[\eg][]{KG03}. The maximum Lorentz
factors in the front of the outflow are higher than 500. In some of
the type-B models regions in the fireball with $\Gamma > 1000$ are
found at the end of our simulations. The terminal value of the Lorentz
factor should depend on the ambient density of the merger site. Our
results are obtained for a fireball that accelerates into a $r^{-3.4}$
density profile (chosen mainly to overcome numerical problems in the
construction of the initial configuration). However, the real
situation may be more complex, e.g., a steeper density gradient might
connect the torus ``surface'' (or a small cloud of baryons that
surrounds the BH-torus system as the relic of the preceding, violent
merger history) with the much lower and almost constant density of the
interstellar medium.
 
With a half-opening angle of $\theta_{\mathrm{j}} \sim
5\gra$--$10\gra$ ($15\gra$--$25\gra$) the collimated ultrarelativistic
outflows of type-B models having $\Gamma$\simgt100 ($\Gamma$\simgt10) at
$t=0.5$ seconds after the onset of the energy release by the GRB
engine, cover a fraction $f_{\Omega} = 1-\cos{\theta_\mathrm{j}}\sim
0.4\%$--1.5\% ($<10\%$) of the sky%
\footnote{The numbers in brackets correspond to outflow with Lorentz
factors \simgt10 at 0.5\,s, which shows still ongoing acceleration so
that much larger Lorentz factors can finally result,
cf. Table~\ref{tab:B-models}.}\addtocounter{footnote}{-1}.  Assuming
equal detectability from all directions within the opening angle this
fraction implies about 100 (more than
10)\footnotemark\addtocounter{footnote}{-1} times more events than
measured gamma-ray bursts. A rate of short GRBs of about 100 per year
therefore requires an event rate per galaxy and year of
$10^{-5}(f_{\Omega}/0.01)^{-1}(N_{\mathrm{g}}/10^9)^{-1}$ where
$N_{\mathrm{g}}$ is the number of visible galaxies.  Comparing with
estimated NS+NS and NS+BH merger rates, which are typically around
$10^{-5}$ per year and galaxy \citep[with about a factor of 10 or more
uncertainty; e.g.,][]{Kaloetal04,Fryeretal99}, we therefore conclude
that a significant fraction of such mergers but probably not all,
should produce GRB viable outflows.

The edges of the ultrarelativistic ($\Gamma > 100$) jet core are very
sharp in terms of the isotropic equivalent energy $E_{\rm
iso}$. Maximum (terminal) Lorentz factors of the order of 1000 suggest
the potential to account for hard GRB spectra. Our simulations
actually showed inhomogeneous and anisotropic, collimated outflows
with lateral variation of the Lorentz factor and of the apparent
isotropic energy. Hence, we do not expect equal observability from all
positions in the beam direction.  The maximum values of the apparent
isotropic energy, $E_{\rm{iso}}$, are found to be up to about
$10^{51}\,$erg at angles \simlt$10\gra$ around the symmetry axis of
the ultrarelativistic outflow, declining towards the outer wings of
less relativistic ejecta. These numbers are obtained for an energy
deposition rate of a few $10^{50}\,$erg$\,$s$^{-1}$ over a period of
typically 100$\,$ms. These are reasonable and not extreme values in
view of model calculations for the energy release by
neutrino-antineutrino-annihilation in case of post-merging BH
accretion \citep{RJ99,Jetal99,SRJ04}. Provided a major fraction of the
energy of the ultrarelativistic fireball gets converted to gamma-rays,
our maximum isotropic equivalent energies are in good agreement with
estimates based on a comparison of the energetics of short and long
GRBs, suggesting an approximate fluence-duration proportionality
\citep{Balazsetal03}. Since long bursts last typically about 50--100
times longer, a similar luminosity (\eg \citealp{MNP94};
$L_{\mathrm{iso}}^{\mathrm{short}}\sim
L_{\mathrm{iso}}^{\mathrm{long}}\sim 10^{51-52}$erg$\,$s$^{-1}$)
implies an apparent energy which is around $10^{51}\,$erg for short
bursts instead of $\sim 10^{53}\,$erg for long ones \citep{FK97}.

Due to the lateral structure of the outflow we also expect some degree
of variability between observed bursts, depending on the viewing angle
relative to the system axis. Such differences will be superimposed on
variations due to intrinsic properties of the binary systems and
remnant BH-torus systems, e.g., associated with different masses and
spins of the merging neutron stars or black holes.  This finding
should be taken into account in studies of the diversity of short
gamma-ray bursts like the recent one by \cite{RR03}.  These authors
also employed the assumption that the ultrarelativistic outflow is
confined as suggested by \cite{LE00}. Our models, however, show a much
different hydrodynamic scenario for the fireball evolution and
collimation, in which the outflow-torus interaction, relativistic
shock effects, and Kelvin-Helmholtz instabilities play a crucial
role. We therefore emphasize that conclusions based on grounds of the
simplified picture developed by \cite{LE00} should be drawn with
caution. Finally, we repeat that it is impossible to estimate the
energetics of ultrarelativistic outflows without performing
simulations that follow the complex hydrodynamics phenomena which
develop in response to the deposition of energy in the vicinity of the
BH-torus system.  A static analysis of time slices for mass
distribution and energy deposition by $\nu\bar\nu$-annihilation
\citep[\eg][]{RR02} can therefore be misleading, in particular with
respect to the outflow energetics and asymptotic Lorentz factor which
are crucial for judging the viability of $\nu\bar\nu$-annihilation for
powering GRBs.

Our simulations are only a preliminary step towards fully
self-consistent models. So far we have investigated the relativistic
hydrodynamic flow that is triggered by the deposition of energy near
the BH-torus system. However, this energy release was prescribed
according to a defined functional behavior instead of being linked to
the neutrino emission of the evolving accretion torus. We are in the
process of removing this limitation by preparing two-dimensional
relativistic simulations of the viscosity-driven evolution of BH-torus
systems with a simplified, but consistent treatment of neutrino
transport and $\nu\bar\nu$-annihilation.

\vskip 0.4cm
\begin{acknowledgements}
 We thank M. Ruffert for providing us with data to construct our
initial models.  This work was supported by the
Sonderforschungsbereich 375 ``Astro-Teilchenphysik'' and the
Sonderforschungsbereich-Transregio 7 ``Gravitationswellenastronomie''
of the Deutsche Forschungsgemeinschaft. M.A.A. acknowledges the
partial support of the Spanish Ministerio de Ciencia y Tecnolog\'{\i}a
(AYA2001-3490-C02-C01) and the possibility of making use of the
SGI-Altix computer of the University of Valencia.
\end{acknowledgements}

\bibliographystyle{aa}
\bibliography{general}

\begin{thebibliography}{74}
\expandafter\ifx\csname natexlab\endcsname\relax\def\natexlab#1{#1}\fi

\bibitem[{{Aloy} {et~al.}(2003){Aloy}, {Mart{\'{\i}}}, {G{\'o}mez}, {Agudo},
  {M{\"u}ller}, \& {Ib{\' a}{\~n}ez}}]{Alo03}
{Aloy}, M., {Mart{\'{\i}}}, J., {G{\'o}mez}, J., {et~al.} 2003, Astrophysical
  Journal, 585, L109

\bibitem[{Aloy {et~al.}(2000)Aloy, G{\'o}mez, Ib{\'a}{\~n}ez, Mart{\'{\i}}, \&
  M{\"u}ller}]{Alo00}
Aloy, M.~A., G{\'o}mez, J.-L., Ib{\'a}{\~n}ez, J.~M., Mart{\'{\i}}, J.~M., \&
  M{\"u}ller, E. 2000, Astrophysical Journal, 528, L85

\bibitem[{Aloy {et~al.}(1999)Aloy, Ib{\'a}{\~n}ez, Mart{\'{\i}}, \&
  M{\"u}ller}]{Alo99a}
Aloy, M.~A., Ib{\'a}{\~n}ez, J.~M., Mart{\'{\i}}, J.~M., \& M{\"u}ller, E.
  1999, Astrophysical Journal Supplement Series, 122, 151

\bibitem[{{Aloy} \& {Mart{\'{\i}}}(2002)}]{AM02}
{Aloy}, M.~A. \& {Mart{\'{\i}}}, J.~M. 2002, in LNP Vol. 589: Relativistic
  Flows in Astrophysics, 197--+

\bibitem[{Bal\'azs {et~al.}(2003)Bal\'azs, Bagoly, Horv\'ath, M\'esz\'aros, \&
  M\'esz\'aros}]{Balazsetal03}
Bal\'azs, L., Bagoly, Z., Horv\'ath, I., M\'esz\'aros, A., \& M\'esz\'aros, P.
  2003, AA, 401, 129

\bibitem[{Birkinshaw(1991)}]{Bir91}
Birkinshaw, M. 1991, Monthly Notices the Royal Astronomical Society, 252, 505

\bibitem[{{Blandford} \& {Znajek}(1977)}]{BZ77}
{Blandford}, R.~D. \& {Znajek}, R.~L. 1977, Monthly Notices the Royal
  Astronomical Society, 179, 433

\bibitem[{{Bremer} {et~al.}(1998){Bremer}, {Krichbaum}, {Galama},
  {Castro-Tirado}, {Frontera}, {van Paradijs}, {Mirabel}, {Costa}, {Hanlon}, \&
  {Parmar}}]{BK98}
{Bremer}, M., {Krichbaum}, T.~P., {Galama}, T.~J., {et~al.} 1998, Astronomy and
  Astrophysics, 332, L13

\bibitem[{{Cavallo} \& {Rees}(1978)}]{CR78}
{Cavallo}, G. \& {Rees}, M.~J. 1978, Monthly Notices the Royal Astronomical
  Society, 183, 359

\bibitem[{{Costa} {et~al.}(1997){Costa}, {Frontera}, {Heise}, {Feroci}, {in 't
  Zand}, {Fiore}, {Cinti}, {dal Fiume}, {Nicastro}, {Orlandini}, {Palazzi},
  {Rapisarda}, {Zavattini}, {Jager}, {Parmar}, {Owens}, {Molendi}, {Cusumano},
  {Maccarone}, {Giarrusso}, {Coletta}, {Antonelli}, {Giommi}, {Muller}, {Piro},
  \& {Butler}}]{CX97}
{Costa}, E., {Frontera}, F., {Heise}, J., {et~al.} 1997, NAT, 387, 783

\bibitem[{{Di Matteo} {et~al.}(2002){Di Matteo}, {Perna}, \&
  {Narayan}}]{DMPN02}
{Di Matteo}, T., {Perna}, R., \& {Narayan}, R. 2002, Astrophysical Journal,
  579, 706

\bibitem[{{Drenkhahn}(2002)}]{Drenkhahn02}
{Drenkhahn}, G. 2002, Astronomy and Astrophysics, 387, 714

\bibitem[{{Drenkhahn} \& {Spruit}(2002)}]{DS02}
{Drenkhahn}, G. \& {Spruit}, H.~C. 2002, Astronomy and Astrophysics, 391, 1141

\bibitem[{{Duez} {et~al.}(2004){Duez}, {Liu}, {Shapiro}, \&
  {Stephens}}]{Duezetal04}
{Duez}, M.~D., {Liu}, Y.~T., {Shapiro}, S.~L., \& {Stephens}, B.~C. 2004, PRD,
  69, 104030

\bibitem[{{Duncan} {et~al.}(1986){Duncan}, {Shapiro}, \& {Wasserman}}]{DWS86}
{Duncan}, R.~C., {Shapiro}, S.~L., \& {Wasserman}, I. 1986, APJ, 309, 141

\bibitem[{{Eichler} {et~al.}(1989){Eichler}, {Livio}, {Piran}, \&
  {Schramm}}]{Ei89}
{Eichler}, D., {Livio}, M., {Piran}, T., \& {Schramm}, D.~N. 1989, NAT, 340,
  126

\bibitem[{{Font} \& {Daigne}(2002)}]{FD02}
{Font}, J.~A. \& {Daigne}, F. 2002, Monthly Notices the Royal Astronomical
  Society, 334, 383

\bibitem[{{Frail} {et~al.}(1997){Frail}, {Kulkarni}, {Nicastro}, {Feroci}, \&
  {Taylor}}]{FK97}
{Frail}, D.~A., {Kulkarni}, S.~R., {Nicastro}, S.~R., {Feroci}, M., \&
  {Taylor}, G.~B. 1997, Nature, 389, 261

\bibitem[{{Frail} {et~al.}(2001){Frail}, {Kulkarni}, {Sari}, {Djorgovski},
  {Bloom}, {Galama}, {Reichart}, {Berger}, {Harrison}, {Price}, {Yost},
  {Diercks}, {Goodrich}, \& {Chaffee}}]{Frail01}
{Frail}, D.~A., {Kulkarni}, S.~R., {Sari}, R., {et~al.} 2001, APJ, 562, L55

\bibitem[{{Fryer} {et~al.}(1999){Fryer}, {Woosley}, \&
  {Hartmann}}]{Fryeretal99}
{Fryer}, C.~L., {Woosley}, S.~E., \& {Hartmann}, D.~H. 1999, Astrophysical
  Journal, 526, 152

\bibitem[{{Galama} {et~al.}(1998){Galama}, {Vreeswijk}, {van Paradijs},
  {Kouveliotou}, {Augusteijn}, {Bohnhardt}, {Brewer}, {Doublier}, {Gonzalez},
  {Leibundgut}, {Lidman}, {Hainaut}, {Patat}, {Heise}, {in 't Zand}, {Hurley},
  {Groot}, {Strom}, {Mazzali}, {Iwamoto}, {Nomoto}, {Umeda}, {Nakamura},
  {Young}, {Suzuki}, {Shigeyama}, {Koshut}, {Kippen}, {Robinson}, {de Wildt},
  {Wijers}, {Tanvir}, {Greiner}, {Pian}, {Palazzi}, {Frontera}, {Masetti},
  {Nicastro}, {Feroci}, {Costa}, {Piro}, {Peterson}, {Tinney}, {Boyle},
  {Cannon}, {Stathakis}, {Sadler}, {Begam}, \& {Ianna}}]{GV98}
{Galama}, T.~J., {Vreeswijk}, P.~M., {van Paradijs}, J., {et~al.} 1998, Nature,
  395, 670

\bibitem[{{Goodman}(1986)}]{Go86}
{Goodman}, J. 1986, Astrophysical Journal, 308, L47

\bibitem[{{Hjorth} {et~al.}(2003){Hjorth}, {Sollerman}, {M{\o}ller}, {Fynbo},
  {Woosley}, {Kouveliotou}, {Tanvir}, {Greiner}, {Andersen}, {Castro-Tirado},
  {Castro Cer{\'o}n}, {Fruchter}, {Gorosabel}, {Jakobsson}, {Kaper}, {Klose},
  {Masetti}, {Pedersen}, {Pedersen}, {Pian}, {Palazzi}, {Rhoads}, {Rol}, {van
  den Heuvel}, {Vreeswijk}, {Watson}, \& {Wijers}}]{Hjorth03}
{Hjorth}, J., {Sollerman}, J., {M{\o}ller}, P., {et~al.} 2003, NAT, 423, 847

\bibitem[{{Janka} {et~al.}(1999){Janka}, {Eberl}, {Ruffert}, \&
  {Fryer}}]{Jetal99}
{Janka}, H.-T., {Eberl}, T., {Ruffert}, M., \& {Fryer}, C.~L. 1999,
  Astrophysical Journal, 527, L39

\bibitem[{{Janka} \& {Ruffert}(2002)}]{JR02}
{Janka}, H.-T. \& {Ruffert}, M. 2002, in ASP Conference Series, Vol. 263,
  Stellar Collisions, Mergers and their Consequences, ed. M.~M. Shara, 333+

\bibitem[{{Kalogera} {et~al.}(2004){Kalogera}, {Kim}, {Lorimer}, {Burgay},
  {D'Amico}, {Possenti}, {Manchester}, {Lyne}, {Joshi}, {McLaughlin}, {Kramer},
  {Sarkissian}, \& {Camilo}}]{Kaloetal04}
{Kalogera}, V., {Kim}, C., {Lorimer}, D.~R., {et~al.} 2004, Astrophysical
  Journal, 601, L179

\bibitem[{{Kohri} \& {Mineshige}(2002)}]{KM02}
{Kohri}, K. \& {Mineshige}, S. 2002, Astrophysical Journal, 577, 311

\bibitem[{{Kouveliotou} {et~al.}(1993){Kouveliotou}, {Meegan}, {Fishman},
  {Bhat}, {Briggs}, {Koshut}, {Paciesas}, \& {Pendleton}}]{Ketal93}
{Kouveliotou}, C., {Meegan}, C.~A., {Fishman}, G.~J., {et~al.} 1993,
  Astrophysical Journal, 413, L101

\bibitem[{{Kumar} \& {Granot}(2003)}]{KG03}
{Kumar}, P. \& {Granot}, J. 2003, APJ, 591, 1075

\bibitem[{{Lee} \& {Ramirez-Ruiz}(2002)}]{LR02}
{Lee}, W.~H. \& {Ramirez-Ruiz}, E. 2002, APJ, 577, 893

\bibitem[{{Levinson} \& {Eichler}(2000)}]{LE00}
{Levinson}, A. \& {Eichler}, D. 2000, PRL, 85, 236

\bibitem[{{MacFadyen} \& {Woosley}(1999)}]{MW99}
{MacFadyen}, A. \& {Woosley}, S.~E. 1999, APJ, 524, 262

\bibitem[{{MacFadyen} {et~al.}(2001){MacFadyen}, {Woosley}, \& {Heger}}]{MWH01}
{MacFadyen}, A., {Woosley}, S.~E., \& {Heger}, A. 2001, APJ, 550, 410

\bibitem[{{Mao} {et~al.}(1994){Mao}, {Narayan}, \& {Piran}}]{MNP94}
{Mao}, S., {Narayan}, R., \& {Piran}, T. 1994, APJ, 420, 171

\bibitem[{{M\'esz\'aros}(2002)}]{Me02}
{M\'esz\'aros}, P. 2002, Ann. Rev. Astron. Astroph., 40, 137

\bibitem[{{Mizuno} {et~al.}(2004){Mizuno}, {Yamada}, {Koide}, \&
  {Shibata}}]{Mizuetal04}
{Mizuno}, Y., {Yamada}, S., {Koide}, S., \& {Shibata}, K. 2004, APJ, 606, 395

\bibitem[{{Mochkovitch} {et~al.}(1993){Mochkovitch}, {Hernanz}, {Isern}, \&
  {Martin}}]{MH93}
{Mochkovitch}, R., {Hernanz}, M., {Isern}, J., \& {Martin}, X. 1993, NAT, 361,
  236

\bibitem[{{Morrison} {et~al.}(2004){Morrison}, {Baumgarte}, \&
  {Shapiro}}]{Morrisonetal04}
{Morrison}, I.~A., {Baumgarte}, T.~W., \& {Shapiro}, S.~L. 2004,
  \texttt{arXiv:astro-ph/0401581}, to appear in \emph{ApJ}

\bibitem[{{Narayan} {et~al.}(2001){Narayan}, {Piran}, \& {Kumar}}]{NPK01}
{Narayan}, R., {Piran}, T., \& {Kumar}, P. 2001, APJ, 557, 949

\bibitem[{{Norris} {et~al.}(2000){Norris}, {Marani}, \& {Bonnel}}]{NMB00}
{Norris}, J.~P., {Marani}, G.~F., \& {Bonnel}, J.~T. 2000, APJ, 534, 248

\bibitem[{{Oechslin} {et~al.}(2004){Oechslin}, {Ury{\=u}}, \&
  {Thielemann}}]{Oechslinetal04}
{Oechslin}, R., {Ury{\=u}}, K., \& {Thielemann}, F.~K. 2004, MNRAS, 349, 1469

\bibitem[{{Pacy{\'n}ski}(1986)}]{Pa86}
{Pacy{\'n}ski}, B. 1986, APJ, 308, L43

\bibitem[{{Pacy{\'n}ski} \& {Wiita}(1980)}]{PW80}
{Pacy{\'n}ski}, B. \& {Wiita}, P.~J. 1980, AA, 88, 23

\bibitem[{{Piran}(1999)}]{Piran99}
{Piran}, T. 1999, Phys. Rep., 314, 575

\bibitem[{{Piran}(2000)}]{Piran00}
{Piran}, T. 2000, Phys. Rep., 333, 529

\bibitem[{{Piro} {et~al.}(1998){Piro}, {Heise}, {Jager}, {Costa}, {Frontera},
  {Feroci}, {Muller}, {Amati}, {Cinti}, {dal Fiume}, {Nicastro}, {Orlandini},
  \& {Pizzichini}}]{PX98}
{Piro}, L., {Heise}, J., {Jager}, R., {et~al.} 1998, AA, 329, 906

\bibitem[{{Pons} {et~al.}(2000){Pons}, {Mart{\'{\i}}}, \& {M{\"u}ller}}]{Po00}
{Pons}, J.~A., {Mart{\'{\i}}}, J.~M., \& {M{\"u}ller}, E. 2000, J. Fluid Mech.,
  422, 125

\bibitem[{{Popham} {et~al.}(1999){Popham}, {Woosley}, \& {Fryer}}]{PWF99}
{Popham}, R., {Woosley}, S.~E., \& {Fryer}, C. 1999, APJ, 518, 356

\bibitem[{{Proga} {et~al.}(2003){Proga}, {MacFadyen}, {Armitage}, \&
  {Begelman}}]{Petal03}
{Proga}, D., {MacFadyen}, A.~I., {Armitage}, P.~J., \& {Begelman}, M.~C. 2003,
  APJ, 599, L5

\bibitem[{{Qian} \& {Woosley}(1996)}]{QW96}
{Qian}, Y.-Z. \& {Woosley}, S.~E. 1996, APJ, 471, 331

\bibitem[{{Rezolla} \& {Zanotti}(2002)}]{RZ02}
{Rezolla}, L. \& {Zanotti}, O. 2002, PRL, 114, 501

\bibitem[{{Rossi} {et~al.}(2002){Rossi}, {Lazzati}, \& {Rees}}]{RLR02}
{Rossi}, E., {Lazzati}, D., \& {Rees}, M.~J. 2002, MNRAS, 332, 945

\bibitem[{{Rosswog} \& {Liebend{\"o}rfer}(2003)}]{RL03}
{Rosswog}, S. \& {Liebend{\"o}rfer}, M. 2003, MNRAS, 342, 673

\bibitem[{{Rosswog} \& {Ramirez-Ruiz}(2002)}]{RR02}
{Rosswog}, S. \& {Ramirez-Ruiz}, E. 2002, MNRAS, 336, 7

\bibitem[{{Rosswog} \& {Ramirez-Ruiz}(2003)}]{RR03}
{Rosswog}, S. \& {Ramirez-Ruiz}, E. 2003, MNRAS, 343, L36

\bibitem[{{Rosswog} {et~al.}(2003){Rosswog}, {Ramirez-Ruiz}, \&
  {Davies}}]{RRD03}
{Rosswog}, S., {Ramirez-Ruiz}, E., \& {Davies}, M.~B. 2003, MNRAS, 345, 1077

\bibitem[{{Rosswog} {et~al.}(2004){Rosswog}, {Speith}, \& {Wynn}}]{RSW04}
{Rosswog}, S., {Speith}, R., \& {Wynn}, G.~A. 2004, MNRAS, 345, mNRAS accepted.
  astro-ph/0403500

\bibitem[{{Ruffert} \& {Janka}(1999)}]{RJ99}
{Ruffert}, M. \& {Janka}, H.-T. 1999, AA, 344, 573

\bibitem[{{Ruffert} \& {Janka}(2001)}]{RJ01}
{Ruffert}, M. \& {Janka}, H.-T. 2001, AA, 380, 544

\bibitem[{{Ruffert} {et~al.}(1996{\natexlab{a}}){Ruffert}, {Janka}, \&
  {Sch{\"a}fer}}]{RJS96}
{Ruffert}, M., {Janka}, H.-T., \& {Sch{\"a}fer}, G. 1996{\natexlab{a}}, AA,
  311, 532

\bibitem[{{Ruffert} {et~al.}(1996{\natexlab{b}}){Ruffert}, {Janka},
  {Takahashi}, \& {Sch{\"a}fer}}]{Retal97}
{Ruffert}, M., {Janka}, H.-T., {Takahashi}, K., \& {Sch{\"a}fer}, G.
  1996{\natexlab{b}}, AA, 319, 122

\bibitem[{{Salmonson}(2000)}]{Salmonson00}
{Salmonson}, J. 2000, APJ, 544, L115

\bibitem[{{Salmonson}(2003)}]{Salmonson03}
{Salmonson}, J. 2003, APJ, 592, 1002

\bibitem[{{Sari} \& {Piran}(1997)}]{SP97}
{Sari}, R. \& {Piran}, T. 1997, APJ, 485, 270

\bibitem[{{Setiawan} {et~al.}(2004){Setiawan}, {Ruffert}, \& {Janka}}]{SRJ04}
{Setiawan}, S., {Ruffert}, M., \& {Janka}, H.-T. 2004, MNRAS, 352, 753

\bibitem[{{Shibata} {et~al.}(2003){Shibata}, {Taniguchi}, \&
  {Ury{\=u}}}]{STU03}
{Shibata}, M., {Taniguchi}, K., \& {Ury{\=u}}, K. 2003, PRD, 68, 084020

\bibitem[{{Shibata} \& {Ury{\=u}}(2000)}]{SU00}
{Shibata}, M. \& {Ury{\=u}}, K. 2000, PRD, 61, 064001

\bibitem[{{Stanek} \& {et al.}(2003)}]{Stanek03}
{Stanek}, K.~Z. \& {et al.} 2003, APJ, 591, L17

\bibitem[{{Thompson} {et~al.}(2001){Thompson}, {Burrows}, \& {Meyer}}]{TBM01}
{Thompson}, T.~A., {Burrows}, A., \& {Meyer}, B.~S. 2001, APJ, 562, 887

\bibitem[{{van Paradijs} {et~al.}(1997){van Paradijs}, {Groot}, {Galama},
  {Kouveliotou}, \& {Strom}}]{PG97}
{van Paradijs}, J., {Groot}, P.~J., {Galama}, T., {Kouveliotou}, C., \&
  {Strom}, R.~G. 1997, NAT, 386, 686

\bibitem[{{Witti} {et~al.}(1994){Witti}, {Janka}, \& {Takahashi}}]{WJT94}
{Witti}, J., {Janka}, H.-T., \& {Takahashi}, K. 1994, AA, 286, 841

\bibitem[{{Woosley}(1993)}]{Wo93}
{Woosley}, S.~E. 1993, Astrophysical Journal, 405, 273

\bibitem[{{Zhang} \& {Meszaros}(2002)}]{ZM02}
{Zhang}, B. \& {Meszaros}, P. 2002, APJ, 571, 876

\bibitem[{{Zhang} \& {MacFadyen}(2003)}]{ZWM03}
{Zhang}, W.~{Woosley}, S.~E. \& {MacFadyen}, A.~I. 2003, APJ, 586, 356

\end{thebibliography}

\end{document}